\UseRawInputEncoding
\documentclass[12pt,preprint]{aastex63}

\shorttitle{FLARE EXPANSION AND RARE RADIO BURSTS} \shortauthors{Zemanov\'a et al.}
\begin{document}

\title{FLARE EXPANSION TO A MAGNETIC ROPE ACCOMPANIED BY RARE RADIO BURSTS}

\author[0000-0002-7565-5437]{Alena Zemanov\'a}
\affiliation{Astronomical Institute of the Academy of Sciences of the Czech Republic\\ 
Fri\v{c}ova 298, CZ-25165 Ond\v{r}ejov, Czech Republic}
 \email{alena.zemanova@asu.cas.cz}

\author[0000-0002-3963-8701]{Marian Karlick\'y}
\affiliation{Astronomical Institute of the Academy of Sciences of the Czech Republic\\
Fri\v{c}ova 298, CZ-25165 Ond\v{r}ejov, Czech Republic}

\author[0000-0001-9559-4136]{Jana Ka\v{s}parov\'a}
\affiliation{Astronomical Institute of the Academy of Sciences of the Czech Republic\\
Fri\v{c}ova 298, CZ-25165 Ond\v{r}ejov, Czech Republic}

\author[0000-0003-1308-7427]{Jaroslav Dud\'{\i}k}
\affiliation{Astronomical Institute of the Academy of Sciences of the Czech Republic\\
Fri\v{c}ova 298, CZ-25165 Ond\v{r}ejov, Czech Republic}

\begin{abstract}
We present multispectral analysis (radio, H$\alpha$, UV/EUV, and
hard X-ray) of a confined flare from 2015 March 12. This flare started within
the active region NOAA 12 297 and then it expanded into a large preexisting
magnetic rope embedded with a cold filament. The expansion started
with several brightenings located along the rope. This process was
accompanied by a group of slowly positively drifting bursts in the 0.8--2 GHz
range. The frequency drift of these bursts was 45 -- 100 MHz s$^{-1}$.
One of the bursts had an S-like form. During the brightening of the
rope we observed an unique bright EUV structure transverse to
the rope axis. The structure was observed in a broad range of
temperatures and it moved along the rope with the velocity of about
240 km s$^{-1}$. When the structure dissipated, we saw a plasma
further following twisted threads in the rope. The observed slowly
positively drifting bursts were interpreted considering particle beams and we
show that one with the S-like form could be explained by the beam
propagating through the helical structure of the magnetic rope. The bright
structure transverse to the rope axis was interpreted considering
line-of-sight effects and the dissipation-spreading process, which we found
to be more likely.
\end{abstract}

\keywords{plasmas -- Sun: flares -- Sun: radio radiation}

\section{INTRODUCTION}

Solar flares are generally classified as either eruptive or confined
\citep{Fletcher2011}. In X-ray or EUV observations the eruptive flares are
characterized by a cusp-shaped magnetic structure above the flare arcade, while
the confined flares have only a simple loop structure. The eruptive flares are
usually related to an unstable and rising magnetic flux rope below which the
current sheet is formed, where the magnetic reconnection takes place. The
eruptive flares are connected with coronal mass ejections while confined flares
are not. The eruptive flares were often described by the "standard" CSHKP flare
model ~\citep{Carmichael1964,Sturrock1966,Hirayama1974,Kopp1976}, while
recently a more general 3-dimensional model involving slipping magnetic
reconnection was introduced \citep{Aulanier2012_I, Aulanier2013_II,
Janvier2013_III}. Contrary to that, the confined flares are explained by the
loop flare model, in which the energy release occurs inside the
loop~\citep{Spicer1977} or by the model describing interaction of
current-carrying loops \citep{Melrose1997}. Recently, two types of confined
flares were introduced by \cite{Li2019}. The confined flares of the first type
are characterized by more complex magnetic configuration with strong shears,
they involve a stable filament and slipping magnetic reconnection. The second
type of confined flares has an unstable filament that erupts but is still
confined by a strong strapping field and the reconnection occurs in an antiparallel
magnetic field below the filament. Thus, the confined flares could be
associated with appearance of a cusp structure as the eruptive ones; however,
such observations are still rare in the literature
\citep{Liu2014,Gou2015,Hernandez-Perez2019}.

The flaring process is not stationary but involves expansion of the flaring structure
into the neighboring regions. For example, the well-known flare-ribbon
motions in the perpendicular and/or parallel direction to flare ribbons are the most
common types of expansions \citep[e.g.,][]{Fletcher2011,Qiu2017}, when surrounding
coronal structures undergo magnetic reconnection progressively.
Sometimes, the flaring activity can lead to the so-called domino effect, when
after a flare at one location, other flares in the close locations appear
\citep{Zuccarello2009}. Flare ribbons observed at a remote site from
the main flaring loops are also observed in the connection with circular ribbon
flares (e.g. \cite{Masson2009,Wang2012,Liu2019,Liu2020,Devi2020}). Such flares
are connected to the fan-spine magnetic topology, which is also intrinsic to
coronal jets \citep{Pariat2015, Wyper2016a,Wyper2016b}. But an expansion of a
flare into the preexisting magnetic rope as in the present case is a rare
phenomenon according to our knowledge.

Radio emission is an integral part of solar flares and observed
in various frequency ranges and on time scales from fractions of seconds up to
hours. During flares different types of radio bursts (II, III and IV) and their
fine structures can be observed \citep{Krueger1979, McLean1985}. Among them the
most common are type III bursts that are observed at frequencies from about
100\,kHz to 1\,GHz and drift to lower frequencies. They are considered to be a
signature of the electron beams propagating through the solar atmosphere to the
interplanetary space. A review of type III bursts is provided by
\cite{Reid2014}.

Surveys of radio bursts in decimetric wavelengths is presented in papers by
\cite{Isliker1994} and \cite{Jiricka2001}, within 1--3\,GHz and 0.8--2.0\,GHz
frequency ranges, respectively. Some of these bursts are still not well
understood. This is a case of the slowly positively drifting bursts
(SPDBs). They appear in groups or as single bursts, with a duration of an
individual burst from 1 to several seconds and their frequency drift is lower
than about 100\,MHz s$^{-1}$ \citep{Jiricka2001}.
The SPDBs seem to be similar to the reverse type III bursts
\citep{Aschwanden2002} but their frequency drift is much smaller. The majority of
observed SPDBs are connected to solar flares \citep{Jiricka2001}, and they appear many
times at the very beginning of the flares \citep{Benz1986,
Kotrc1999, Kaltman2000, Karlicky2018}. \cite{Kaltman2000} reported on several
SPDBs observed during 3 solar flares in the 0.8--2\,GHz frequency range. They
found frequency drifts of the observed SPDBs to be within the
20--180\,MHz\,s$^{-1}$ range. \cite{Kotrc1999} studied one of those flares. By
combining the radio and spectral plus imaging H$\alpha$ observations, they
explained the observed SPDBs as radio emission generated by downwards
propagating shock waves. Based on numerical simulations of the formation of
thermal fronts in solar flares, \cite{Karlicky2015} proposed that SPDBs
observed in the 1--2 GHz range could be a signature of a thermal front.
Furthermore, \cite{Karlicky2018} reported about the observation of an SPDB
(1.3--2.0\,GHz) observed during the impulsive phase of an eruptive flare. They
found time coincidence between the SPDB occurrence, an appearance of an UV/EUV
multi-thermal plasma blob moving down along the dark H$\alpha$ loop at
approximately 280\,km\,s$^{-1}$, and the observed change of H$\alpha$ profile
at the footpoint of that dark loop. Combining these observations they concluded
that observed SPDB was likely generated by the thermal front formed in front of
the falling EUV blob.

Moreover, in the 230--350\,MHz range \cite{Benz1986} showed five slowly drifting
type III bursts with very low frequency drifts (--16 to --41 MHz\,s$^{-1}$),
connected to hard X-ray flare emissions. They argued that these
bursts cannot be explained by the electron beams considered for typical type III
bursts. They proposed that possible exciters could be proton beams. \cite{Simnett1995}
studied the role of protons during the flares and he argued that the bursts with slow
drift rates and duration of seconds can be manifestations of the proton beams with
energies well below the $\gamma$-ray production threshold.

In the present paper we analyze an M1.4 class flare of 2015 March 12, which
started in the active region (AR) NOAA 12 297 and then it expanded into a large
preexisting neighboring magnetic rope embedded with a cold filament. This
expansion was connected with a formation of the cusp structure
and an unusual EUV bright structure transverse to the rope axis. During this
process rare SPDBs, including a unique one with an S-like form, were observed.
Therefore, we analyze this flare in detail, especially the rare process of
plasma expansion into the preexisting magnetic rope, and investigate the
circumstances under which SPDBs occur and what agent can be responsible for
their origin.

The paper is structured as follows. In Section 2, we present the data used in this
study, Section 3 describes the evolution of the flare in radio, H$\alpha$, UV/EUV,
and hard X-rays. In Section 4 we provide interpretation and discussion and finally
in Section 5 we present our conclusions.\\

\section{DATA}

In this study we used full disk Sun images provided by the Atmospheric Imaging Assembly (AIA)
\citep{Lemen2012} on a board the {\it Solar Dynamic Observatory} (SDO) \citep{Pesnell2012} acquired
during the 11:30--13:00\,UT time interval.  AIA provides images of the Sun in 7 EUV filters
(94, 131, 171, 193, 211, 304, and 335\,\AA), plus two UV filters (1600 and 1700\,\AA), and WL
continuum. The cadence of the data was 12s and spatial resolution of 1.5\arcsec. The AIA data
were processed using the standard Solar Software (SSW) routine {\it aia\_prep.pro} and the
stray-light has been removed by the method of \cite{Poduval2013}.

\textit{Reuven Ramaty High Energy Solar Spectroscopic Imager} (\textit{RHESSI}) observed the
impulsive phase of the M1.4 class flare during the time interval 12:08-12:14\,UT and part of
its late gradual phase at 12:34--13:10\,UT. \textit{RHESSI} data were used to synthetize hard
X-ray images by VIS\_CS algorithm \citep{Felix2017} in standard energy channels: 6--12, 12--25,
25--50\,keV with temporal cadence of 20\,s. We also used the soft X-ray observations made by
{\it Geostationary Operational Environmental Satellite 15 (GOES 15)}\footnote{www.nasa.gov/content/goes-overview/index.html}.

In radio band we analyzed observations from the 800--2000\,MHz Ond\v{r}ejov radiospectrograph
with the time resolution 0.01\,s \citep{Jiricka1993,Jiricka2008}.

To show the line-of-sight magnetic field of the AR NOAA 12 297, we used the full disk magnetograms
acquired by the Helioseismic and Magnetic Imager (HMI) on \textit{SDO} during 11:30--13:00\,UT time
interval. We used full disk magnetograms with cadence of 45s and spatial resolution of 1\arcsec.

For complementary imaging observations we used data produced by an X-ray Telescope (XRT) on board
the {\it Hinode} mission and groundbased synoptic H$\alpha$ images provided by Global Oscillation
Network Group\footnote{https://gong.nso.edu/} (GONG) Program. The XRT observed flaring activity
within 11:39--12:28\,UT time interval, with spatial resolution of about 2\arcsec. The XRT data were
processed by standard SSW routine {\it xrt\_prep.pro} but their coalignment was manually corrected
to match AIA 131\,\AA~images. The GONG H$\alpha$ network obtains the synoptic images using H$\alpha$
filters with bandpass of 0.5--0.6\,\AA~and the nominal resolution of 2$\arcsec$. It provides H$\alpha$
data covering the whole AIA time interval from several stations but not always with a good seeing
conditions. In this study we used H$\alpha$ data observed at Cerro Tololo station which were already
preprocessed by the provider\footnote{https://www.nso.edu/data/nisp-data/h-alpha/} but
lack the images during 12:19:34--12:25:34\,UT time interval.
Further, the H$\alpha$ images were manually coaligned with \textit{SDO}/AIA 304\,\AA~filter images.\\

\section{THE EVOLUTION OF M1.4 FLARE}

\begin{figure*}
\centering
\includegraphics[width=8.6cm,clip,viewport=0 0 495 350]{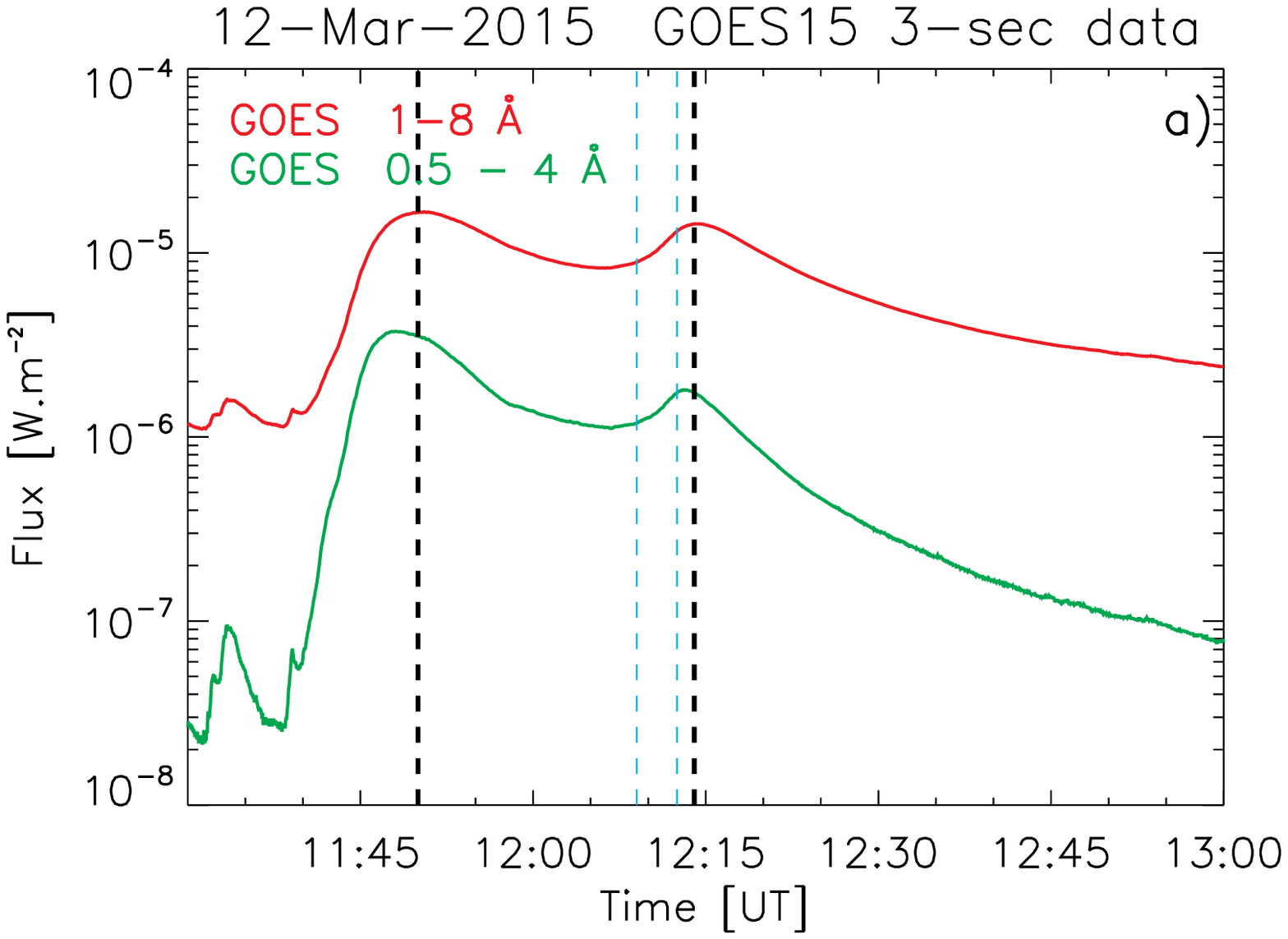}
\includegraphics[width=8.6cm,clip,viewport=0 0 495 350]{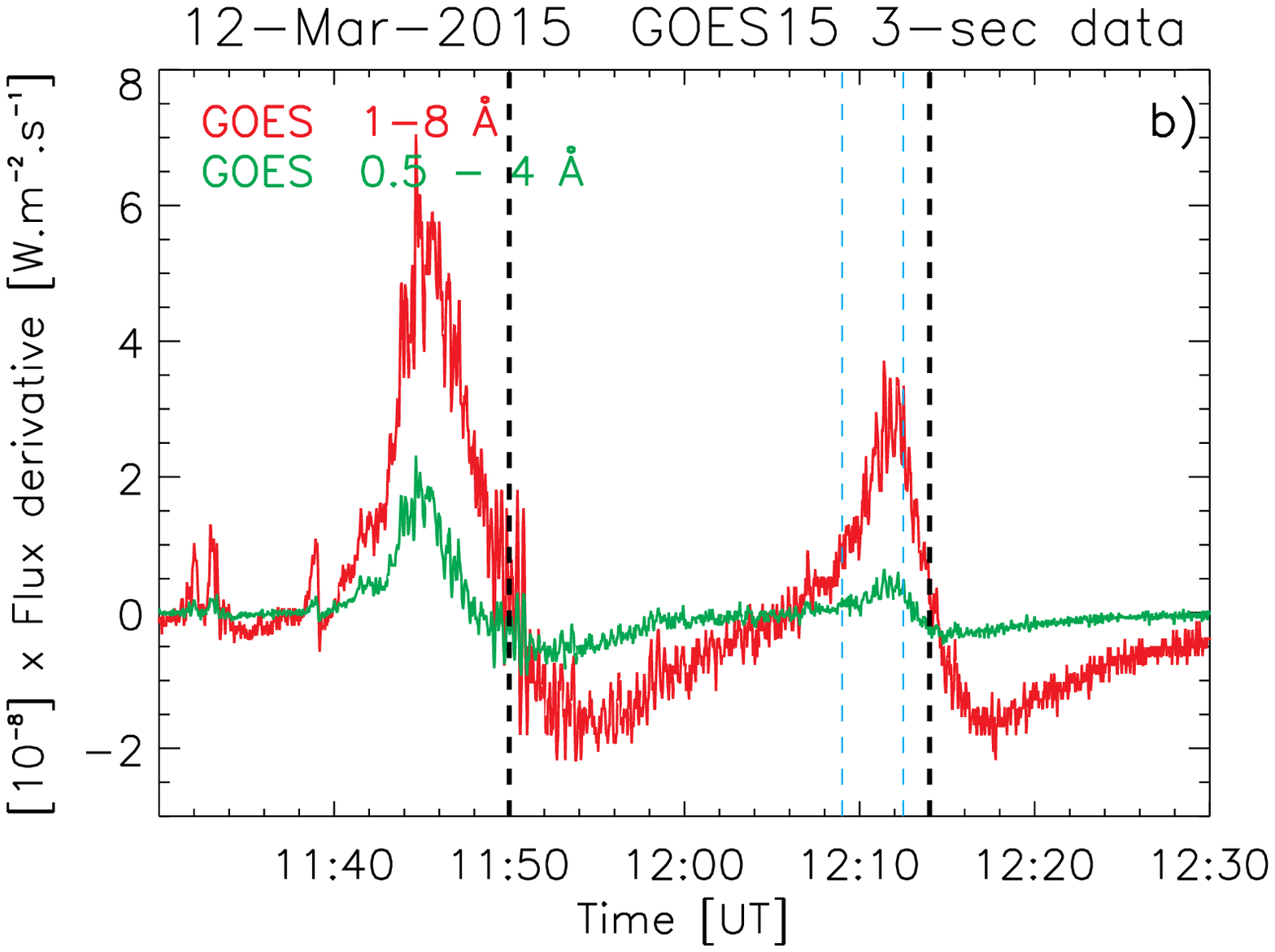}
\caption{(a) Soft X-ray light curves for two M-class flares on 2015 March 12 for
both GOES15 channels and (b) their derivatives. The black dashed lines show
maxima of the flares and the dashed blue lines indicate the time interval when
SPDBs were observed (12:09:00--12:12:30\,UT).}
    \label{fig_goes}
\end{figure*}

In the AR NOAA 12 297 on 2015 March 12 two subsequent flares were observed:
M1.6 SOL2015–03–12T11:50:00 and M1.4 SOL2015–03–12T12:14:00. The M1.4 flare, appeared during
a gradual phase of the previous M1.6 class flare, see Figure~\ref{fig_goes}, where the GOES 15
X-ray fluxes and their time derivatives are plotted. While the first flare was spectroscopically
analyzed in several papers using the Interface Region Imaging Spectrograph (IRIS) \citep{Brannon2016,
Tian2016,Tian2018}, the second flare, according to our knowledge, has not been studied yet. This flare
started within the same AR but later it expanded into a large preexisting magnetic rope outside
the AR. This expansion was associated with rarely observed slowly positively drifting bursts in
the 0.8--2.0\,GHz range with one having an S-like form and an unusual EUV structure transverse
to the rope axis. In the following, the evolution of the M1.4 flare in radio, H$\alpha$, EUV,
and X-rays is described.\\

\subsection{Radio observations}\label{sec:radio}

Figure~\ref{fig_radio}(a) shows the 800--2000 MHz radio spectrum observed during the 12:09--12:13\,UT
time interval. The first weak bursts appeared in the 0.95--1.1 GHz range at about 12:09:16 UT, i.e. at the
beginning of the flare impulsive phase. Then, a group of the slowly positively drifting bursts (SPDBs) was
observed at about 12:10:20 -- 12:12:30\,UT in the whole 0.8--2 GHz range. These bursts are rare as shown in
the statistics of radio bursts and their fine structures \citep{Jiricka2001} and their origin is unknown so
far. They remind the reverse (RS) type III bursts but their frequency drift is much smaller. While the typical
frequency drift of the RS type III bursts in the present frequency range is about $\pm$ 1 GHz s$^{-1}$
\citep{Aschwanden2002}, the frequency drift of those SPDBs is in the 45--100 MHz\,s$^{-1}$ interval. Moreover,
in comparison to the RS type III bursts which appear on the spectrum more or less as a straight line, some of
the SPDBs observed here are curved. For example, see the burst marked by S in Figure~\ref{fig_radio}(b) at
12:10:34 -- 12:10:41\,UT in the 1000--1700 MHz range, which has an unusual S-like form. Such a shape of SPDB
was observed for the first time. Further in the text we will refer to this burst as to the Burst S. Note,
that these are not S-bursts (Solar Storm burst) observed in metric frequency range \citep{McConnell1980, McConnell1982}.\\

\subsection{Flare evolution in EUV, X-rays, and H$\alpha$}\label{sec:euv}

As our radio spectrum does not provide us with spatial information
about radio source locations, we turn to investigate the
origin of these specific SPDBs by following the evolution of the M1.4 flare in EUV,
X-rays, and H$\alpha$ images to search for any exceptional phenomena that could
be connected to the observed bursts.

\begin{figure*}
\centering
     \includegraphics[width=9.7cm,clip,viewport=3 0 410 260]{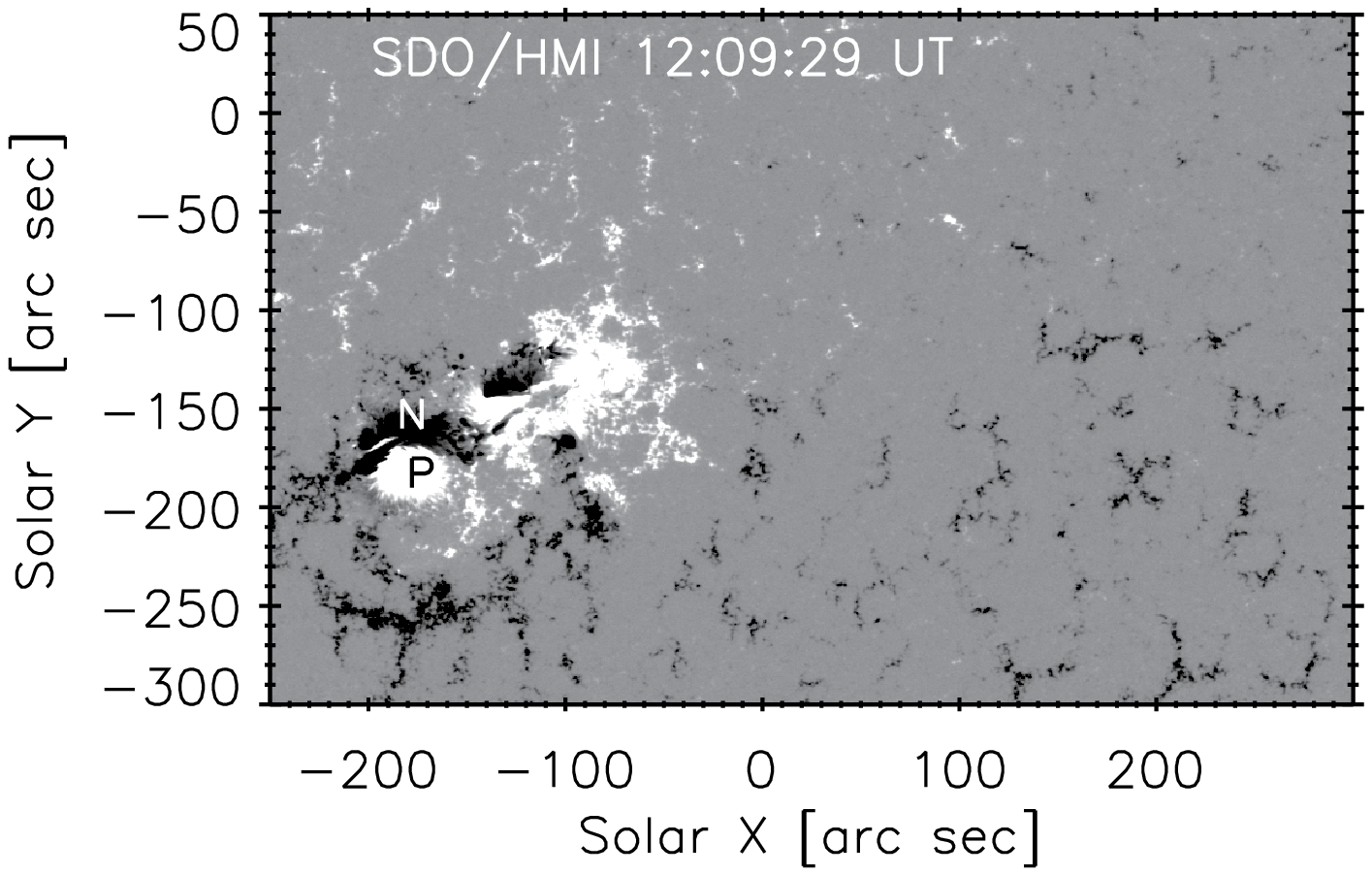}\put(-20,163){\textcolor{white}{\bf a)}}
     \includegraphics[width=7.9cm,clip,viewport=78 0 410 260]{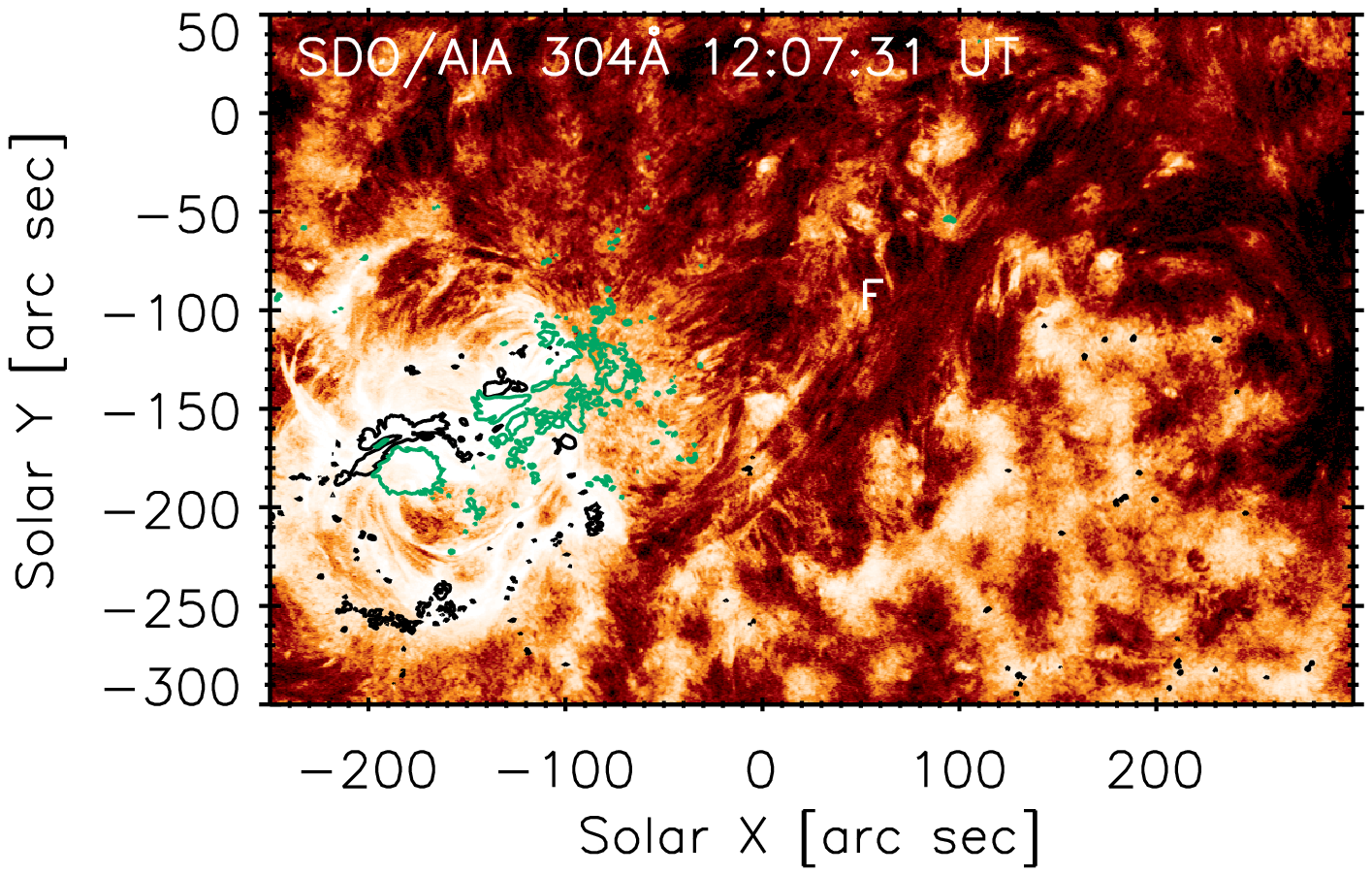}\put(-20,163){\textcolor{white}{\bf b)}}
\caption{Panel (a) shows the \textit{SDO}/HMI line-of-sight magnetic field at 12:09:29\,UT with a big spot of positive polarity
    P and negative polarity N. Panel (b) shows the histogram equalized image of the
    \textit{SDO}/AIA 304\,\AA~filter showing the location of the dark filament marked by F. Contours of the line-of-sight magnetic
    field are overlaid in green and black ($\pm$500G, respectively).}
    \label{fig_mf}
\end{figure*}

\begin{figure*}
        \centering
        \includegraphics[width=9.7cm,clip,viewport=3 50 410 260]{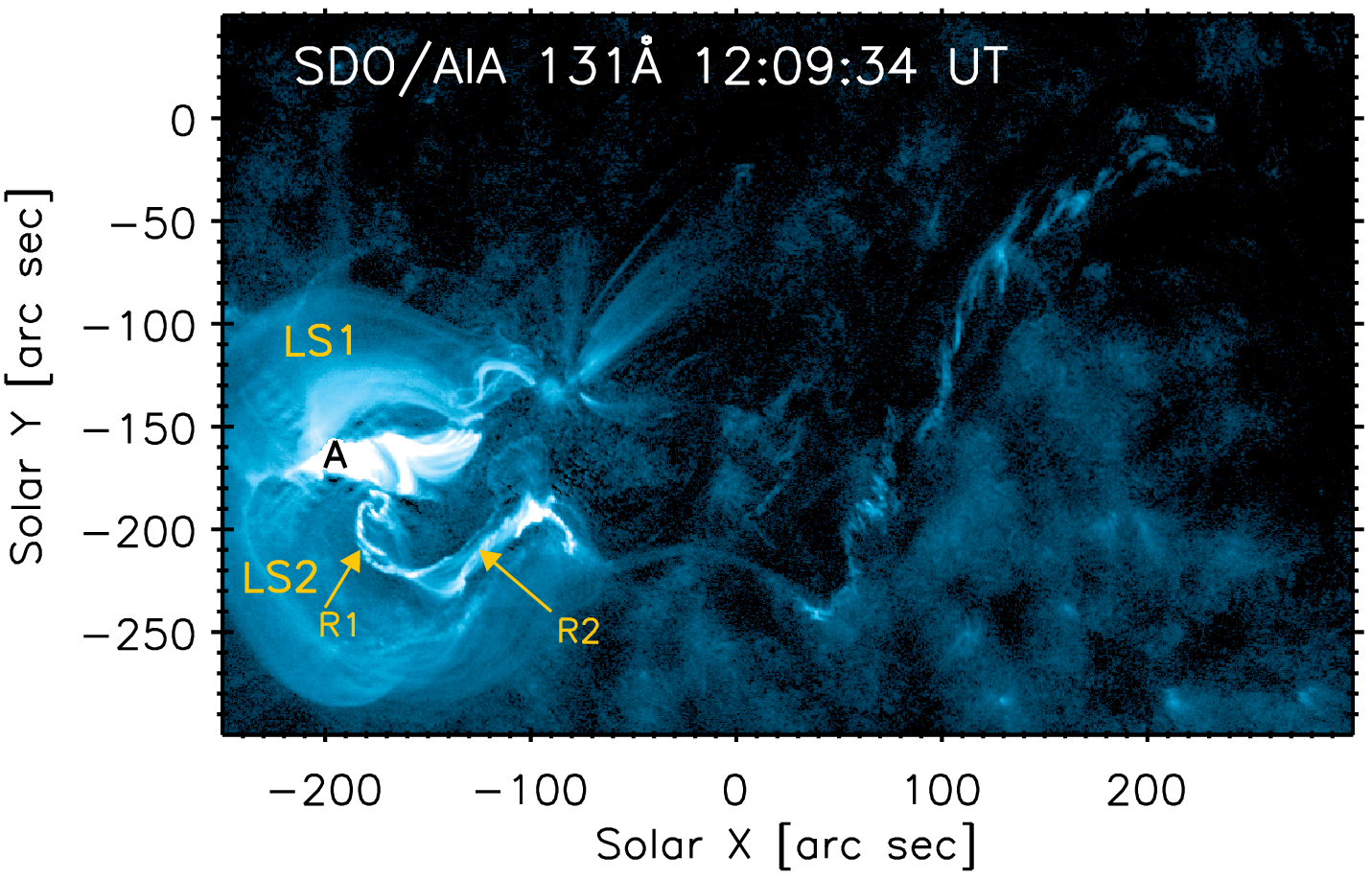}
        \put(-20,130){\textcolor{white}{\bf a)}}
        \includegraphics[width=7.9cm,clip,viewport=78 50 410 260]{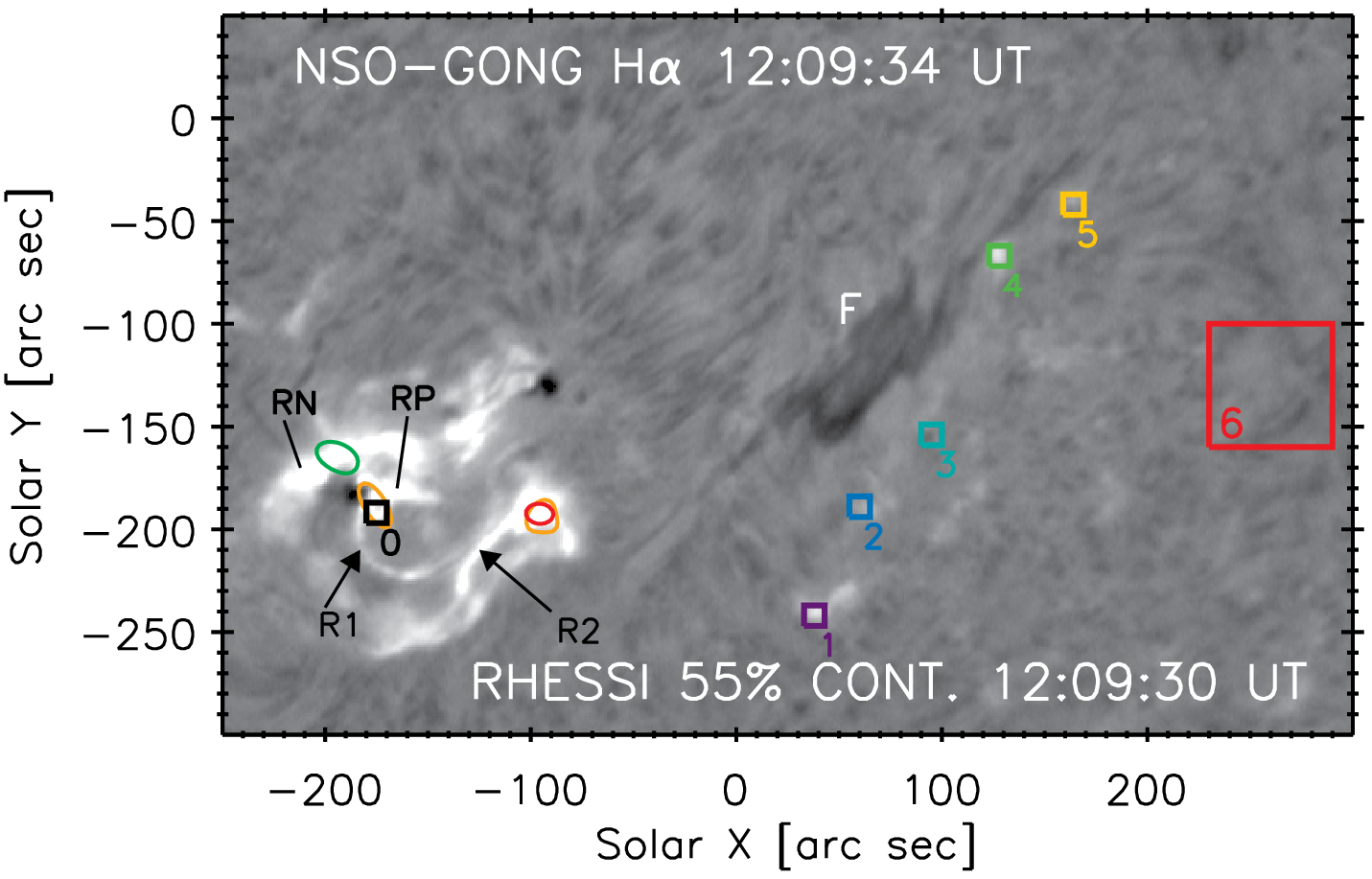}
        \put(-20,130){\textcolor{white}{\bf b)}}

        \includegraphics[width=9.7cm,clip,viewport=3 50 410 260]{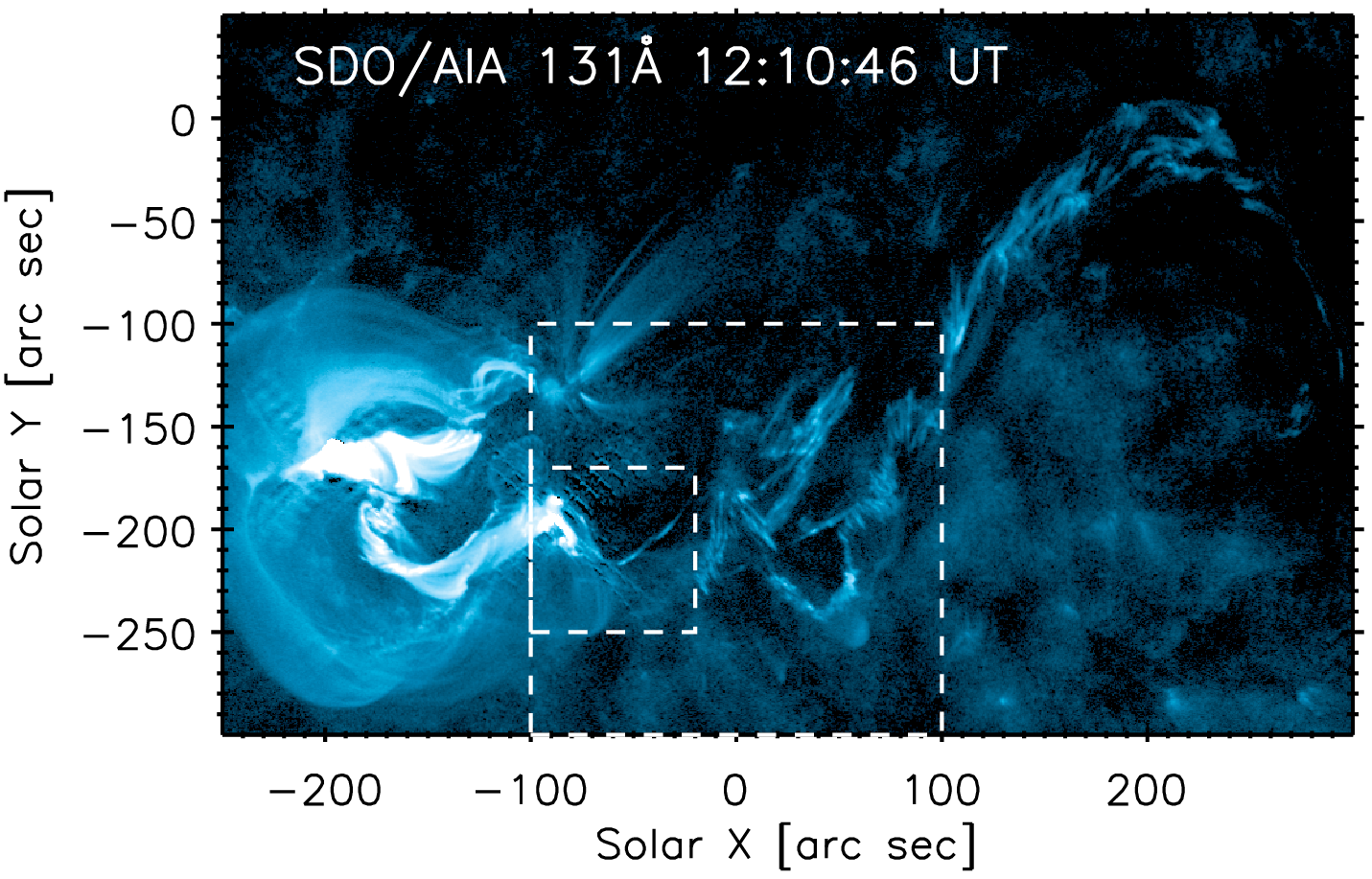}
        \put(-20,130){\textcolor{white}{\bf c)}}
        \includegraphics[width=7.9cm,clip,viewport=78 50 410 260]{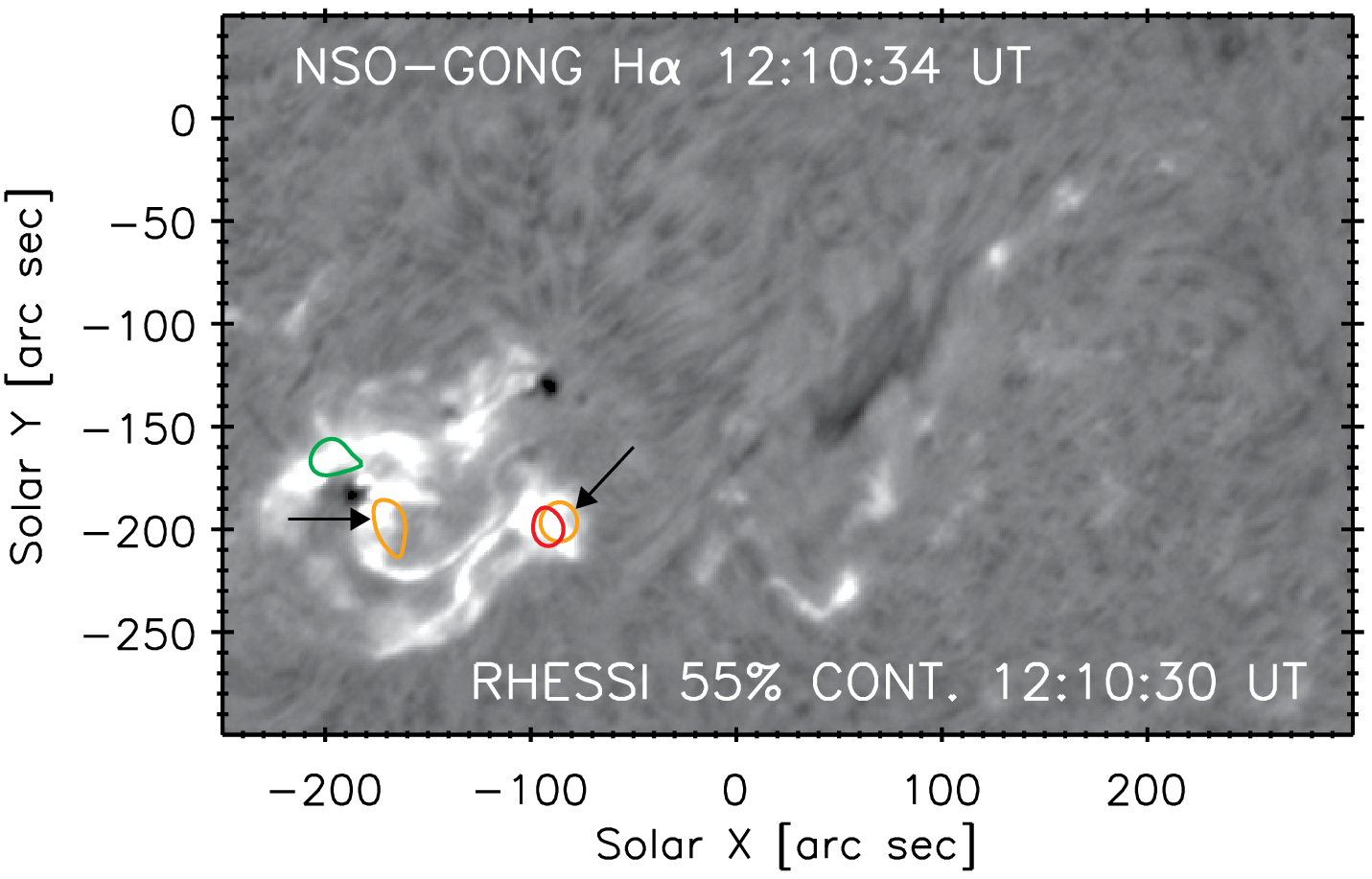}
        \put(-20,130){\textcolor{white}{\bf d)}}

        \includegraphics[width=9.7cm,clip,viewport=3 50 410 260]{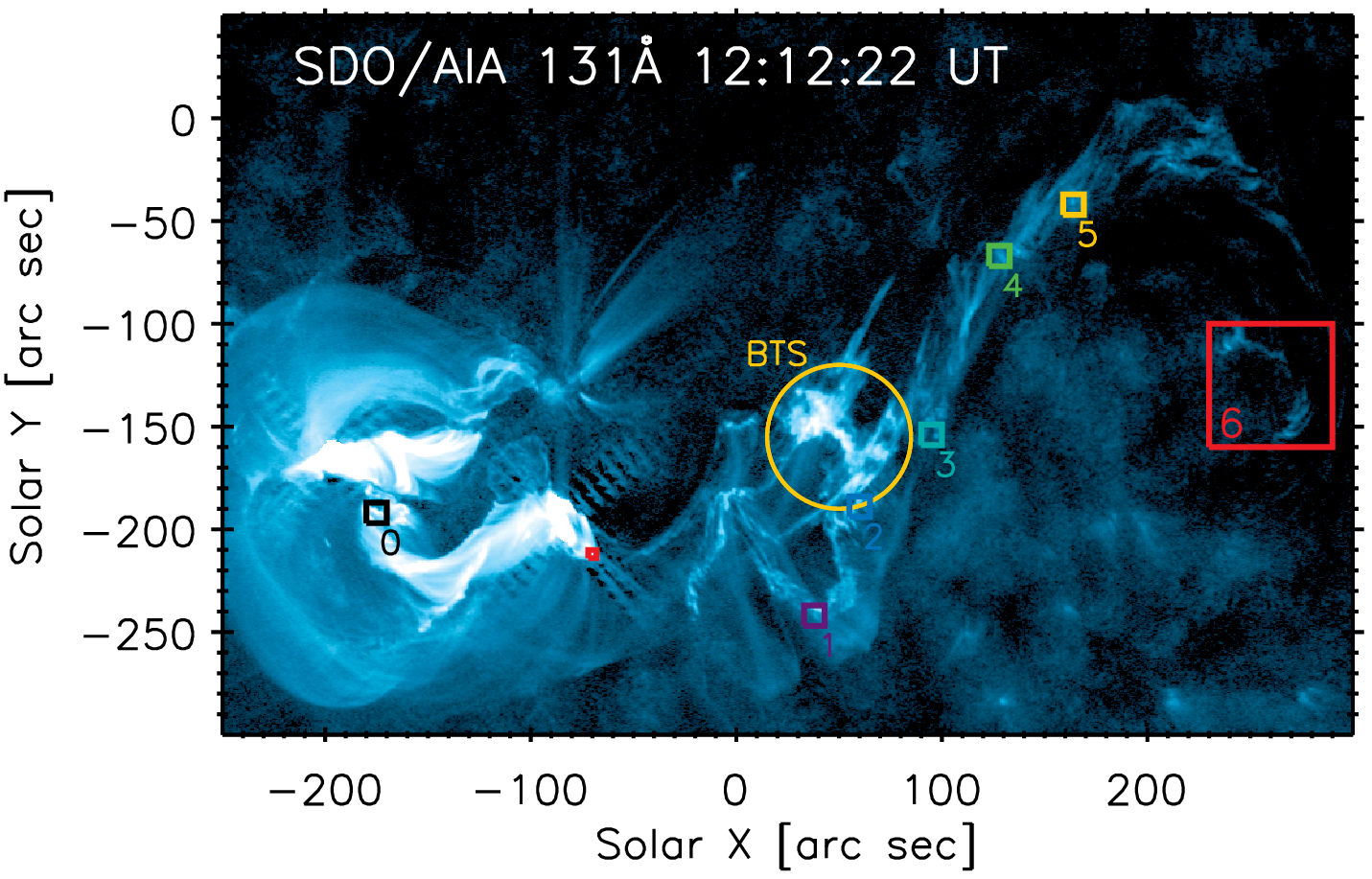}
        \put(-20,130){\textcolor{white}{\bf e)}}
        \includegraphics[width=7.9cm,clip,viewport=78 50 410 260]{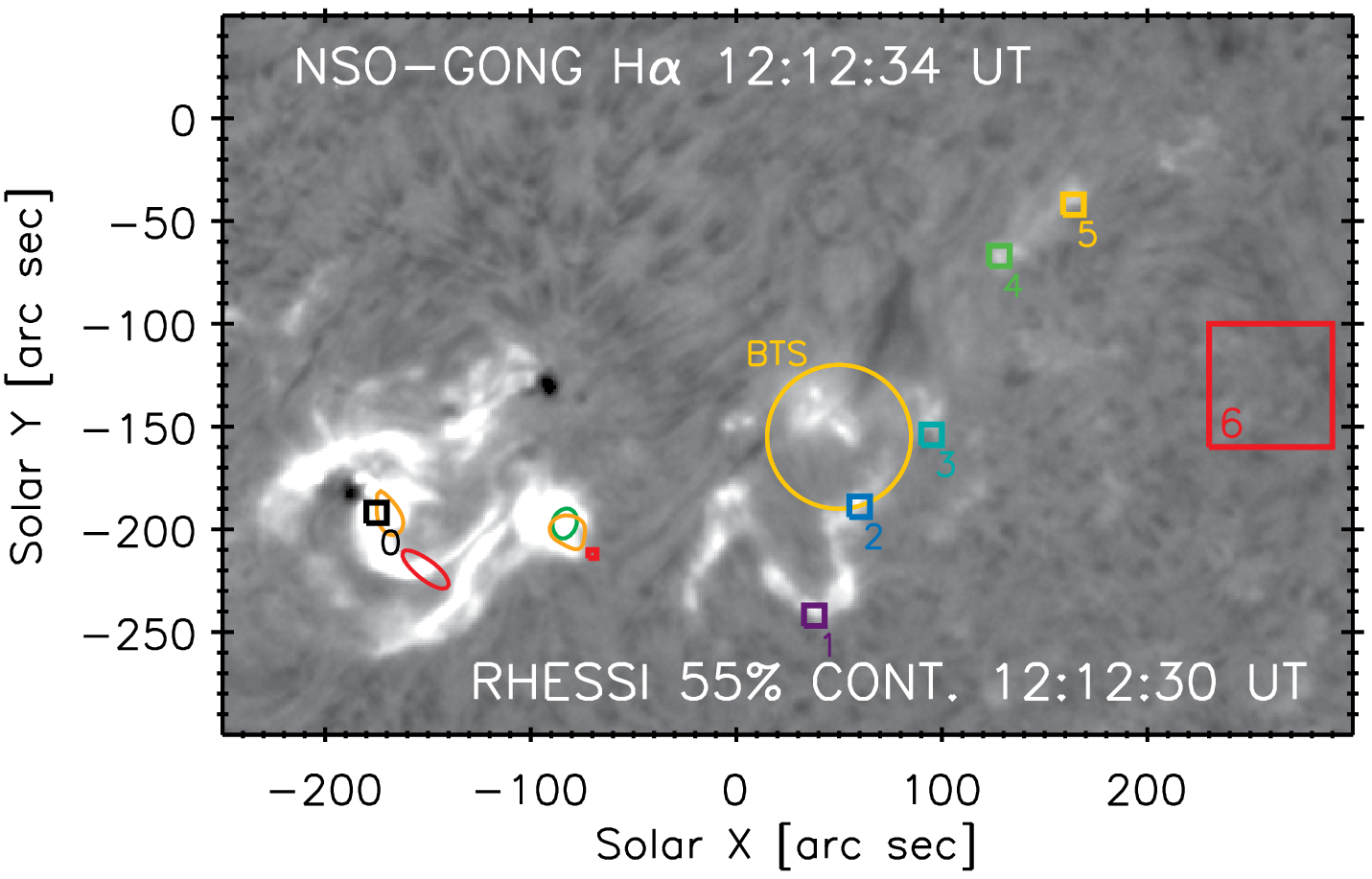}
        \put(-20,130){\textcolor{white}{\bf f)}}

        \includegraphics[width=9.7cm,clip,viewport=3 0 410 260]{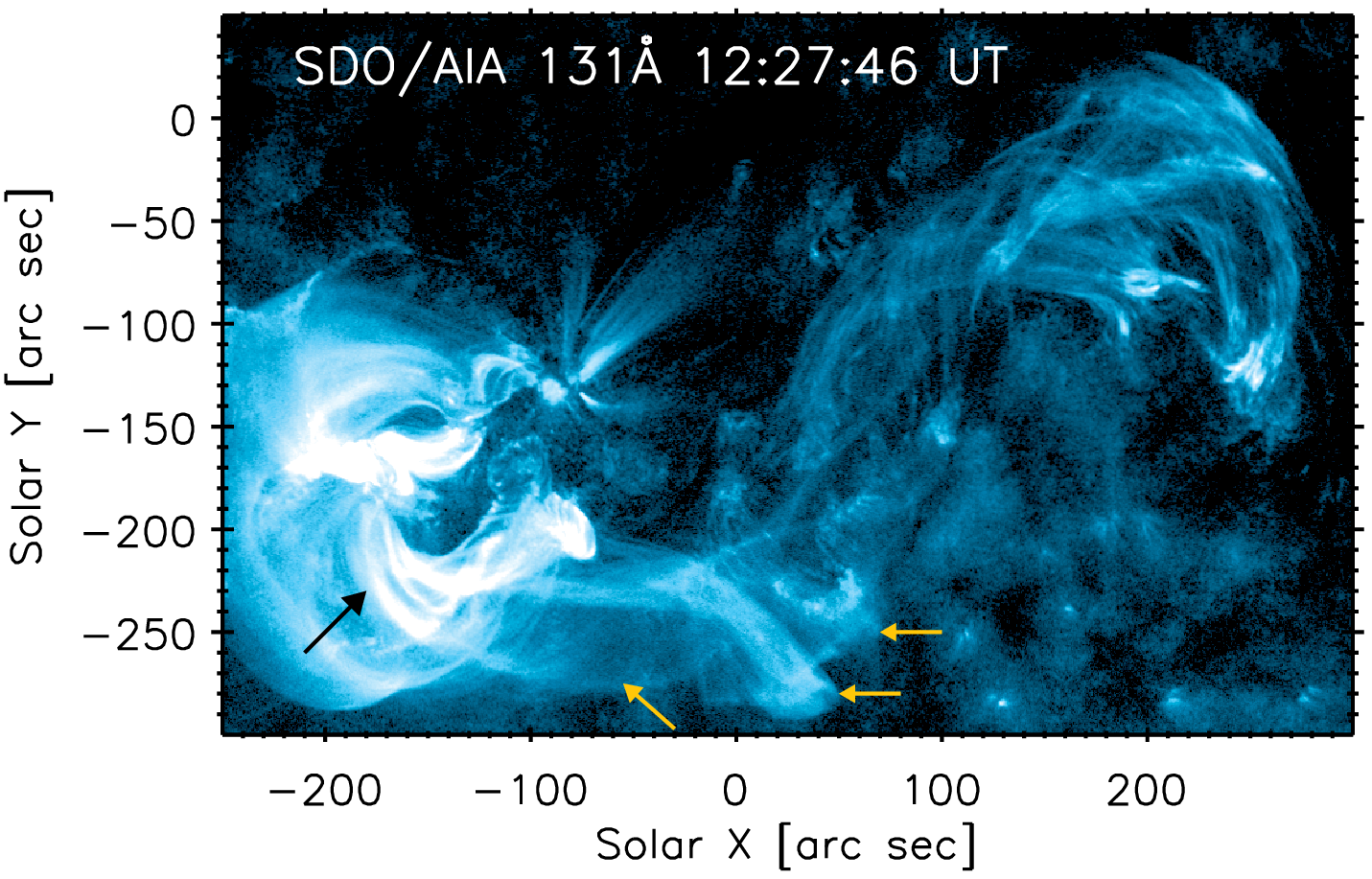}
        \put(-20,163){\textcolor{white}{\bf g)}}
        \includegraphics[width=7.9cm,clip,viewport=78 0 410 260]{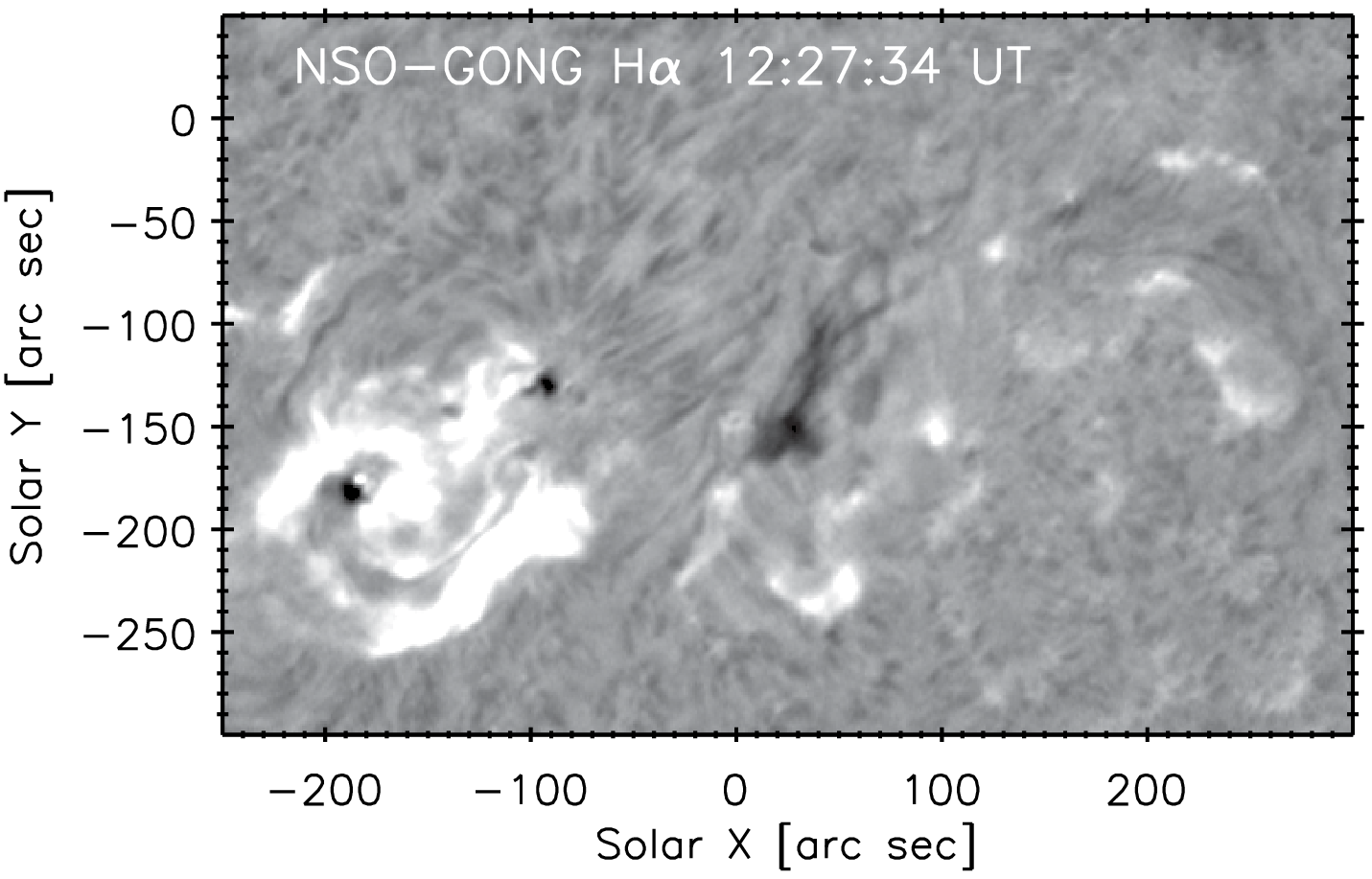}
        \put(-20,163){\textcolor{white}{\bf h)}}

    \caption{Evolution of the M1.4 flare. X-ray sources are shown by green (6--12\,keV),
    orange (12--25\,keV), and red (25--50\,keV) contours. 
    Those located in hooks of R1 and R2 are marked by black arrows in (d).
    Colored Regions 1--6 show locations of brightenings along the F and Region 0 is located at
    the hook of R1. Large dashed rectangle in (c) is FOV for Figures~\ref{fig_reco},~\ref{fig_evol2},
    and~\ref{fig_multi_t}, the small one is FOV for Figure~\ref{fig_cusp}.
    Tiny red square marks the position of the cusp (see Figure~\ref{fig_em}(a)).
    The arrows in (g) point to the flare loops, see Sec. \ref{sec:euv}. An animation of this figure is available.
    It contains 131\AA~and H$\alpha$. Both run approximately from 12:00--13:00\,UT. Ribbons R1, R2 and Regions 0--6 are labeled there.}
    \label{fig_evol1}
\end{figure*}

The M1.4 flare occurred during the gradual phase of the M1.6 flare. The M1.6
flare occurred at the northern part of AR NOAA 12 297 (Figure~\ref{fig_mf}).
During its peak phase an arcade (A) of bright flare loops appeared in EUV
(Figure~\ref{fig_evol1}(a)). In H$\alpha$ (Figure~\ref{fig_evol1}(b)) we observed
two bright ribbons belonging to the arcade A: RN located in negative polarity N
and RP in positive polarity spot P (see Figure~\ref{fig_mf}(a)). We observed also
two large systems of flare loops LS1 and LS2 associated with this flare. They were
rooted at the east side of A, having their conjugate footpoints connected to two
distinct locations within the AR (Figure~\ref{fig_evol1}(a)). These fading loops
and A of M1.6 flare were present during the whole M1.4
flare.

The M1.4 flare occurred at the southern part of the AR (Figure~\ref{fig_evol1}(a) and (b)).
Its impulsive phase was accompanied by SPDBs (Figure~\ref{fig_goes}(b)). It involved
a magnetic rope of the huge EUV filament (F) (Figure~\ref{fig_mf}(b)) lying westward of
AR NOAA 12 297. The F is observed in \textit{SDO}/AIA 304\,\AA~filter (Figure~\ref{fig_mf}(b))
as well as in H$\alpha$ (Figure~\ref{fig_evol1}) but in 304\,\AA~it was much more extended.

At the beginning of M1.4 flare a new pair of J-shaped ribbons, R1 and R2, appeared to the
southwest of A. They were connected by an EUV loop that started to rise (see the the animation
accompanying Figure~\ref{fig_evol1}). The R1 was located in positive polarity of sunspot P,
starting very close to RP of A, and the negative polarity ribbon R2 was located about
100$\arcsec$ west from it (Figure~\ref{fig_evol1}(b)). At about 12:09\,UT we observed a brightening
at the elbow of the hook of R2 (Figure~\ref{fig_evol1}(a), accompanied by a \textit{RHESSI}
X-ray source (Figure~\ref{fig_evol1}(b)). The X-ray source continually moved along the hook of
R2 toward its tip (Figures~\ref{fig_evol1}(b)--(d)--(f)). Simultaneously, a brightening at
the conjugate hook of R1 appeared and was also accompanied by moving X-ray source
(Figure~\ref{fig_evol1}(b)--(d)--(f)). For clarity, the observed X-ray sources located
within the hooks of J-shaped ribbons R1 and R2 are pointed by black arrows in
Figure~\ref{fig_evol1}(d). The X-ray source observed in the hook of R1
(Figure~\ref{fig_evol1}(b)--(d)--(f), orange contour) did not move as much as the one in R2.
An additional soft X-ray source located over the ribbon RN (Figure~\ref{fig_evol1}(b) and (d),
green contour) did not belong to the M1.4 flare. It represents emission from hot flare loops belonging to A.

Slightly earlier, from about 12:08\,UT, several regions bright in EUV started to appear along F,
lying westward of AR NOAA 12 297 (Figures~\ref{fig_evol1}(a) and (b) and see the animation
accompanying Figure~\ref{fig_evol1}).

After 12:10\,UT, the loops connecting R1 and R2 became brighter in EUV and we observed a thin
stripe of plasma coming out of the tip of the hook of R2, see the smaller rectangle in
Figure~\ref{fig_evol1}(c). Later, a small cusp appeared at the tip of R2 (Figure~\ref{fig_evol1}(e)).
Its position is marked by a red square in Figures~\ref{fig_evol1}(e) and (f). The large rectangle in
Figure~\ref{fig_evol1}(c) indicates the position of a bulk of heated plasma which brightened over
the F (Figure~\ref{fig_evol1}(d)). Figure~\ref{fig_evol1}(e) shows that heated plasma, located
over F, formed a bright emission structure seen in the 131\,\AA~filter (circle). The structure was narrow
and oriented transverse to the axis of F. We will refer to this structure as to the bright transverse
structure (BTS). Figure~\ref{fig_evol1}(f) shows that part of the BTS cloud be seen in emission also
in H$\alpha$ (circle). At about 12:12:30\,UT, a new hard X-ray source also appeared over R1
(red contour, Figure~\ref{fig_evol1}(f)).

The narrow BTS appeared to move along the filament axis. Simultaneously, a new ribbon appeared
at the end of EUV filament (red square of Region 6, Figure~\ref{fig_evol1}(e)). Following 12:12:30\,UT
the narrow BTS started to dissipate and increased its width. In the gradual phase of the flare we
already observed a bulk of heated plasma following the twisted filament threads (Figure~\ref{fig_evol1}(g))
and several new flare loops were created: within the AR (black arrow), below the cusp and also longer ones
at other places (yellow arrows). In H$\alpha$ we observed changes in appearance of dark filament threads
compared to the situation at the beginning of the flare (see Figure~\ref{fig_evol1}(h) and (b)
and see the animation accompanying Figure~\ref{fig_evol1}).

As has been described in subsection~\ref{sec:radio}, the main part of SPDBs was observed during
the 12:10:20--12:12:30\,UT time interval (Figure~\ref{fig_radio}(a)). This corresponds well to the
appearance of the cusp and formation of BTS. Therefore, we focus on these phenomena in the following subsection.\\

\subsubsection{The cusp, BTS, and flare loops}\label{subsec:euv}

\begin{figure*}
        \centering
        \includegraphics[width=5.28cm,clip,viewport=0 50 350 318]{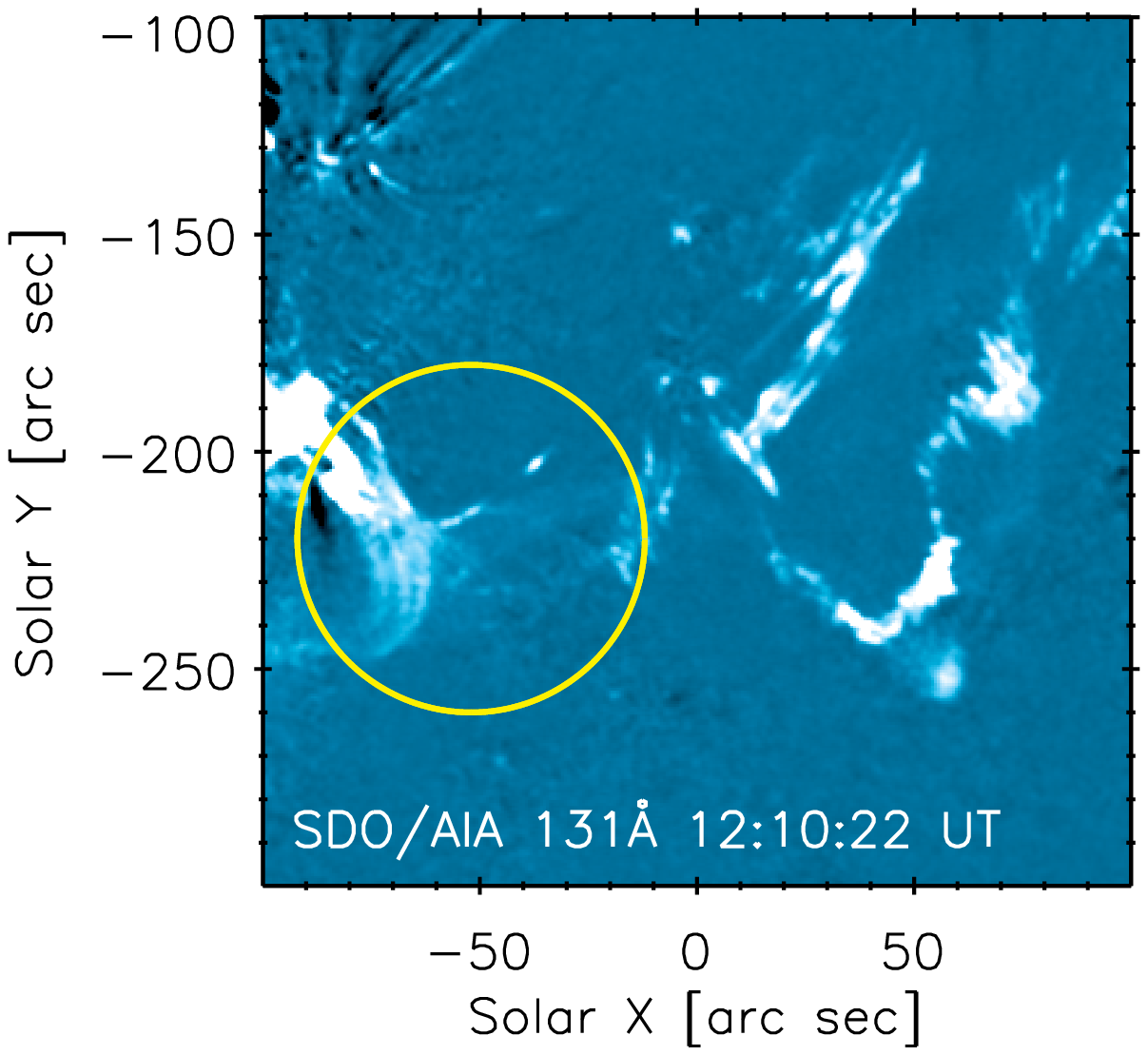}
        \put(-110,103){\textcolor{white}{\bf a)}}
        \includegraphics[width=4.1cm,clip,viewport=78 50 350 318]{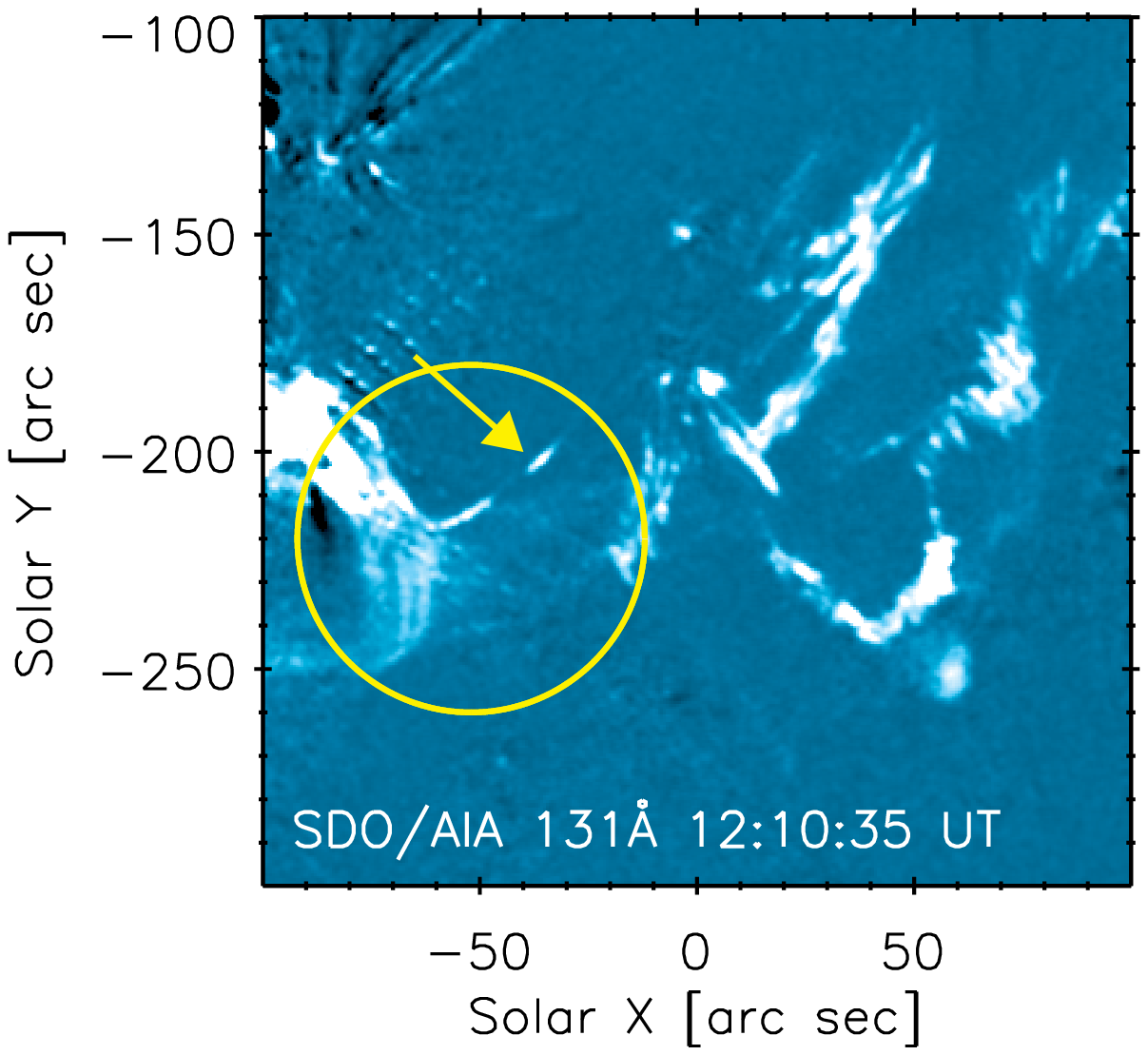}
        \put(-110,103){\textcolor{white}{\bf b)}}
        \includegraphics[width=4.1cm,clip,viewport=78 50 350 318]{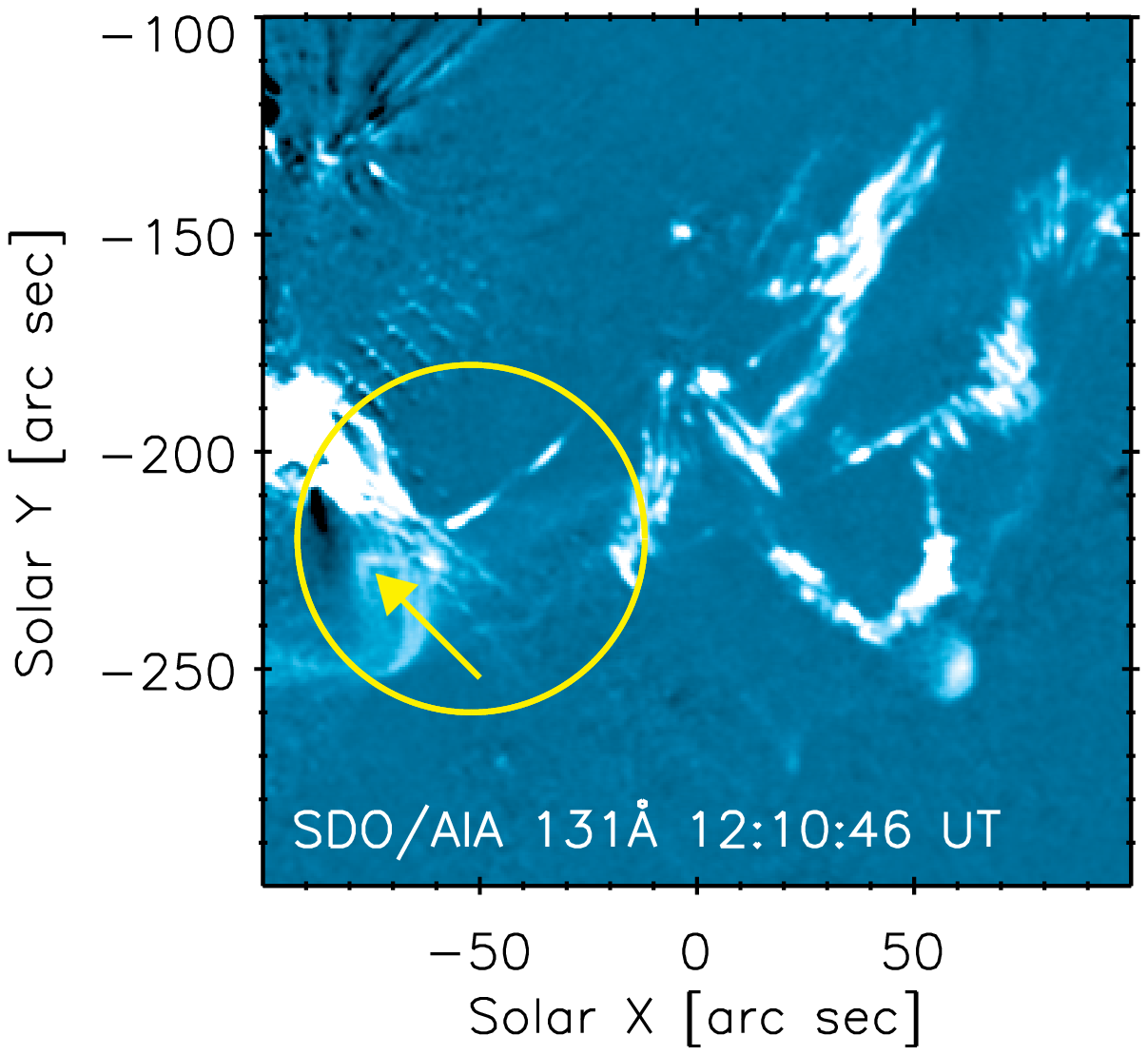}
        \put(-110,103){\textcolor{white}{\bf c)}}
        \includegraphics[width=4.1cm,clip,viewport=78 50 350 318]{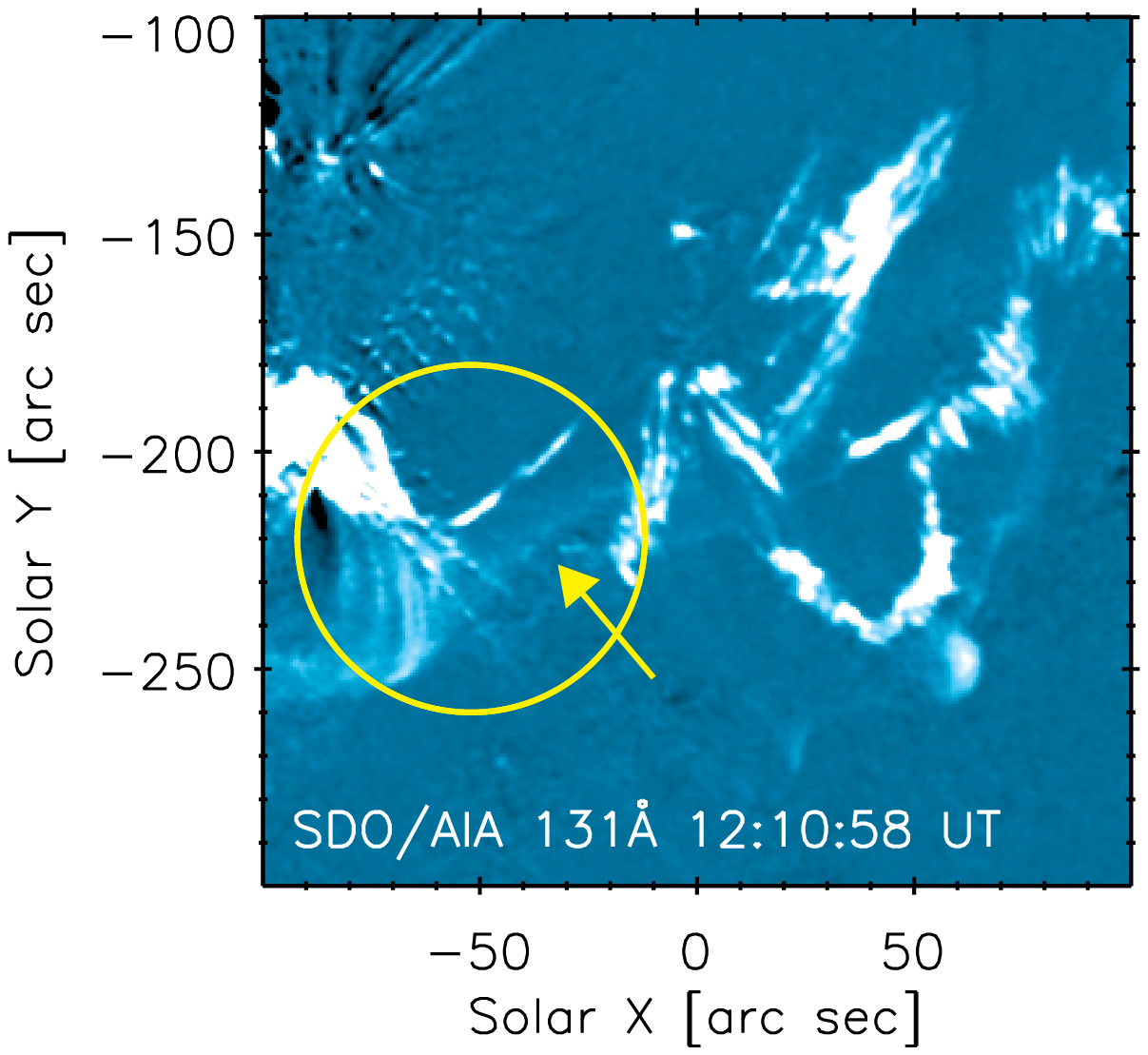}
        \put(-110,103){\textcolor{white}{\bf d)}}

        \includegraphics[width=5.28cm,clip,viewport=0 0 350 318]{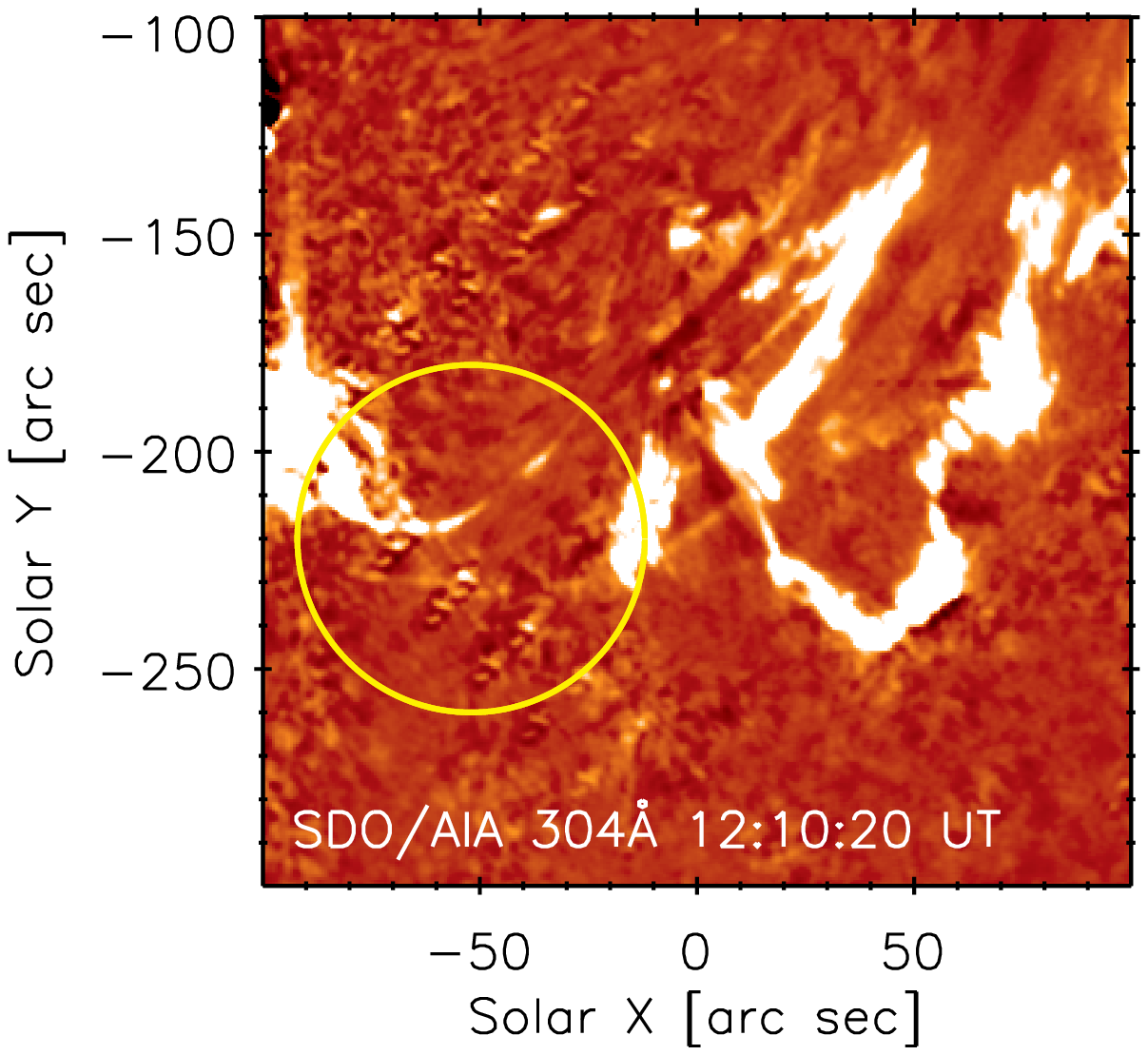}
        \put(-110,125){\textcolor{white}{\bf e)}}
        \includegraphics[width=4.1cm,clip,viewport=78 0 350 318]{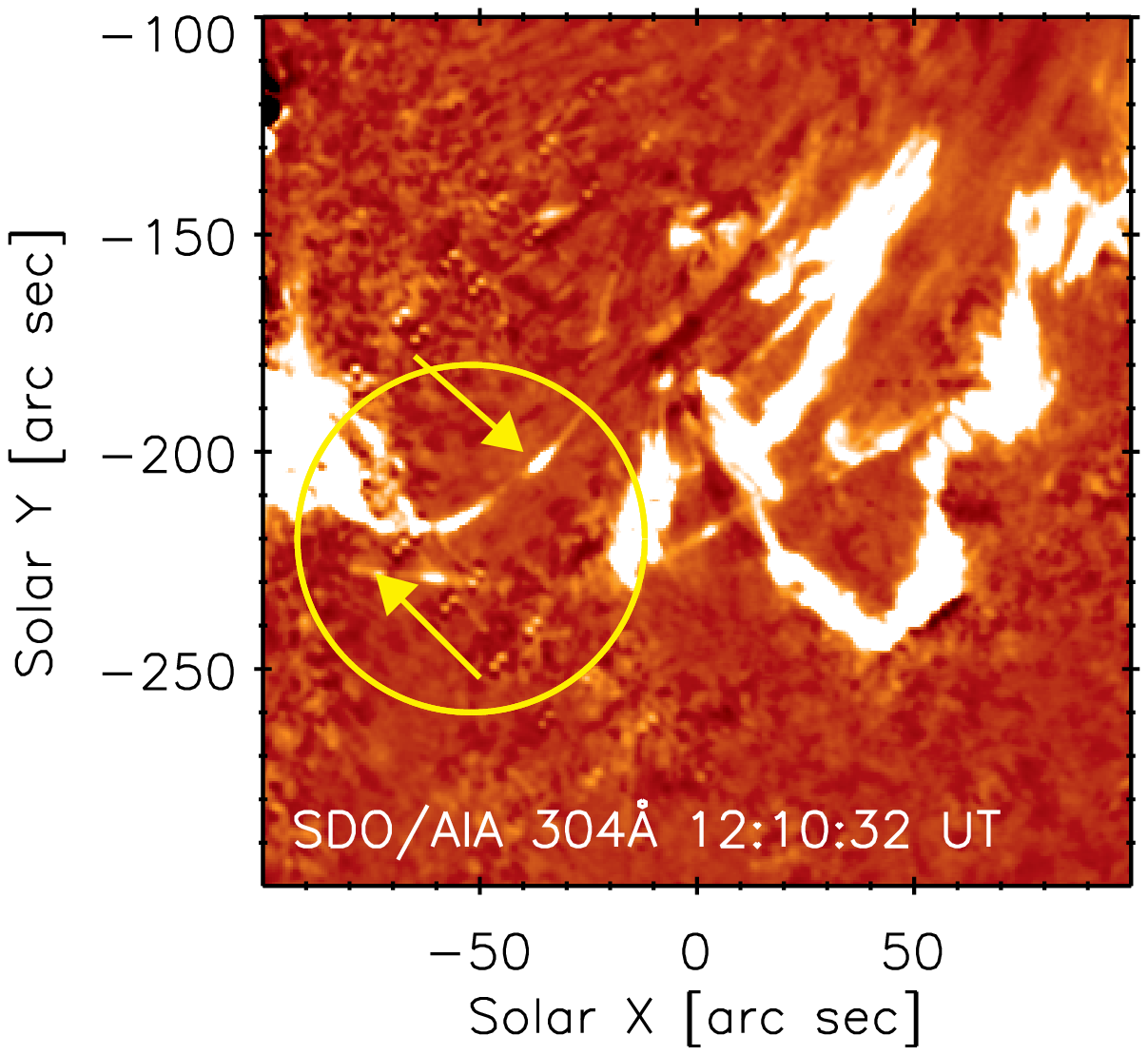}
        \put(-110,125){\textcolor{white}{\bf f)}}
        \includegraphics[width=4.1cm,clip,viewport=78 0 350 318]{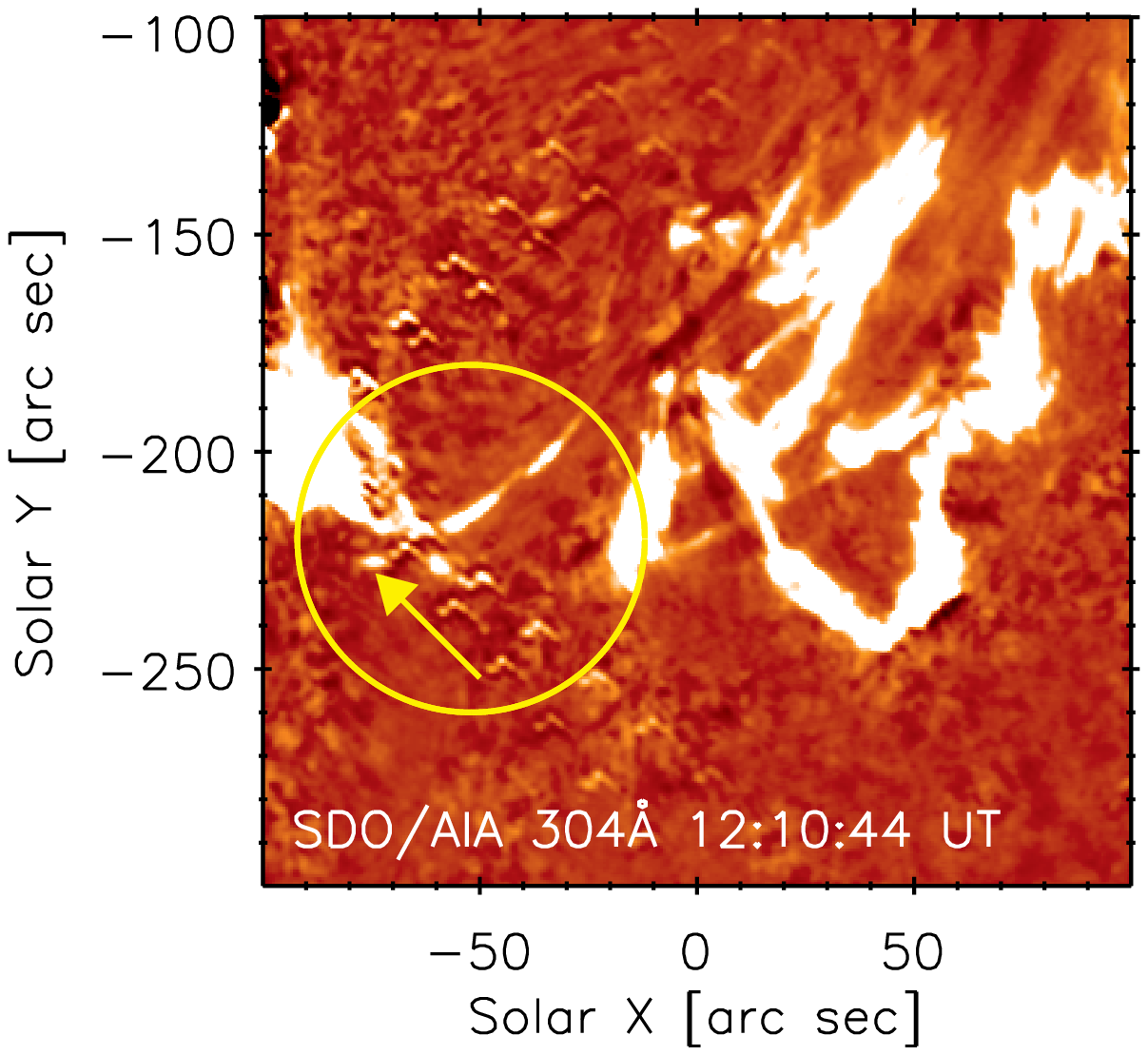}
        \put(-110,125){\textcolor{white}{\bf g)}}
        \includegraphics[width=4.1cm,clip,viewport=78 0 350 318]{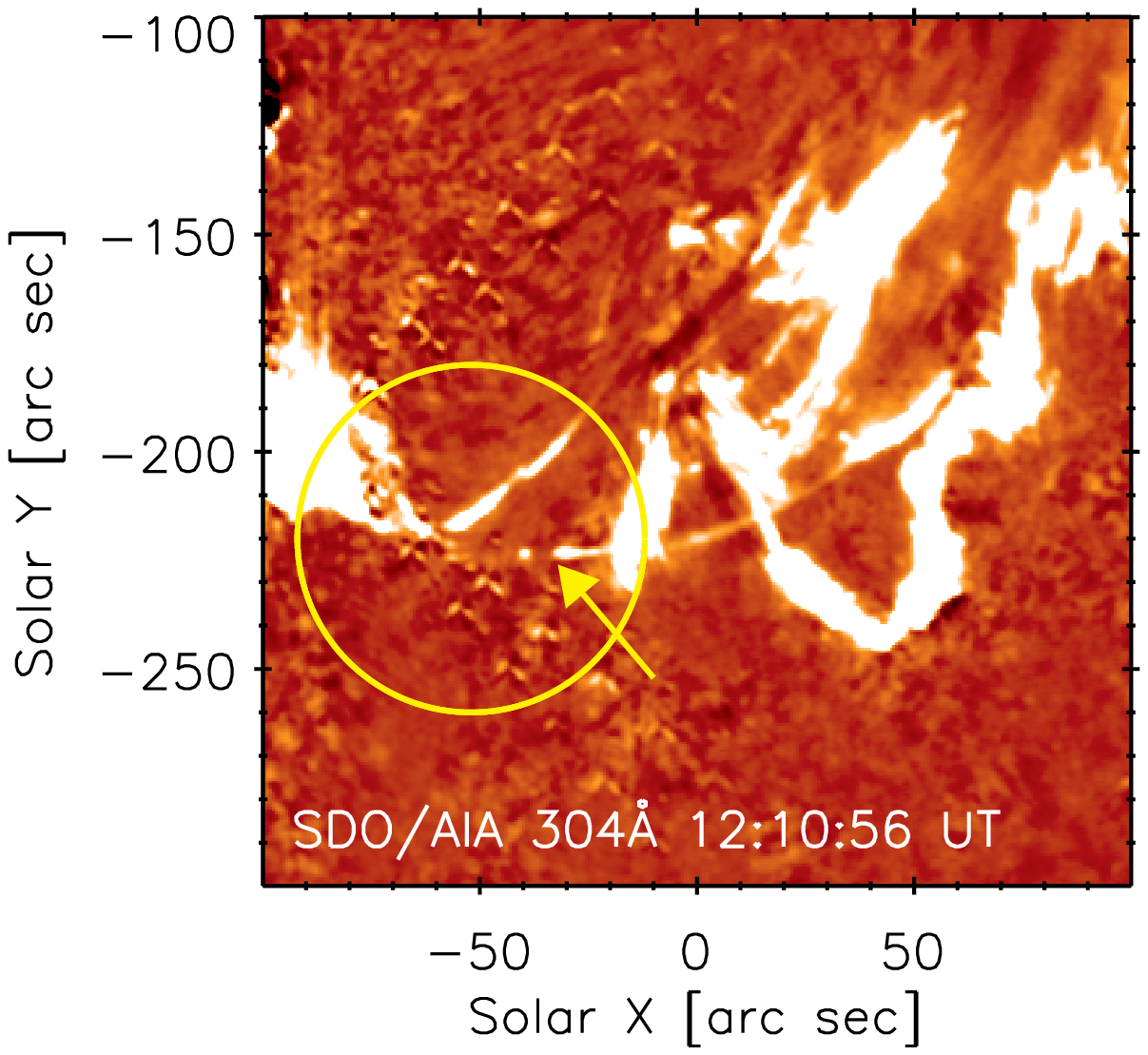}
        \put(-110,125){\textcolor{white}{\bf h)}}

       \caption{Detail of the cusp and plasma ejection during the Burst S.
       (a)--(d) Base difference images in \textit{SDO}/AIA 131\,\AA~filter with a base image at 11:30:10\,UT.
       (e)--(h) Base difference images in \textit{SDO}/AIA 304\,\AA~filter with a base image at 11:30:08\,UT.
        For an explanation see Sec.~\ref{subsec:euv}.} \label{fig_reco}
\end{figure*}

\begin{figure*}
        \centering
        \includegraphics[width=5.28cm,clip,viewport=0 50 350 318]{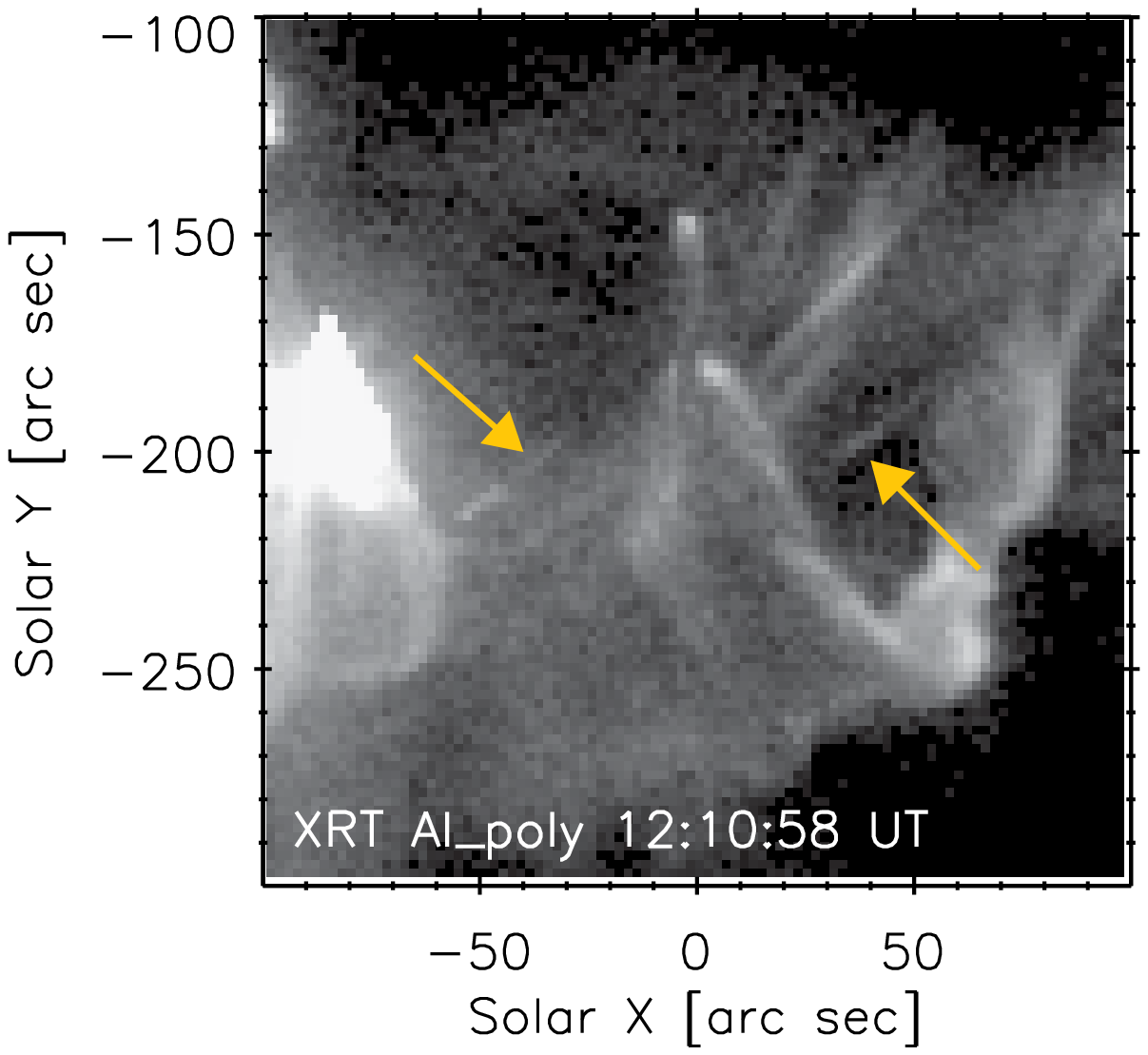}
        \put(-110,103){\textcolor{white}{\bf a)}}
        \includegraphics[width=4.1cm,clip,viewport=78 50 350 318]{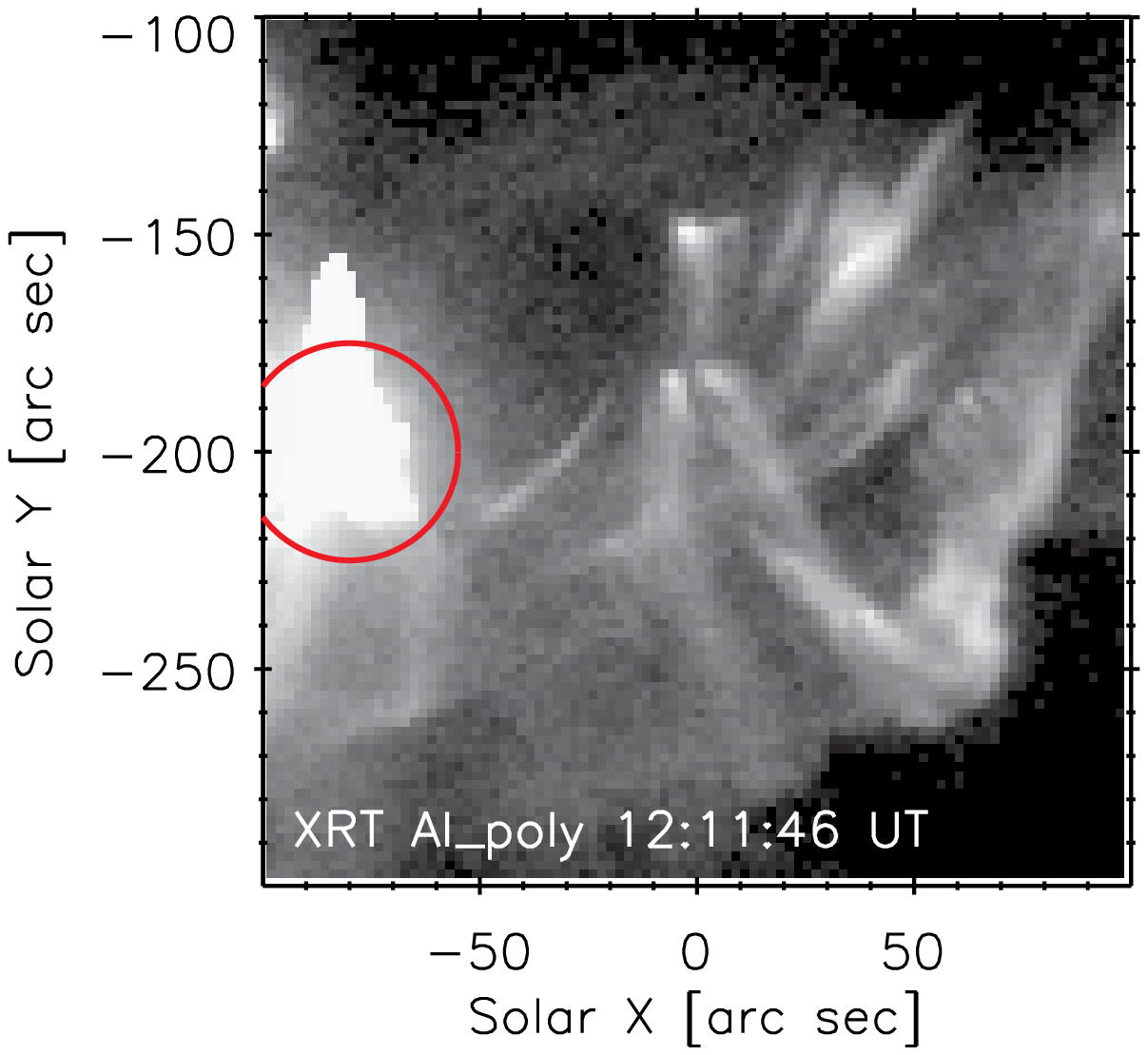}
        \put(-110,103){\textcolor{white}{\bf b)}}
        \includegraphics[width=4.1cm,clip,viewport=78 50 350 318]{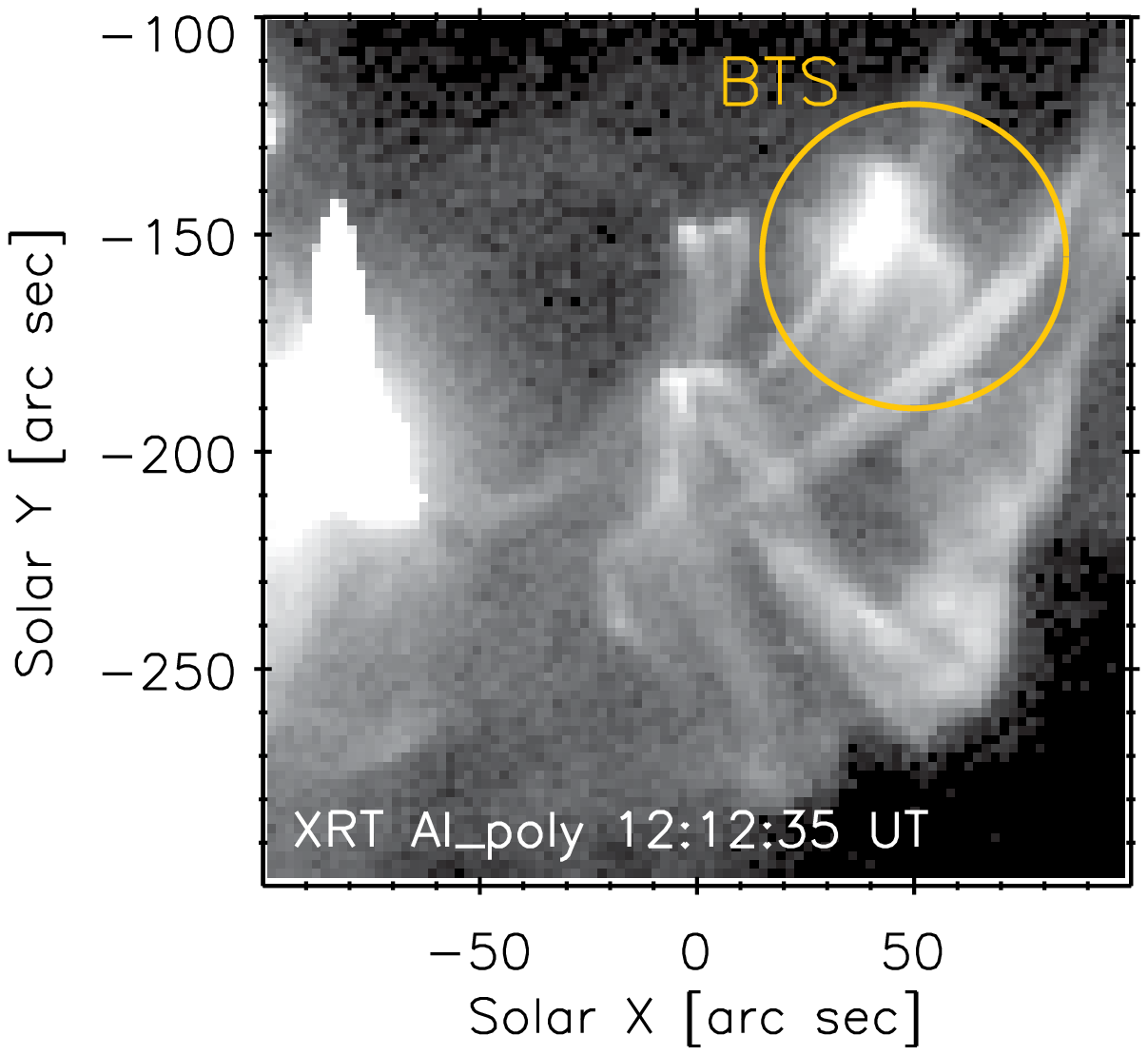}
        \put(-110,103){\textcolor{white}{\bf c)}}
        \includegraphics[width=4.1cm,clip,viewport=78 50 350 318]{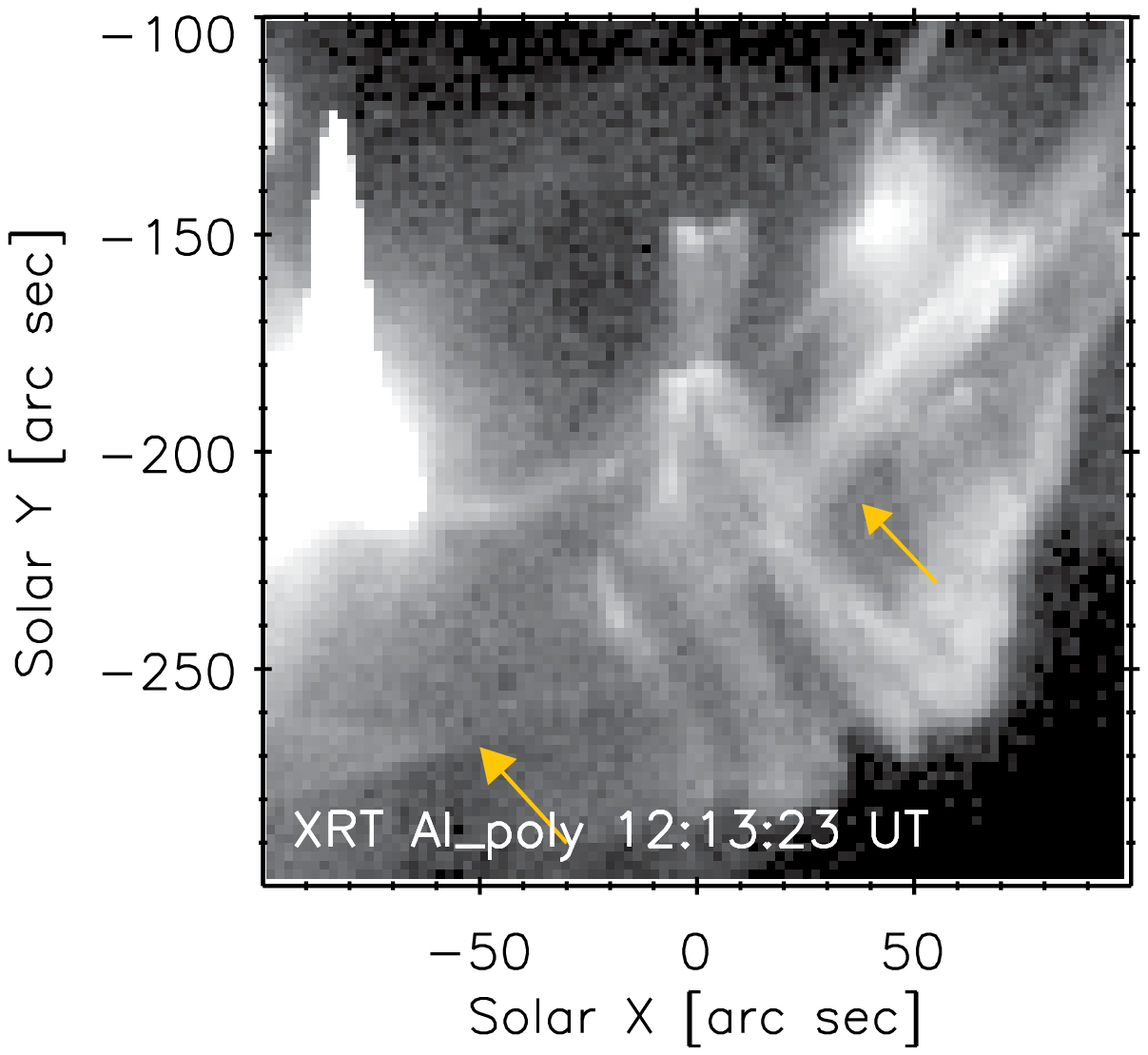}
        \put(-110,103){\textcolor{white}{\bf d)}}

        \includegraphics[width=5.28cm,clip,viewport=0  50 350 318]{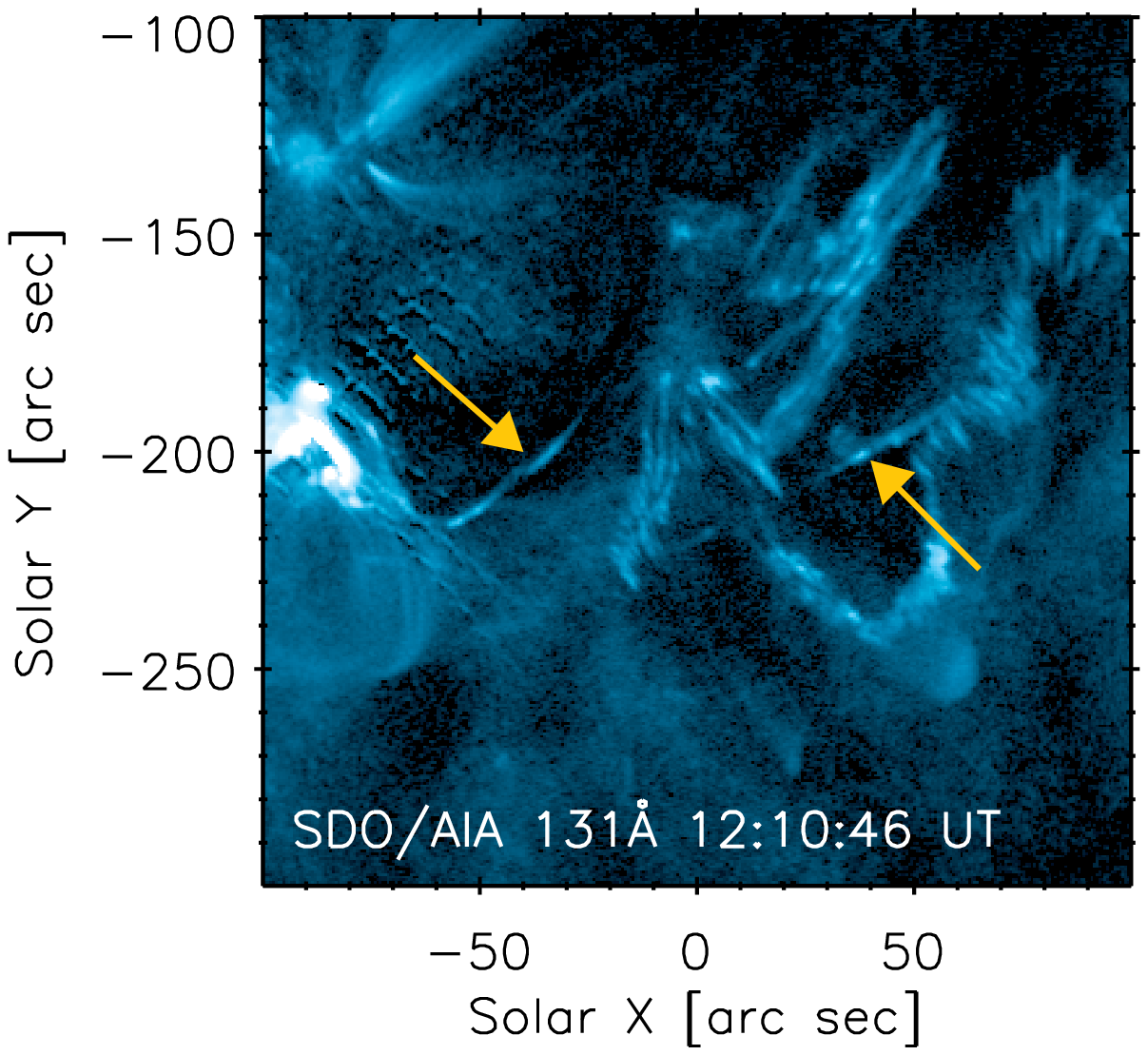}
        \put(-110,103){\textcolor{white}{\bf e)}}
        \includegraphics[width=4.1cm,clip,viewport=78 50 350 318]{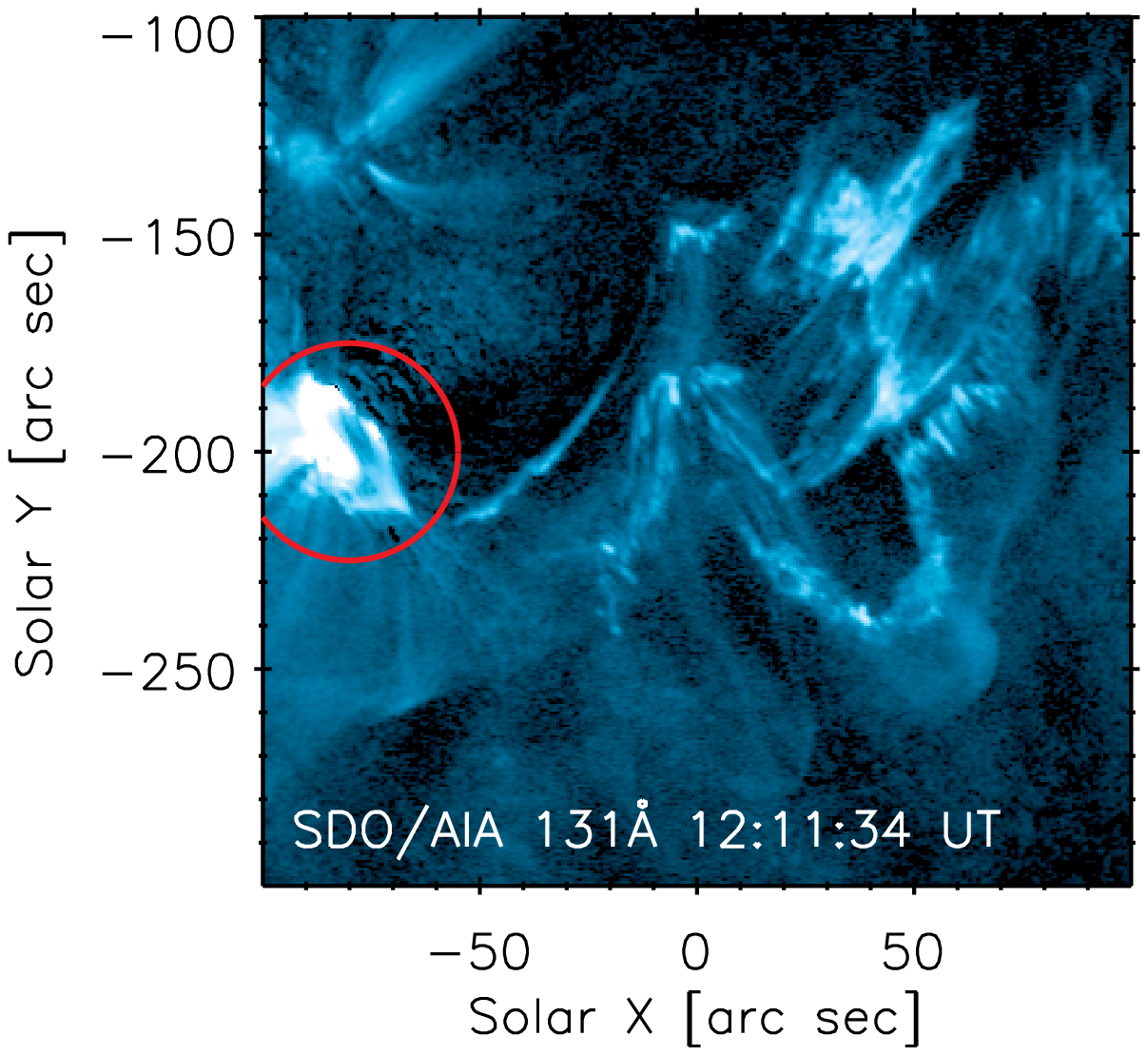}
        \put(-110,103){\textcolor{white}{\bf f)}}
        \includegraphics[width=4.1cm,clip,viewport=78 50 350 318]{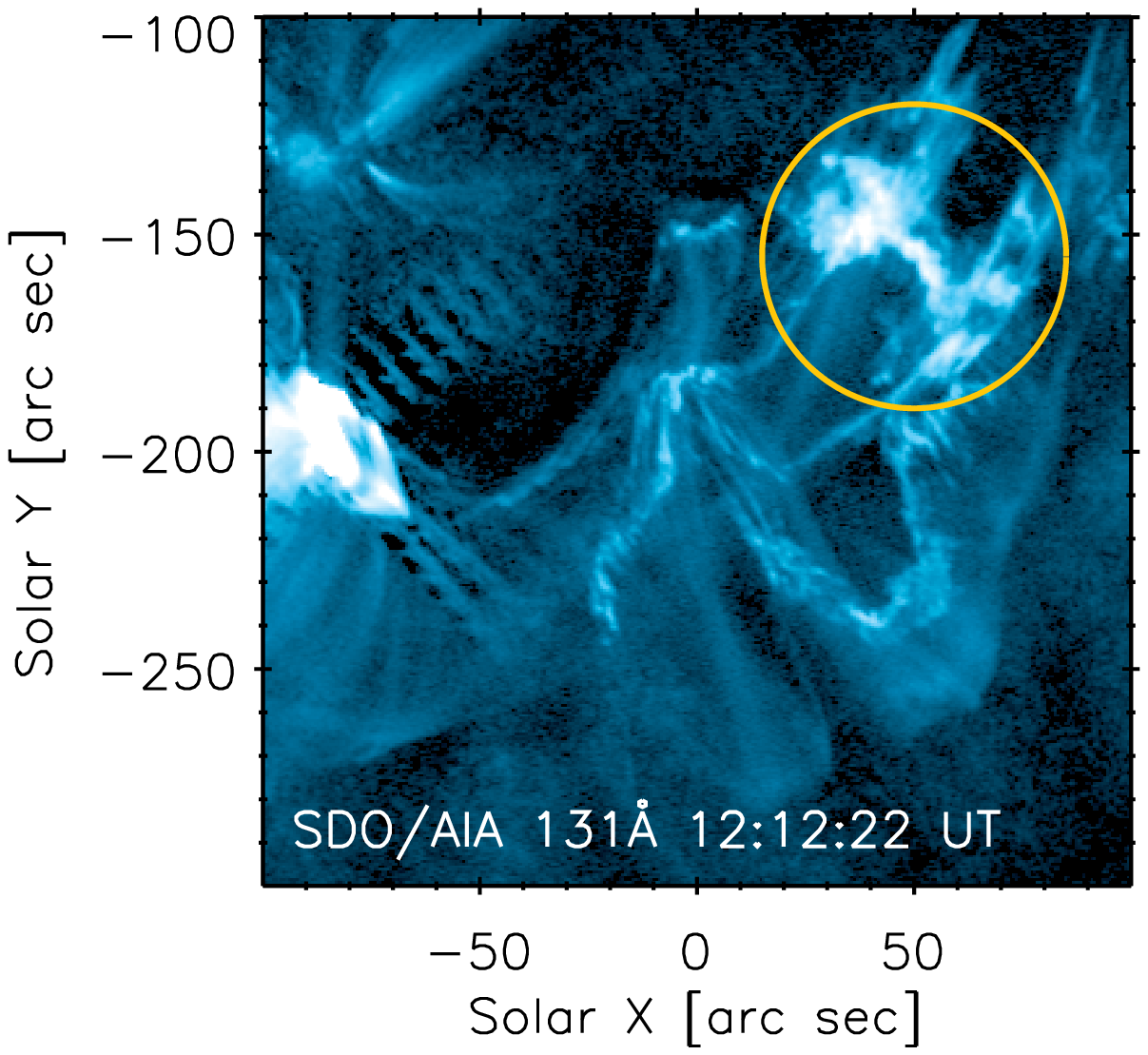}
        \put(-110,103){\textcolor{white}{\bf g)}}
        \includegraphics[width=4.1cm,clip,viewport=78 50 350 318]{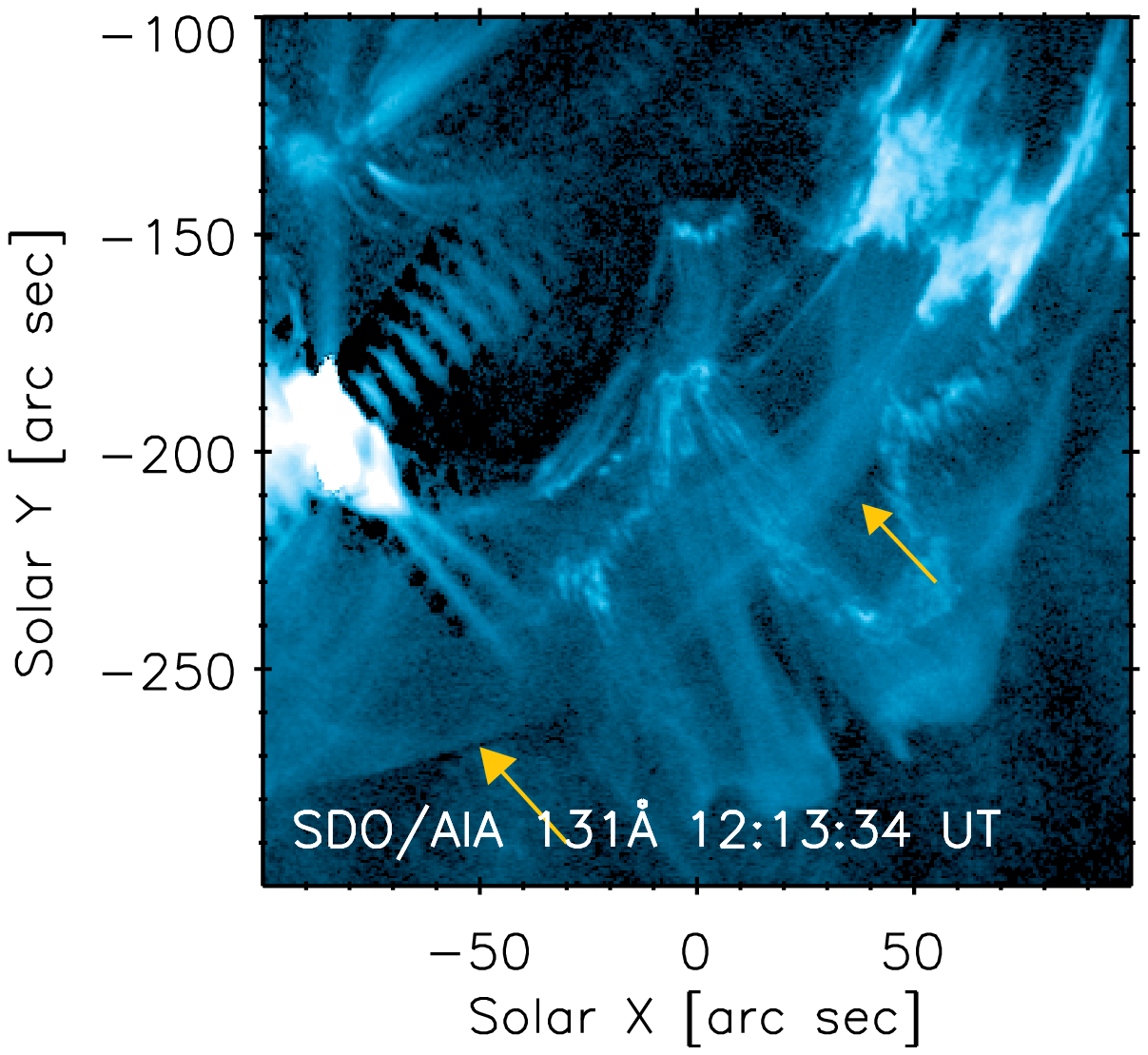}
        \put(-110,103){\textcolor{white}{\bf h)}}

        \includegraphics[width=5.28cm,clip,viewport=0 0 350 318]{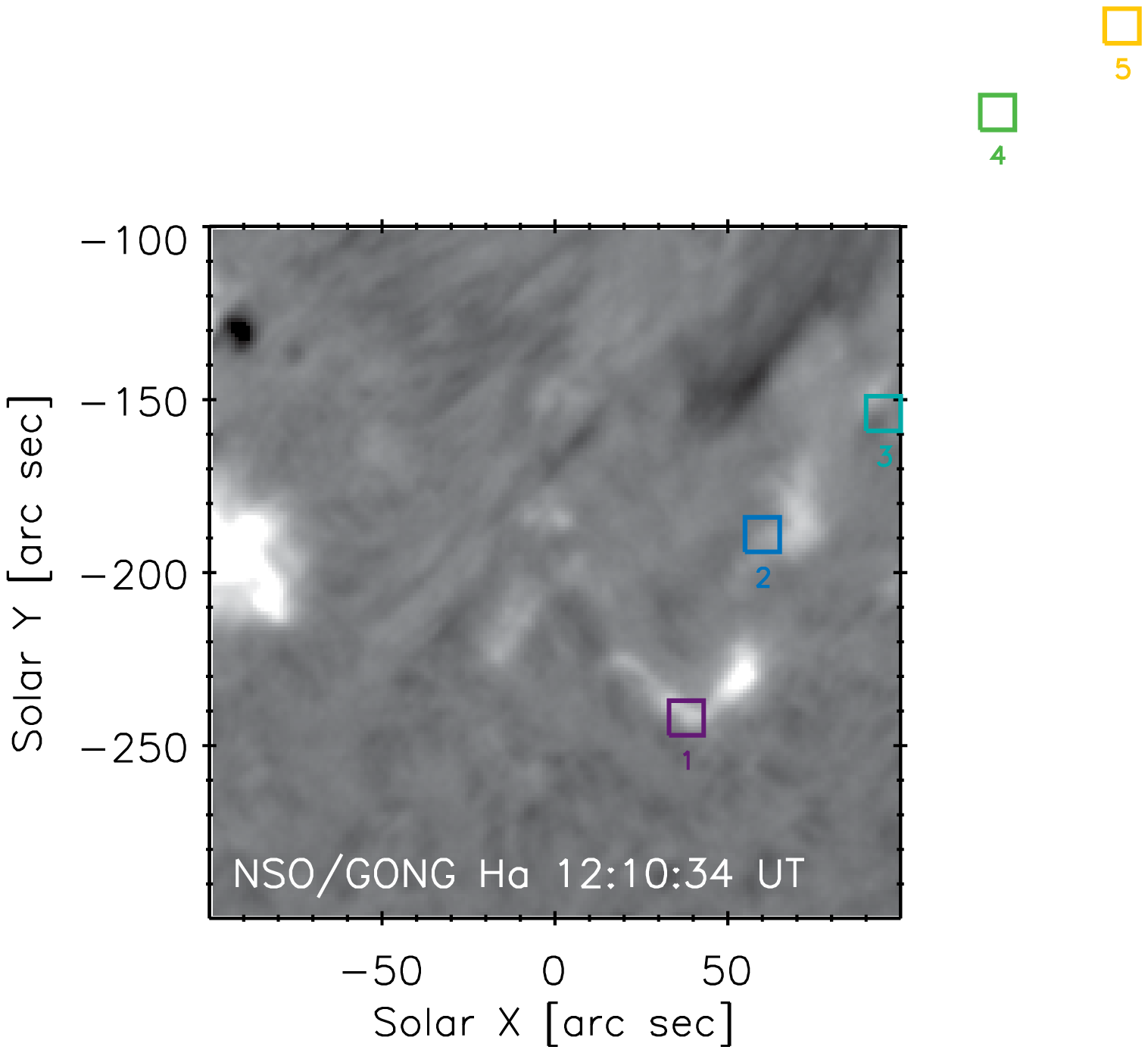}
        \put(-110,125){\textcolor{white}{\bf i)}}
        \includegraphics[width=4.1cm,clip,viewport=78 0 350 318]{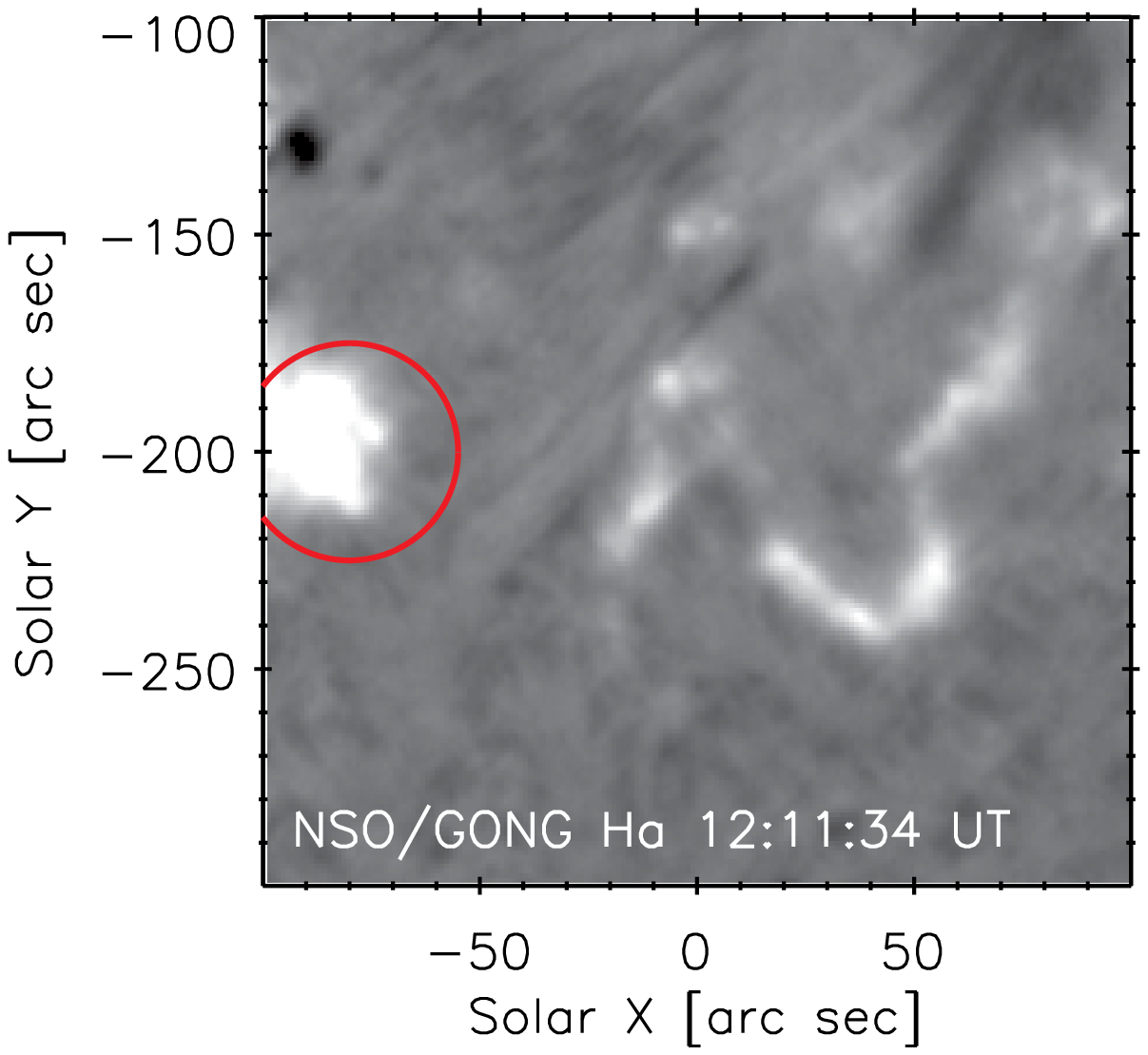}
        \put(-110,125){\textcolor{white}{\bf j)}}
        \includegraphics[width=4.1cm,clip,viewport=78 0 350 318]{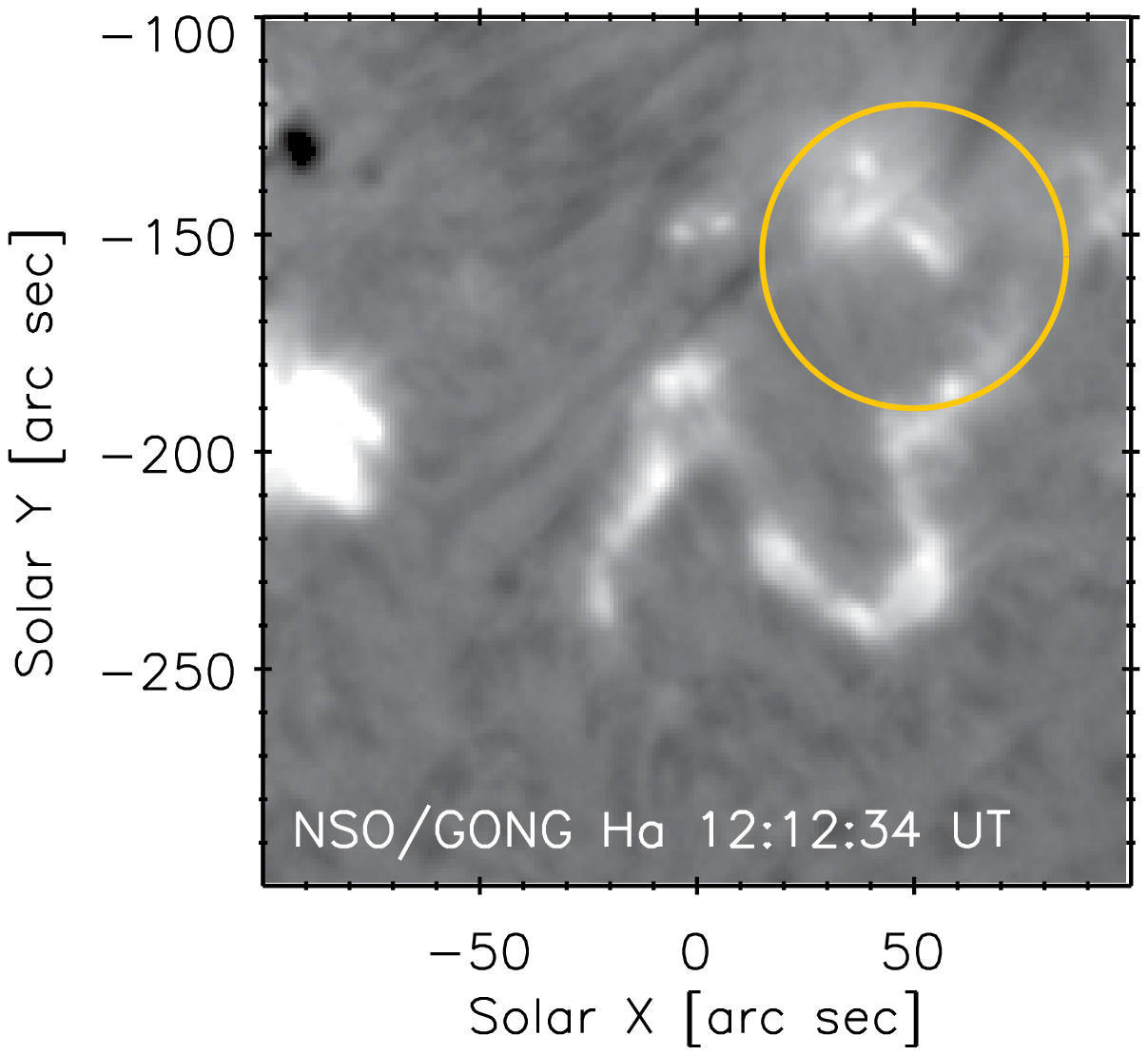}
        \put(-110,125){\textcolor{white}{\bf k)}}
        \includegraphics[width=4.1cm,clip,viewport=78 0 350 318]{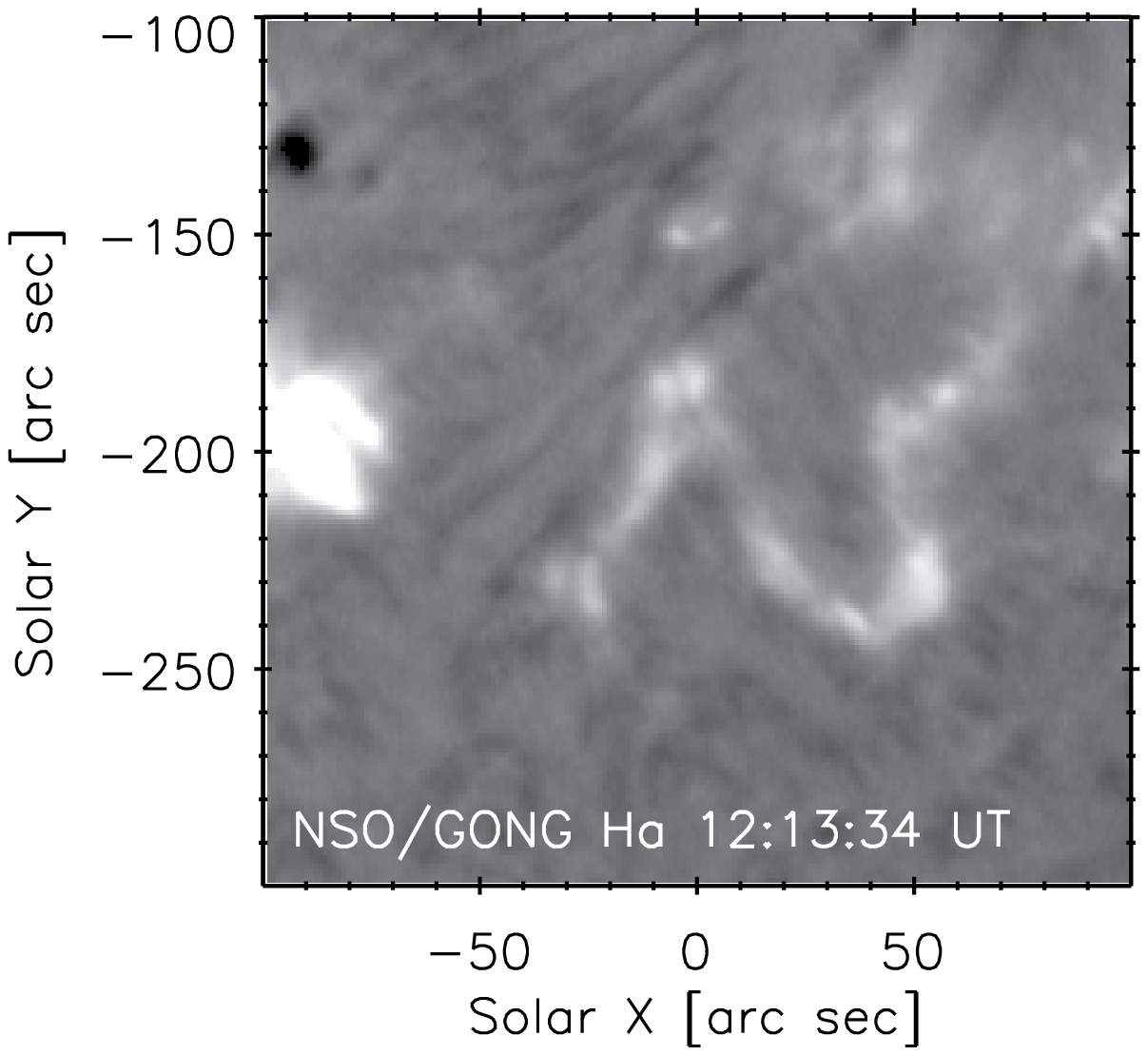}
         \put(-110,125){\textcolor{white}{\bf l)}}

        \caption{Zoomed view of plasma ejection toward the preexisting filament and formation of BTS.
        The FOV is the same as in Figure~\ref{fig_reco}. Top row (a--d) shows formation of the cusp
        and BTS in {\it Hinode}/XRT Al\_poly filter.
        Middle row of (e--h) shows this evolution in \textit{SDO}/AIA 131\,\AA~images.
        Bottom row (i--l) shows the same but in GONG H$\alpha$ filtergrams. Colored squares in (i)
        show positions of brightenings 1--3 along the F. Location of the cusp is marked in red circles
        and the BTS is marked by yellow circles. Arrows in (a) and (e) show plasma coming out
        of the cusp, forming new flare loops after the Burst S. Arrows in (d) and (h) show faint tails of
        hot plasma behind the BTS.} \label{fig_evol2}
\end{figure*}

Formation of the cusp and ejection of plasma toward the F is documented in Figures~\ref{fig_reco}
and \ref{fig_evol2}. In particular, this part of the flare evolution was accompanied by the main
part of SPDBs at 12:10:20 -- 12:12:30\,UT (Figure~\ref{fig_radio}(a)). Among these bursts, the Burst S
with S-like form was observed at 12:10:34 -- 12:10:41\,UT (Figure~\ref{fig_radio}(b)).

Figure~\ref{fig_reco} shows the sequence of \textit{SDO}/AIA base difference images during the time of
the Burst S. Circles show the central part where we can follow the formation of new loops filled with
plasma ejected toward F. On the left side of the circle in Figure~\ref{fig_reco}(a), we observed hot
loops reaching its central area, while on the upper right side we saw a new flare loop to form (upper
arrows in Figure~\ref{fig_reco}(b) and (f)). Formation of another flare loop can be followed in
Figures~\ref{fig_reco}(f)--(h), at the lower side of the circles (arrows). At 12:10:44\,UT, just three
seconds after the Burst S, we observed there two small brightenings there (Figure~\ref{fig_reco}(f) and (g)).
The left one (Figure~\ref{fig_reco}(g), arrow) was located at the area where, in 131\,\AA, we saw bright
loops crossed by other tiny bright loop (Figure~\ref{fig_reco}(c), arrow). Afterwards, both brightenings
disappeared and we observed the second new loop with heated plasma ejected toward F (Figure~\ref{fig_reco}(d) and
(h), arrows). Thus, the Burst S occurred when we observed plasma ejection toward the F.

\begin{figure*}
        \centering
        \includegraphics[width=6.9cm,clip,viewport=3  0 357 357]{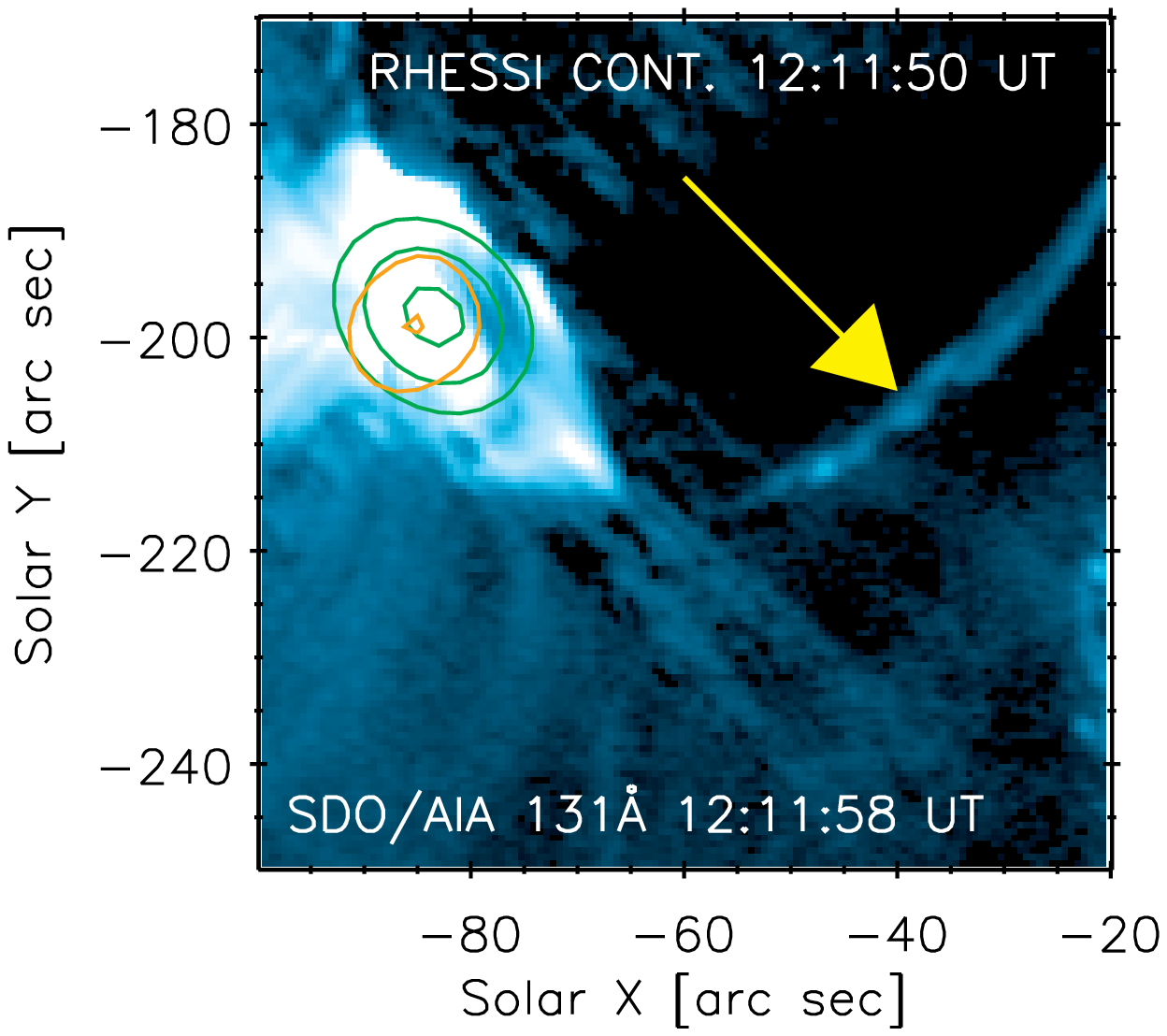}
        \put(-150,162){\textcolor{white}{\bf a)}}
        \includegraphics[width=5.45cm,clip,viewport=78 0 357 357]{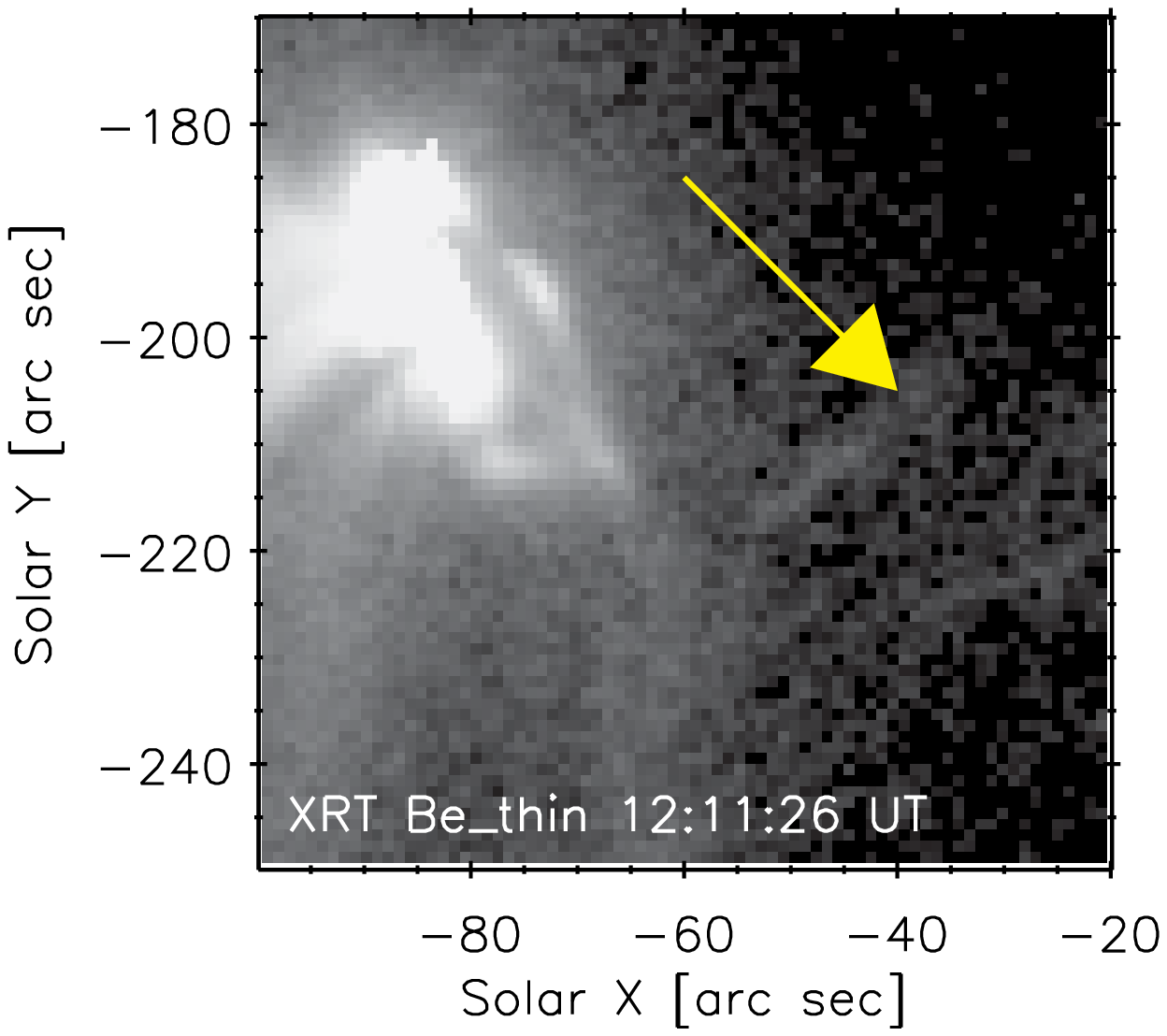}
        \put(-150,162){\textcolor{white}{\bf b)}}
        \includegraphics[width=5.45cm,clip,viewport=78 0 357 357]{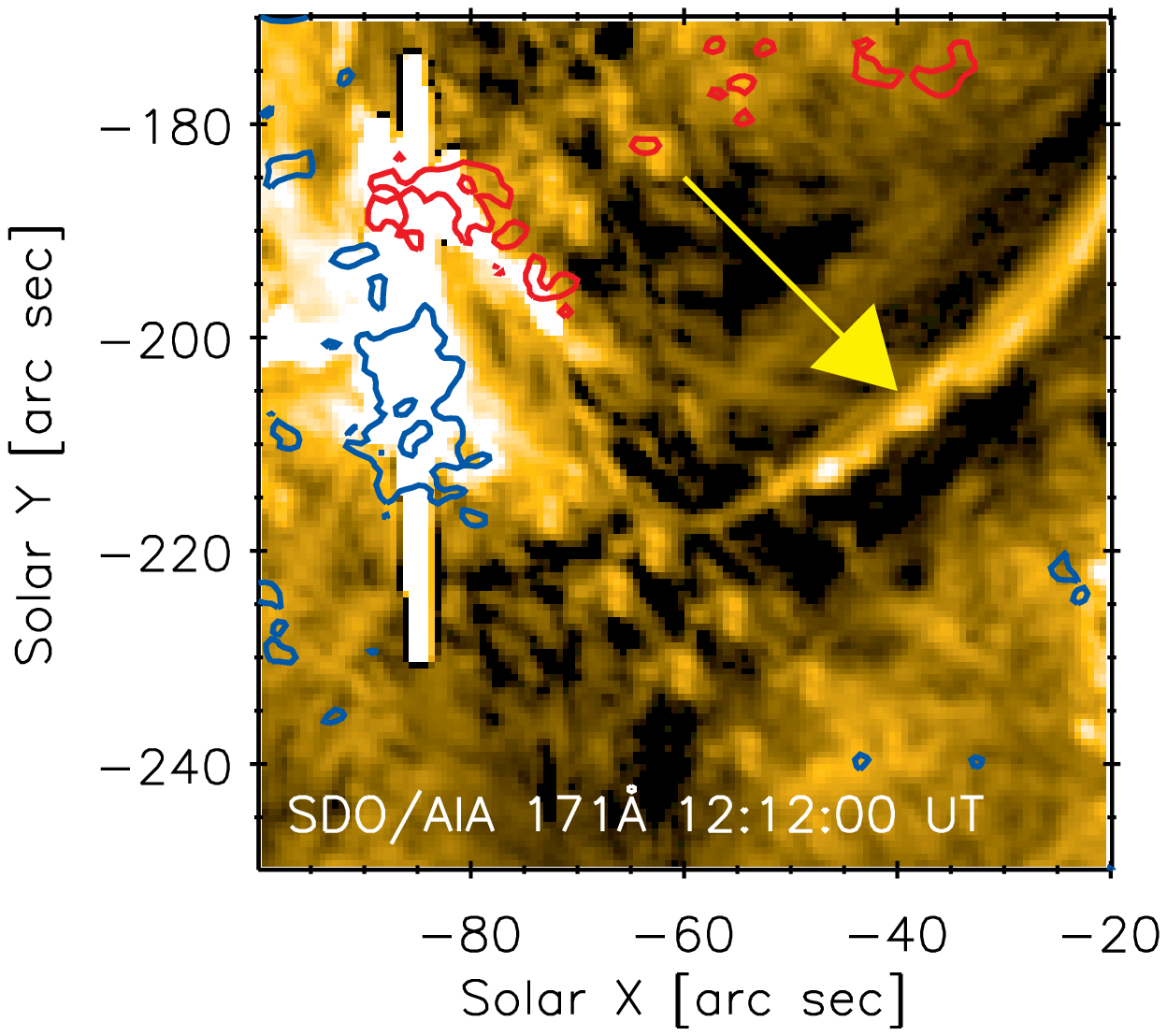}
        \put(-150,162){\textcolor{white}{\bf c)}}
        \caption{Detailed view of the cusp. Hot channels, \textit{SDO}/AIA 131\,\AA~(a) and {Hinode}/XRT
         Be\_thin (b) filters, show the structure in the cusp. The 55, 75, and 95\% contours on panel (a) mark
         the position of the \textit{RHESSI} source at 12:11:50\,UT (green 6--12\,keV, orange 12--25\,keV).
         Panel (c) shows the area of the cusp in the 171\,\AA~\textit{SDO}/AIA filter. Contours depict \textit{SDO}/HMI
         line-of-sight magnetic field of $\pm$300G (red/blue, respectively). The arrows point to the flare loop
         coming out of the cusp, see Sec. \ref{subsec:euv}.
         The FOV for this figure is shown by the small dashed rectangle in Figure~\ref{fig_evol1}(c).}
         \label{fig_cusp}
\end{figure*}

The flare evolution then continues in Figure~\ref{fig_evol2}. Panels (a) and (e) show the plasma coming out
from the forming cusp after the Burst S (arrows). Figure~\ref{fig_evol2}(i) shows the F and H$\alpha$ brightenings
in Regions 1--3. The cusp itself was fully developed a couple of seconds later (Figure~\ref{fig_evol2}(b)--(f)--(j),
red circles). The detailed view of the cusp can be seen in Figure~\ref{fig_cusp}. In the hot channels of \textit{SDO}/AIA
131\,\AA~ ($log\,T\,[K]=7.05-7.15$, \cite{ODwyer2010}) and {\it Hinode}/XRT Be\_thin filter ($log\,T\,[K]=6.50-6.60$,
\cite{ODwyer2014}), a structure can be resolved within the cusp: a bright area at the top and bright kernels in its
legs (Figure~\ref{fig_cusp}(a) and (b)). The X-ray source (Figure~\ref{fig_cusp}(a)) was not located right at the cusp
leg but slightly northeast of it. Figure~\ref{fig_cusp}(c) shows the same area in coronal emission originating at
temperatures $log\,T\,[K]=5.85-6.35$ \citep{ODwyer2010}. No cusp can be seen there. We observed there short flare loops
connecting the ribbons of opposite polarity, and the flare loop coming out of the cusp (Figure~\ref{fig_cusp}(c), arrow).
This flare loop is almost invisible in the Be\_thin filter (Figure~\ref{fig_cusp}(b), arrow), which is probably a consequence
of short exposure time as the XRT was still following the previous flare.

At 12:11:34\,UT in 131\,\AA~(Figure~\ref{fig_evol2}(f)), approximately at coordinate Solar-X=50$\arcsec$, we saw
bright emission of plasma located over the preexisting H$\alpha$ filament F (Figure~\ref{fig_evol2}(j)). This
bright emission started to form narrow BTS, which was moving along the axis of F. About 12:12\,UT the narrow BTS was
already formed (Figure~\ref{fig_evol2}(c)--(g)--(k), yellow circles). Behind the moving and dissipating BTS we
observed a faint tail of hot plasma visible in Al\_poly and 131\,\AA~(Figure~\ref{fig_evol2}(d) and (h), arrows).
Note that the narrow BTS existed approximately during 12:11:30-12:12:30\,UT but we observed a lot of intensive
SPDBs even earlier, at 12:10:50--12:12:10\,UT (Figure~\ref{fig_radio}(a)).

\begin{figure*}
        \centering
        \includegraphics[width=8.5cm,clip,viewport=3 0 410 260]{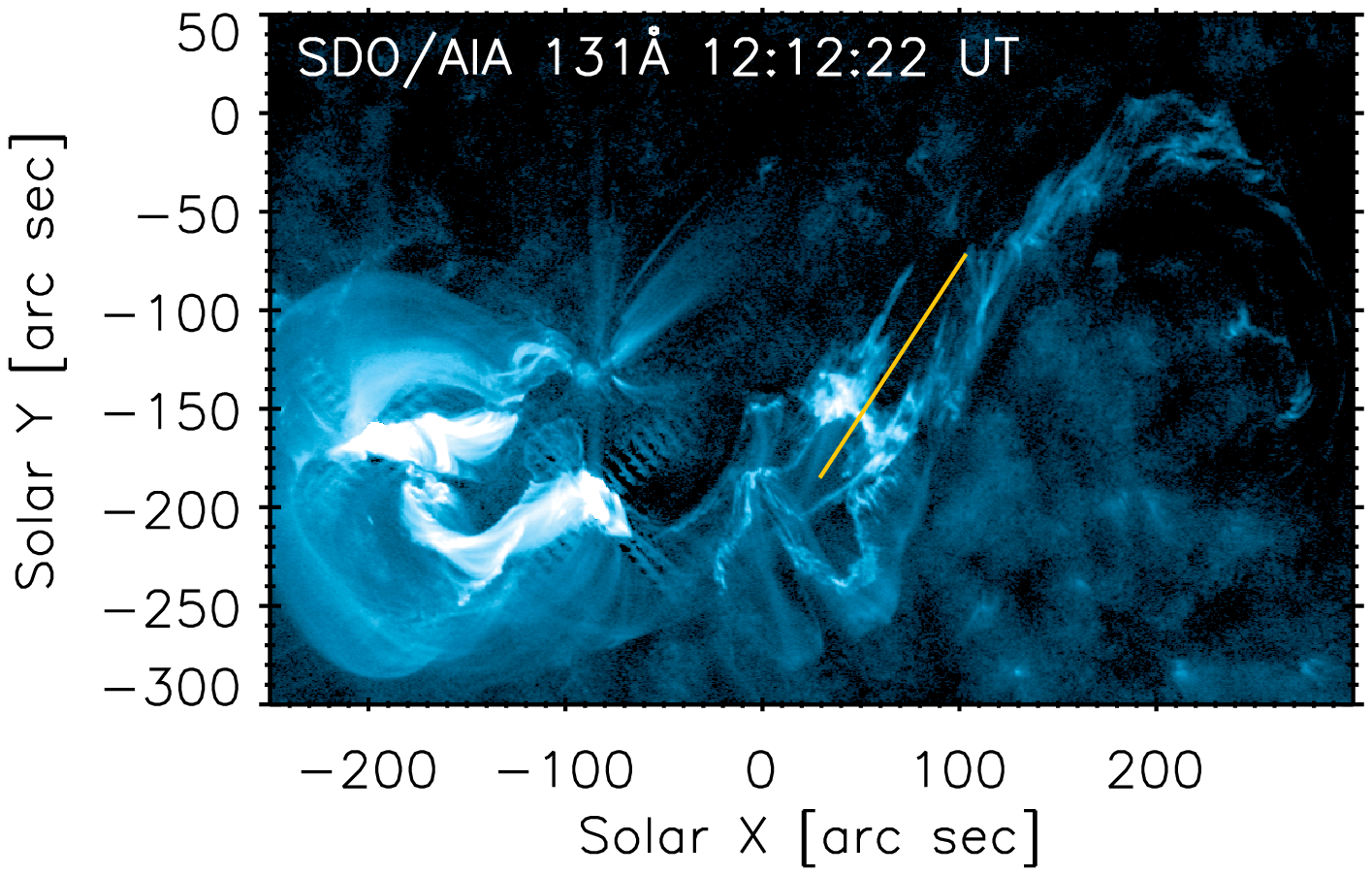}\put(-20,35){\textcolor{white}{\bf a)}}
         \includegraphics[width=9.4cm,clip,viewport=0 -8 440 260]{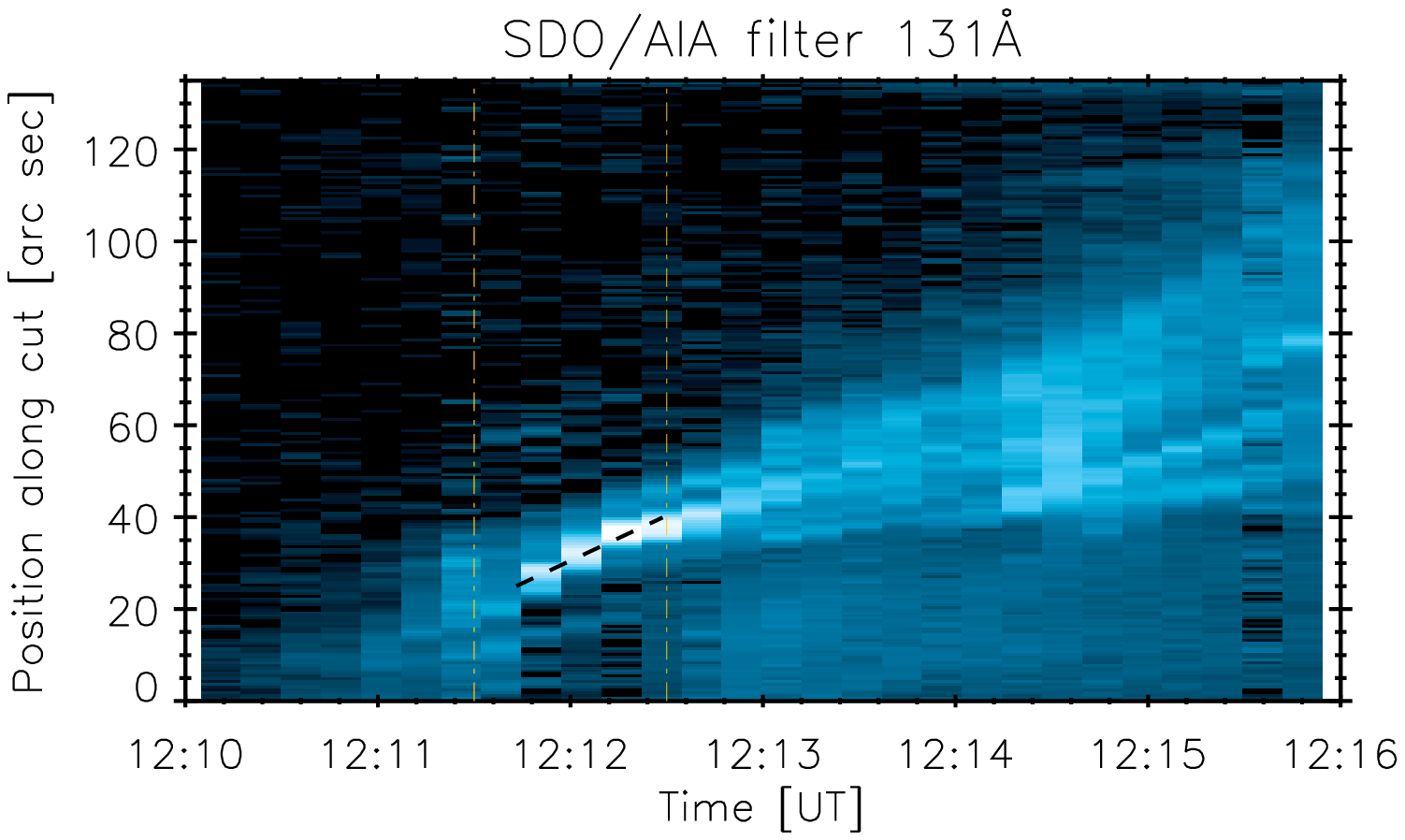}\put(-25,35){\textcolor{white}{\bf b)}}
        \caption{Velocity estimation of BTS. The yellow line running across BTS in (a)
        shows the cut from where the time-distance plot in (b) was constructed.
        Dash-dotted lines mark the 12:11:30--12:12:30\,UT time interval when the BTS was narrow.
        The slope of the bright ridge (black dashed line) was measured to estimate its velocity.}
        \label{fig_vel}
\end{figure*}

We studied the evolution of BTS by the method of time-distance plots (Figure~\ref{fig_vel}). The yellow line
in Figure~\ref{fig_vel}(a) shows a cut used to construct a time-distance plot in Figure~\ref{fig_vel}(b). The BTS
started to form at about 12:11:30\,UT and then it became a bright and narrow structure moving along the F
(see the animation accompanying Figure~\ref{fig_evol1}). After 12:12:30\,UT
it started to dissipate (Figure~\ref{fig_vel}(b)). The projected velocity of narrow BTS was estimated using time-distance
plots as that in Figure~\ref{fig_vel}(b), but constructed also for other \textit{SDO}/AIA EUV filters, plus
1600\,\AA~(not shown here). At all time-distance plots we measured the slope of the bright ridge seen between the dashed
lines at 12:11:30-12:12:30\,UT as is shown in Figure~\ref{fig_vel}(b). The estimated velocities covered the interval
190--292 km\,s$^{-1}$, with an average velocity of 240 $\pm$50 km\,s$^{-1}$.

\begin{figure*}
        \centering
        \includegraphics[width=5.28cm,clip,viewport=0 50 350 318]{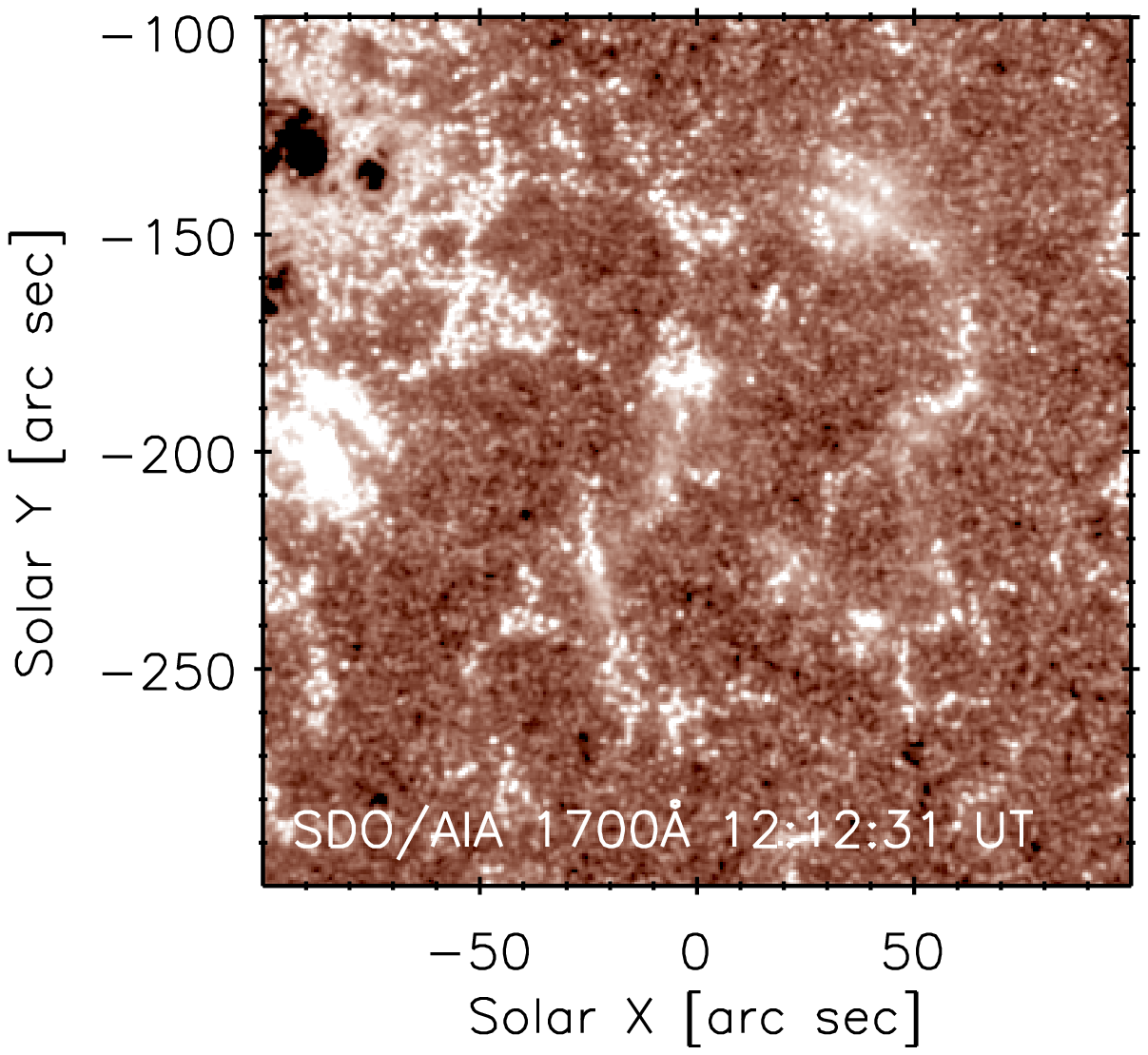}
        \includegraphics[width=4.1cm,clip,viewport=78 50 350 318]{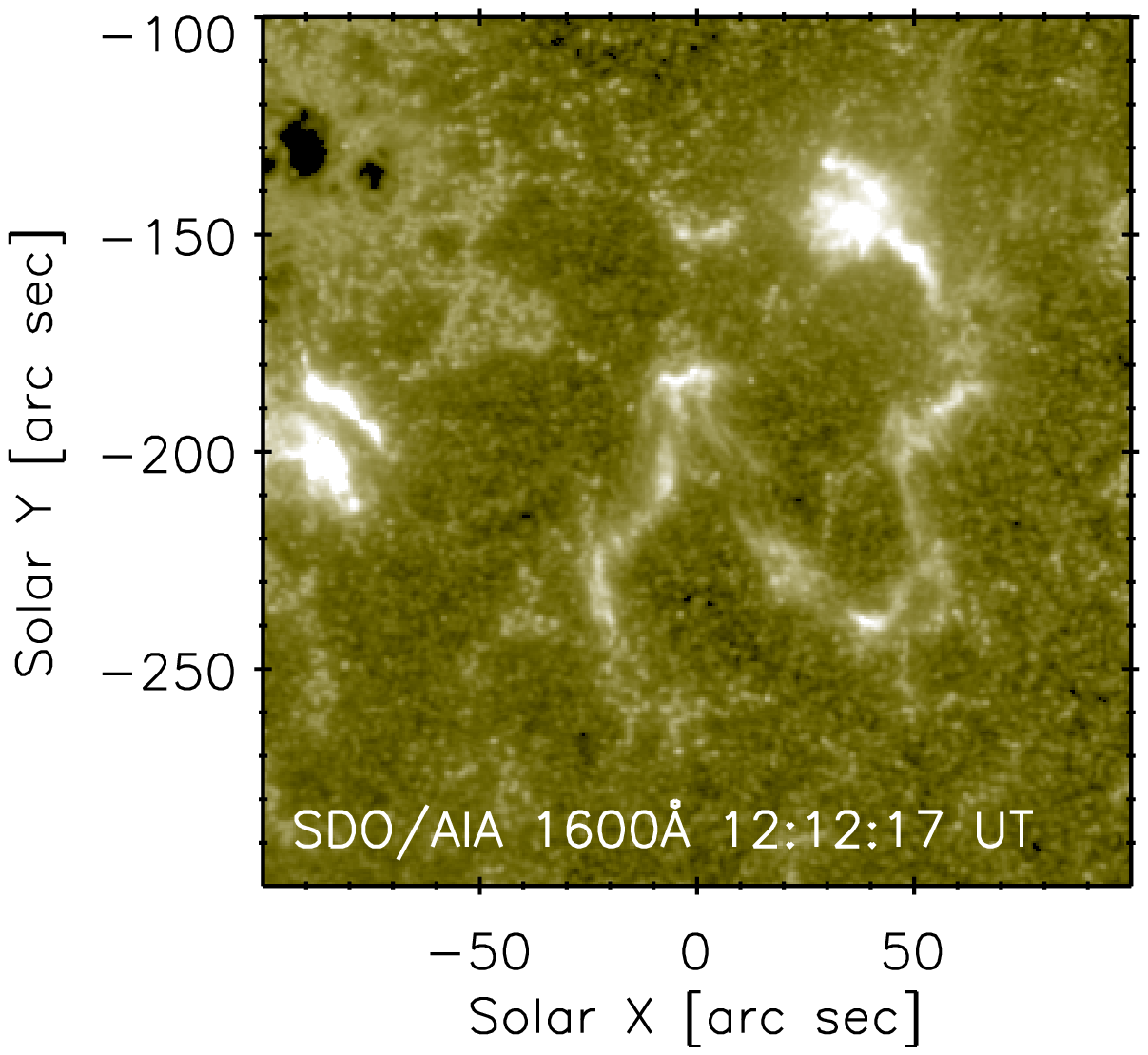}
        \includegraphics[width=4.1cm,clip,viewport=78 50 350 318]{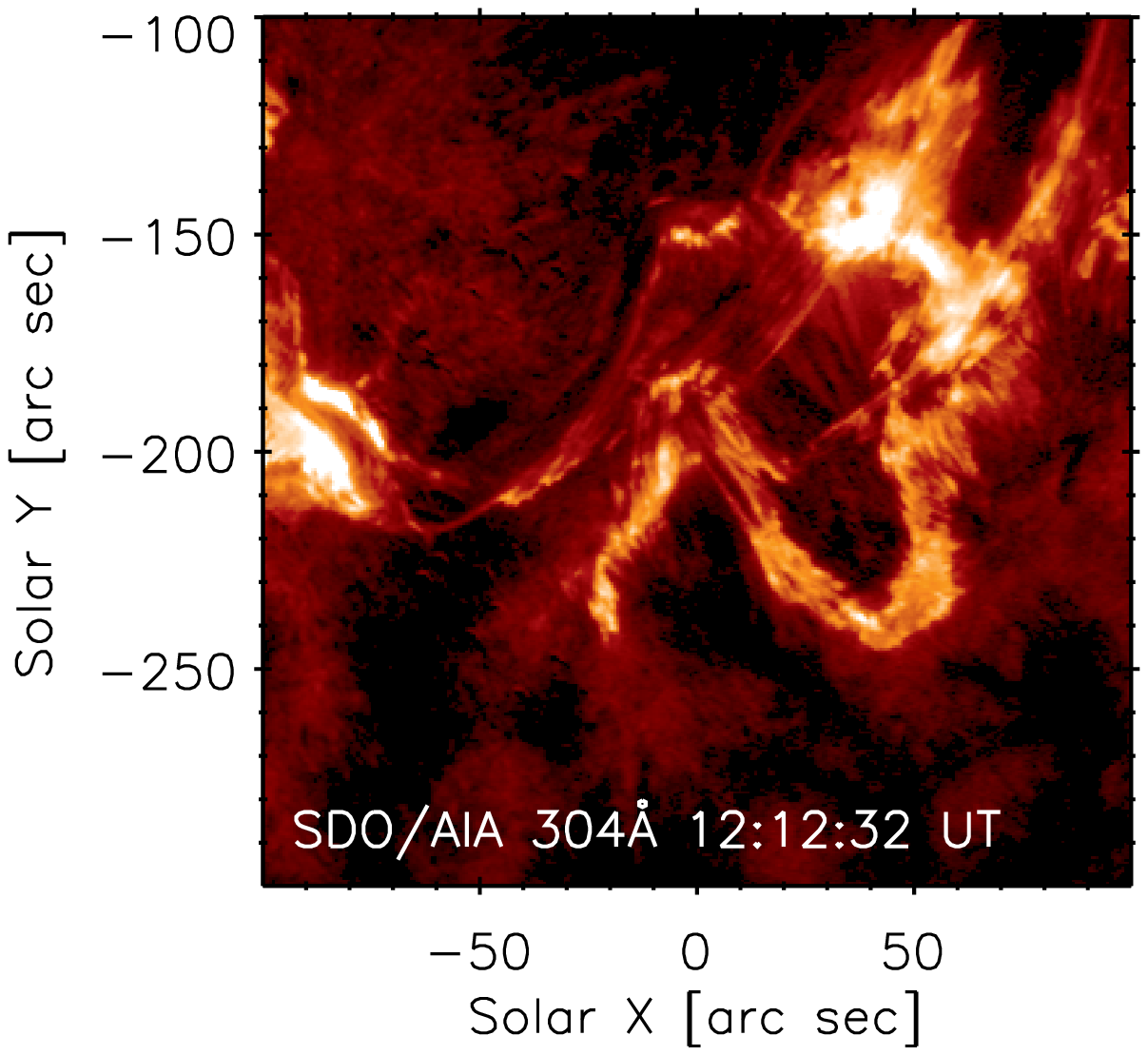}
        \includegraphics[width=4.1cm,clip,viewport=78 50 350 318]{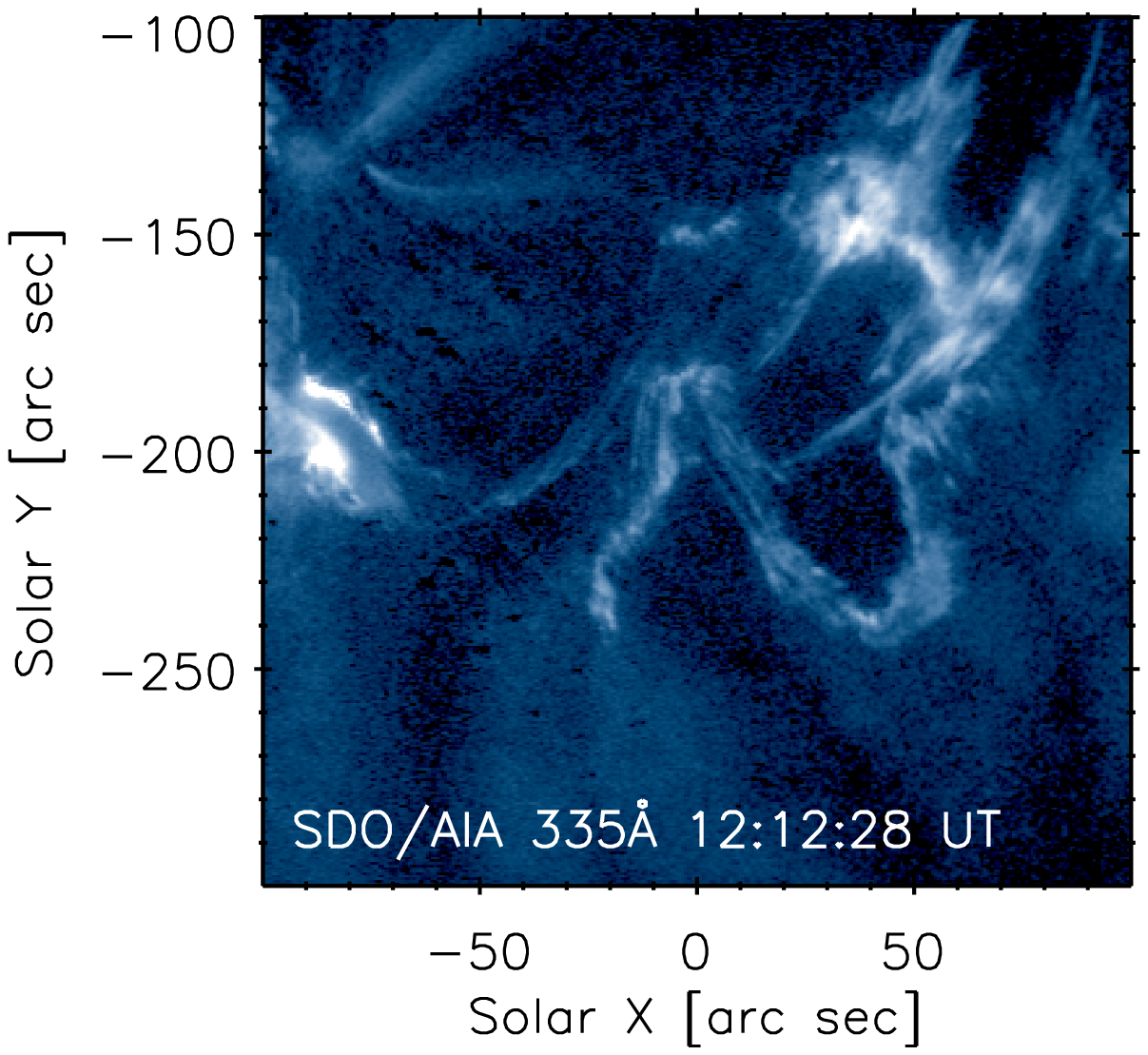}

        \includegraphics[width=5.28cm,clip,viewport=0 0 350 318]{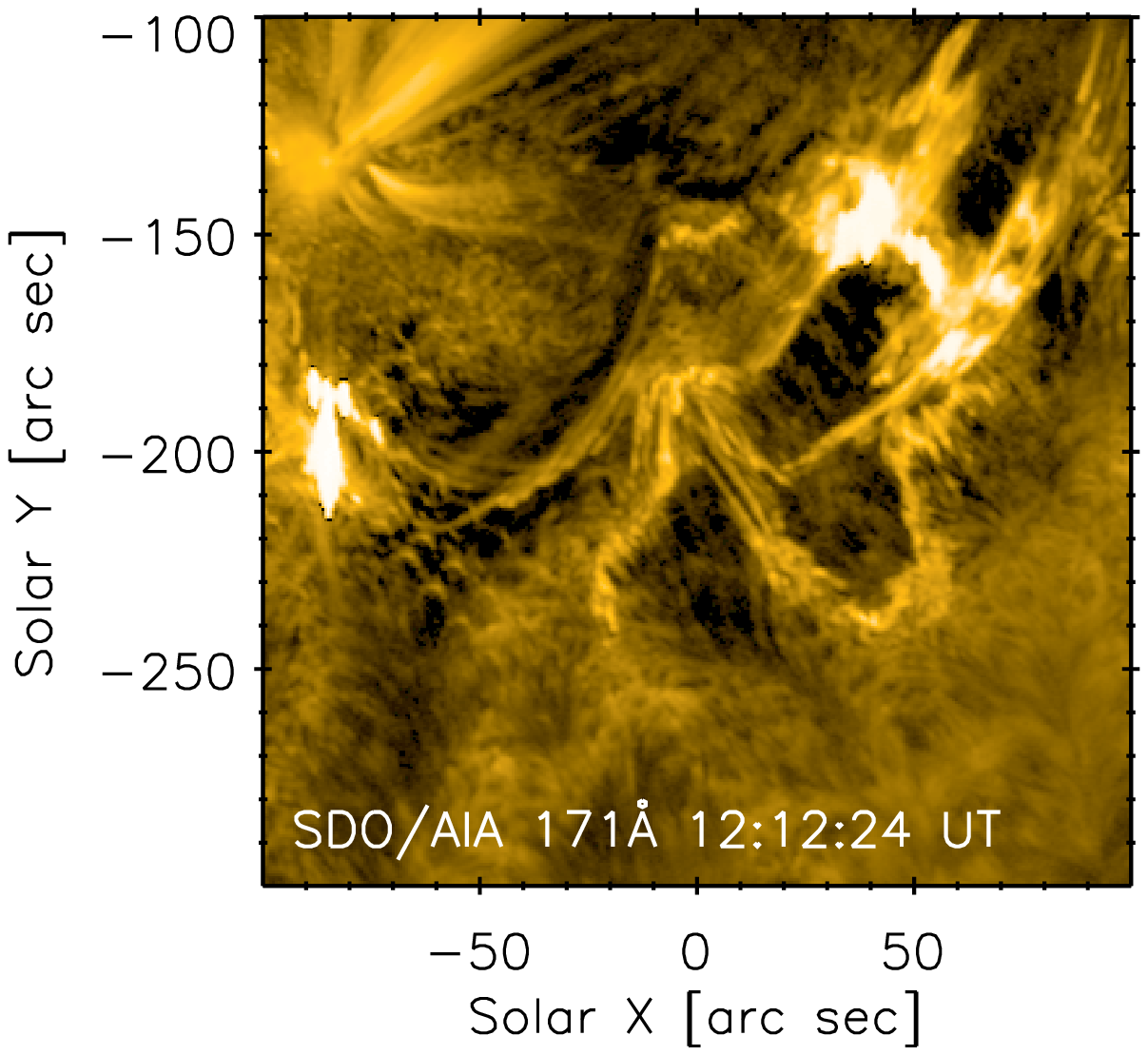}
        \includegraphics[width=4.1cm,clip,viewport=78 0 350 318]{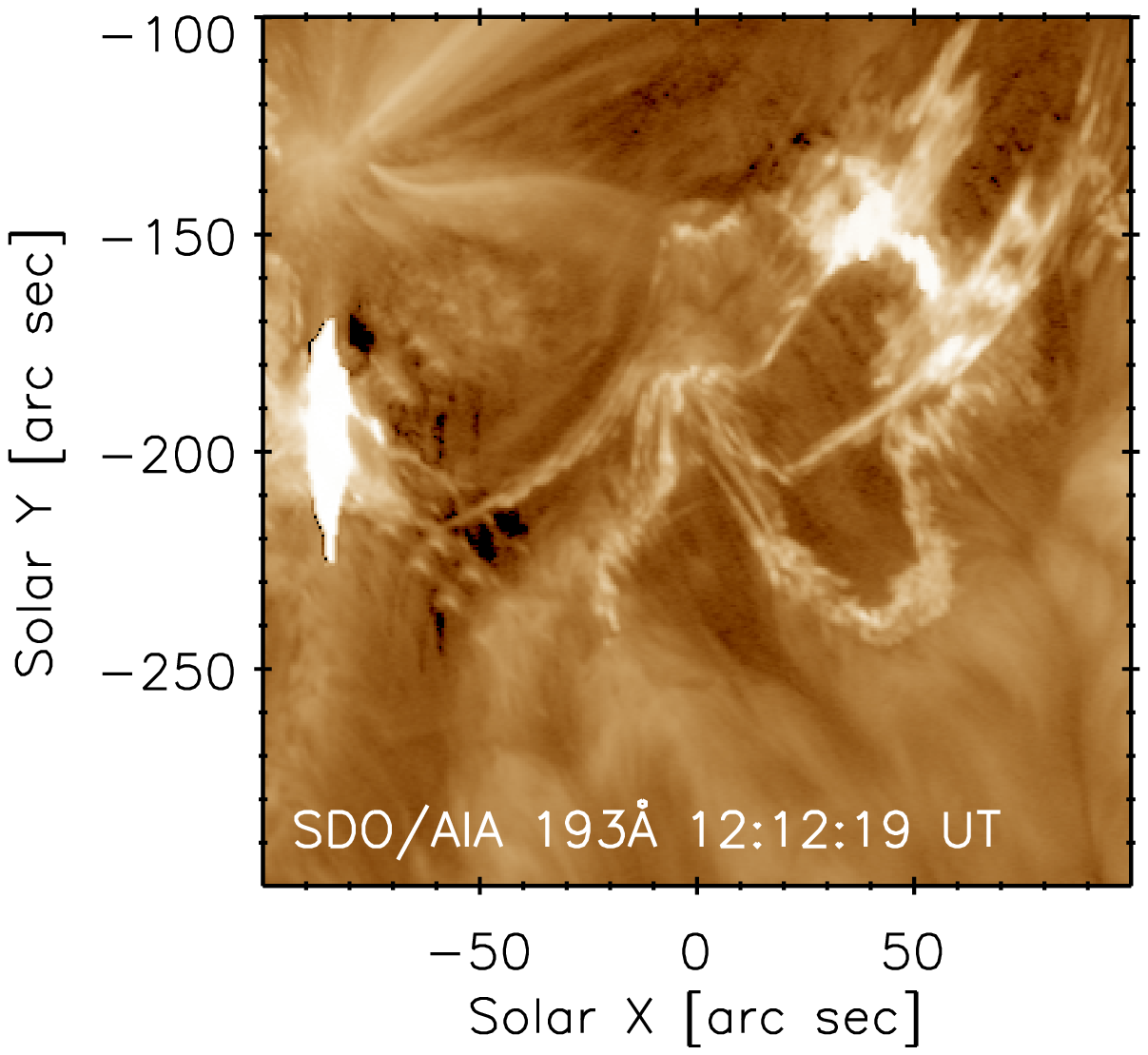}
        \includegraphics[width=4.1cm,clip,viewport=78 0 350 318]{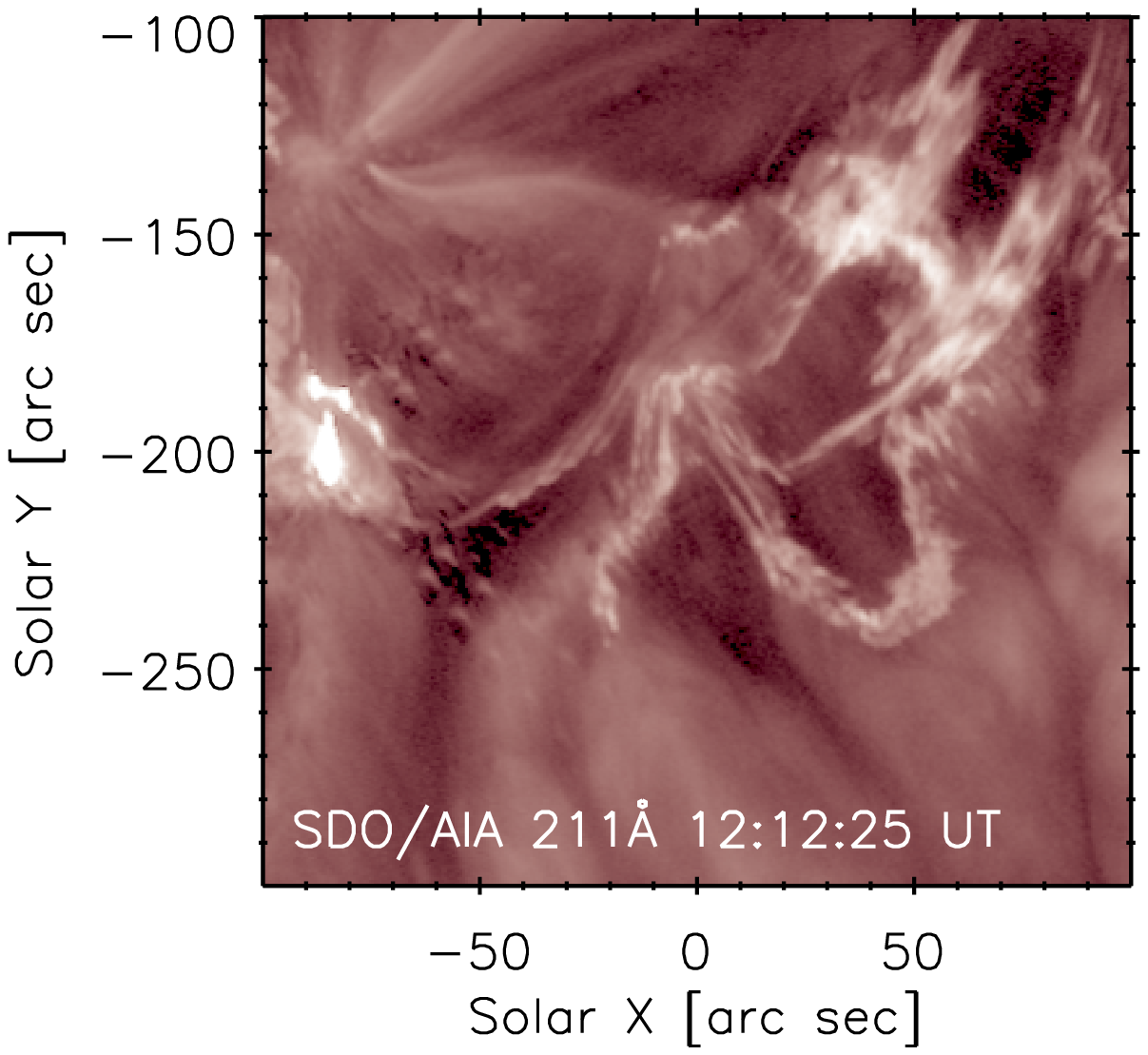}
        \includegraphics[width=4.1cm,clip,viewport=78 0 350 318]{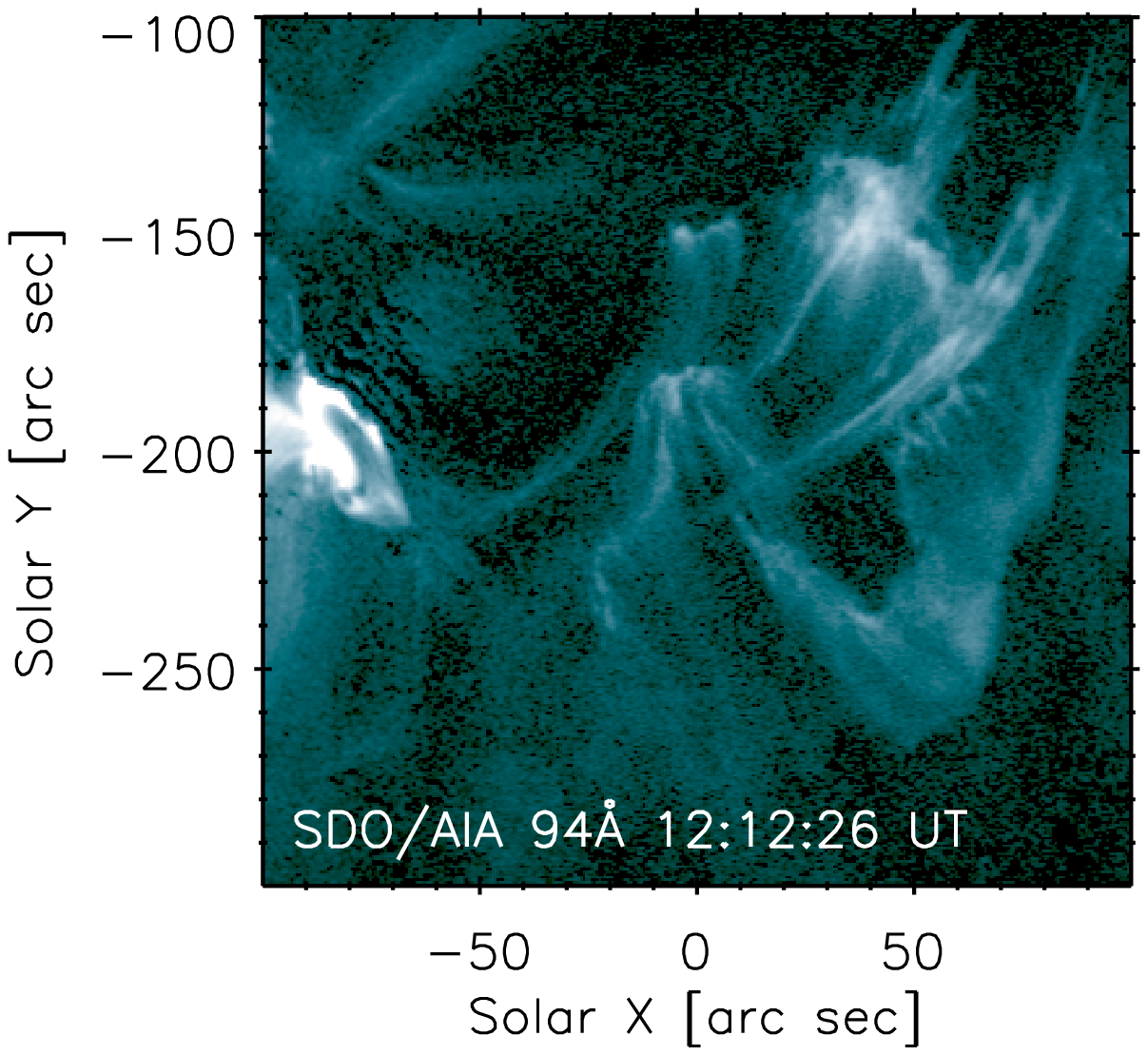}

      \caption{Multi-thermal BTS as seen in other \textit{SDO}/AIA UV/EUV filters at about 12:12:30\,UT. The FOV
      is the same as in Figure~\ref{fig_evol2}.} \label{fig_multi_t}
\end{figure*}

\begin{figure*}
        \centering
        \includegraphics[width=6.8cm,clip,viewport=5 -25 350 318]{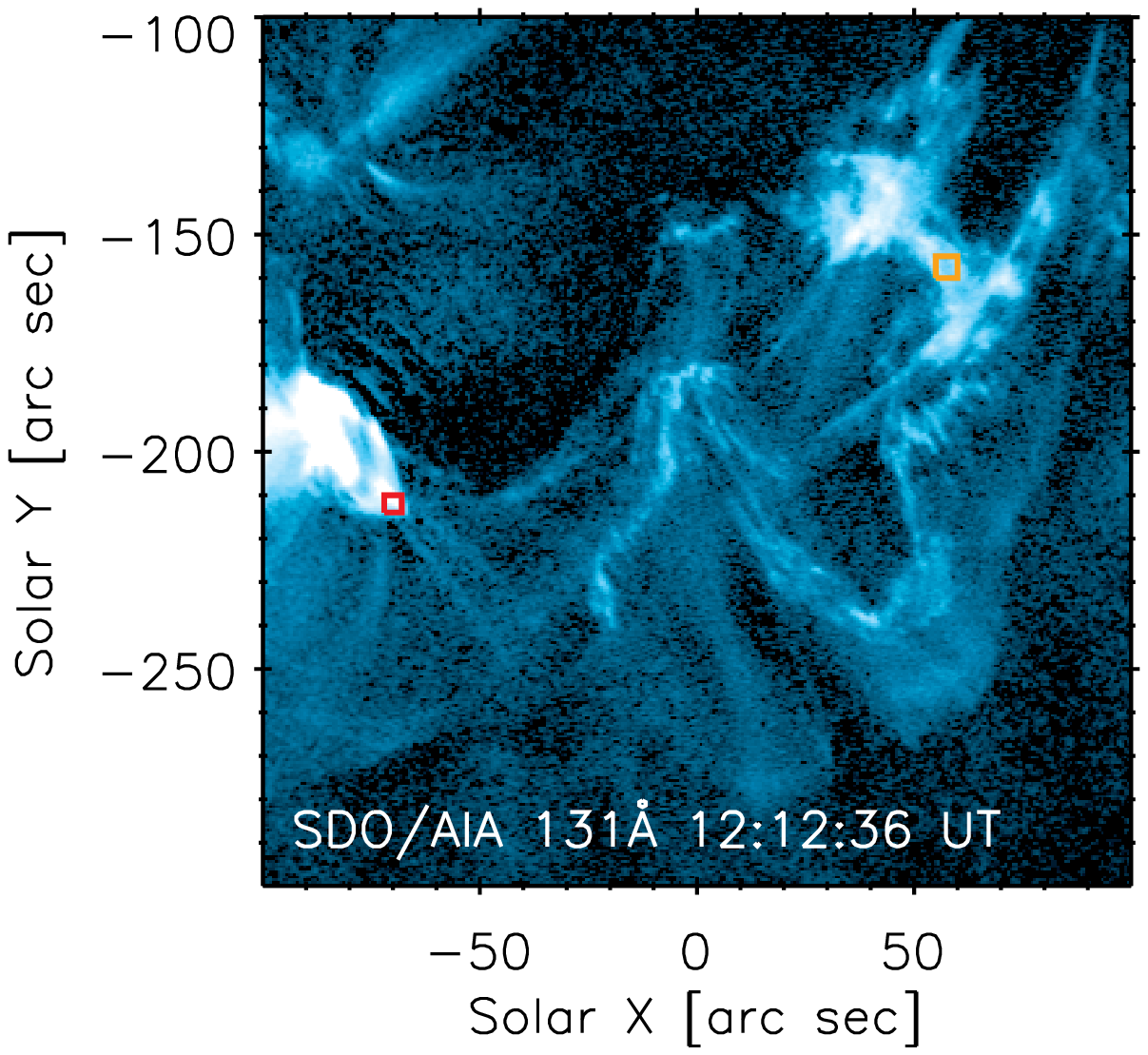}
        \put(-20,180){\textcolor{white}{\bf a)}}
        \includegraphics[width=11cm,clip,viewport=5 0 440 280]{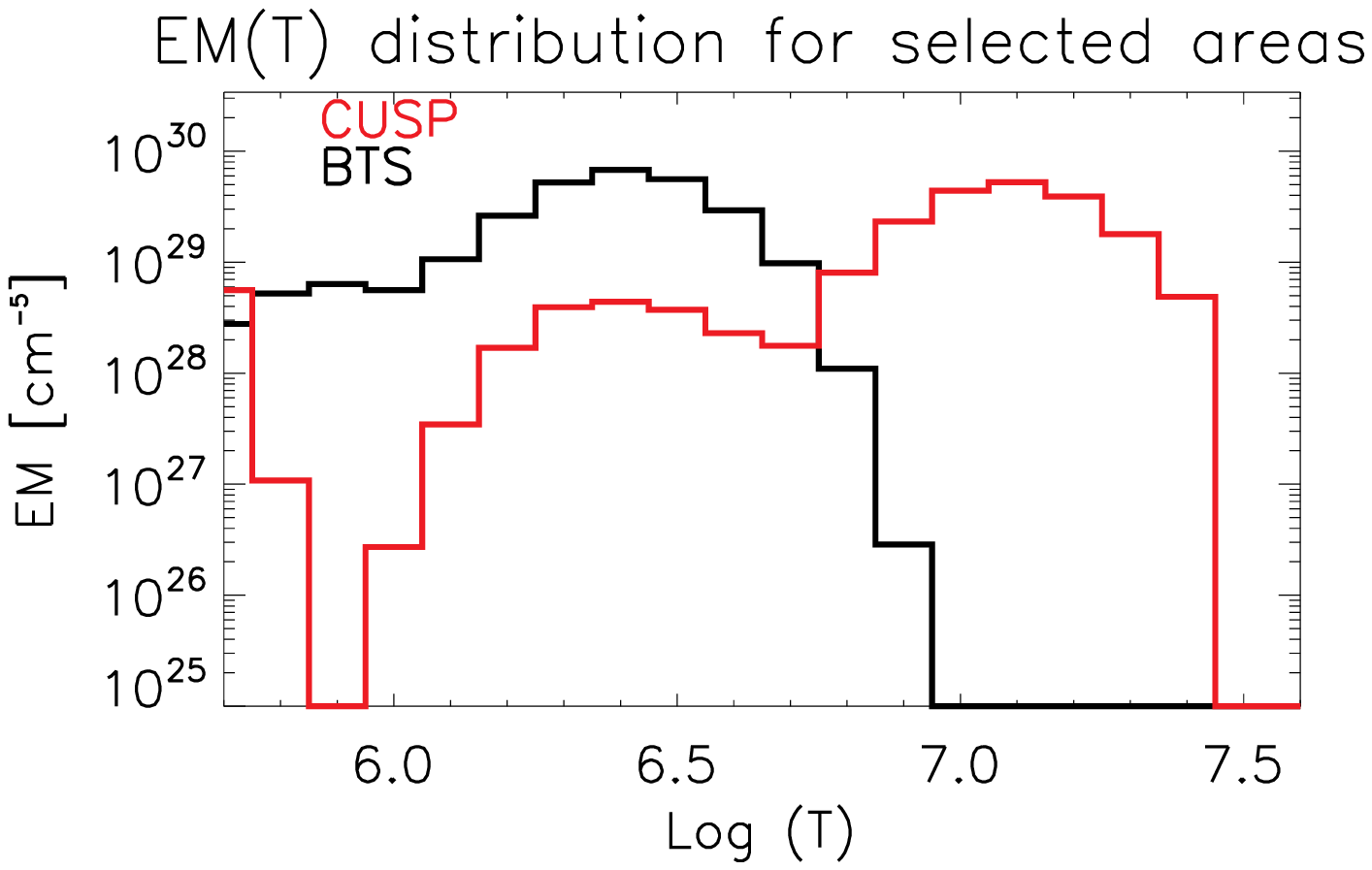}
        \put(-35,158){\textcolor{black}{\bf b)}}
 \caption{Distribution of emission measure per temperature bin at about 12:12:36\,UT,
 calculated using \textit{SDO}/AIA filters. (a) 131\,\AA~filter image showing the areas selected in cusp
 (red square) and in BTS (orange square). (b) Emission measure distribution per temperature bin for
 cusp (red) and BTS (black).} \label{fig_em}
\end{figure*}

The emission of BTS could be seen in all AIA filters as well as in the H$\alpha$ filtergram
(Figure~\ref{fig_multi_t} and Figure~\ref{fig_evol2}(c)--(g)--(k)). This indicates
a presence of multi-thermal plasma. Using the method of \cite{Cheung2015} we estimated
the emission measure per temperature bin $EM_j$ in two small squared areas: in BTS and
in the cusp (Figure~\ref{fig_em}(a)). For this purpose we used images of EUV \textit{SDO}/AIA
filters taken about 12:12:36\,UT, excluding 304\,\AA. The adopted temperature interval
was $log\,T\,[K]=5.70-7.70$, with bin width of 0.1. Although the chosen method of DEM
inversion does not explicitly provide an error of calculated $EM_j$, it can give us a
qualitative result. The average $EM_j$ curves calculated for the areas in the cusp and
BTS are shown in Figure~\ref{fig_em}(b). The black curve represents $EM_j$ distribution
for BTS and has a cutoff at about $log\,T\,[K]=6.90$. The peak appears at $log\,T\,[K]\sim6.40$
but it also shows the presence of plasma with $log\,T\,[K]\lesssim6.00$. This corresponds with
our observations (Figures~\ref{fig_evol2}(c)--(g)--(k) and Figure~\ref{fig_multi_t}). The
red curve is $EM_j$ for the cusp. It shows a small peak at $log\,T\,[K]\sim6.40$ but an
order of magnitude larger amount of plasma at $log\,T\,[K]\sim7.10$. This means that primary
heating occurred at the cusp and hot plasma was ejected toward cold magnetic rope F, where
cooler but still multi-thermal BTS appeared.

After 12:12:30\,UT no SPDBs were observed (Figure~\ref{fig_radio}(a)). When narrow BTS dissipated,
a bulk of heated plasma was seen to move along the axis of F (Figure~\ref{fig_evol2}(h)) and after
12:14\,UT the individual threads  started to appear within that bulk (Figure~\ref{fig_vel}(b)).
Figure~\ref{fig_evol1}(g) shows that in later gradual phase we saw heated plasma following the twisted
threads of the magnetic rope F.

\begin{figure*}
        \includegraphics[width=9.7cm,clip,viewport=3 0 410 260]{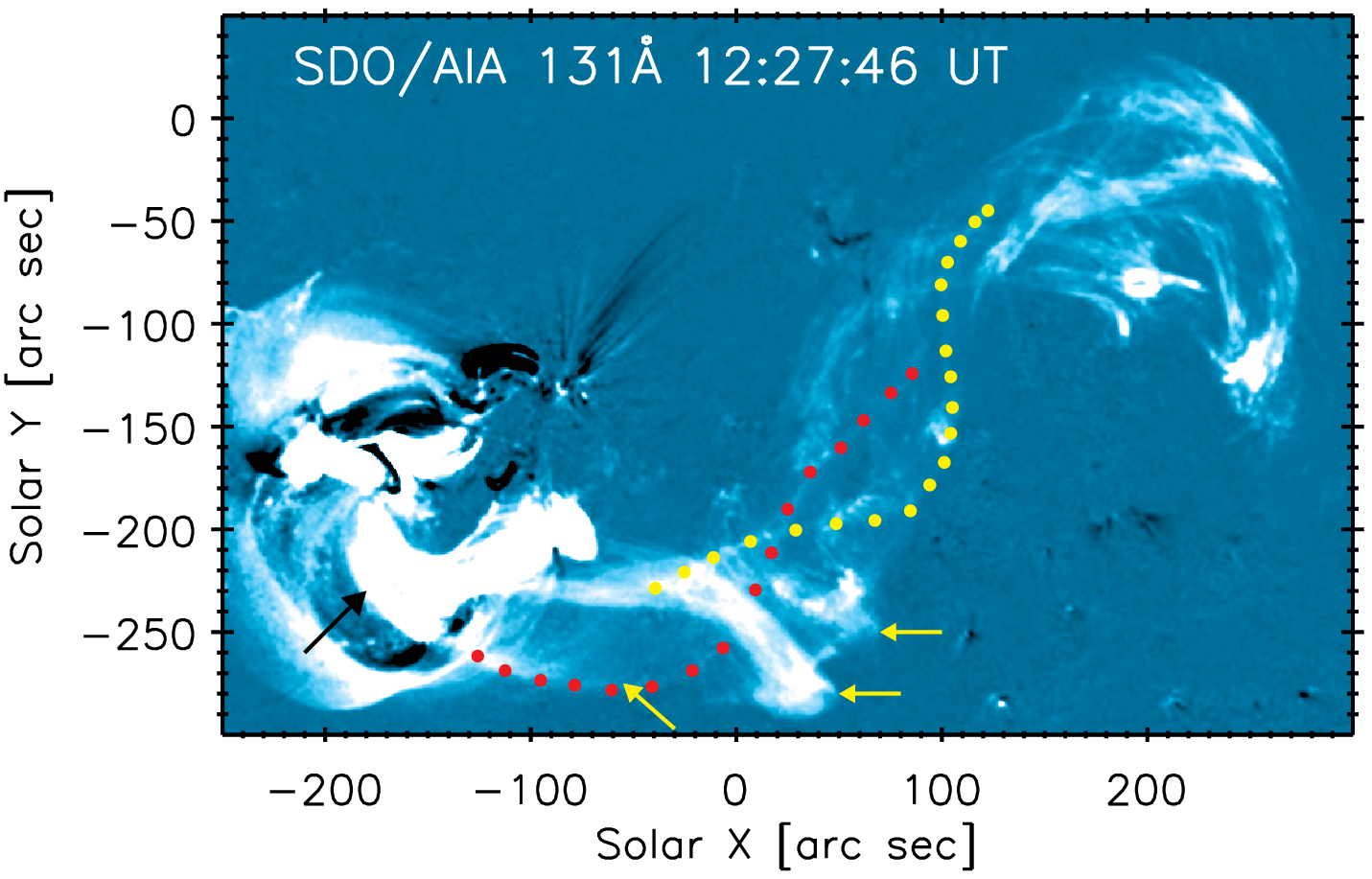}
        \put(-20,163){\textcolor{white}{\bf a)}}
        \includegraphics[width=8.2cm,clip,viewport=66 0 410 260]{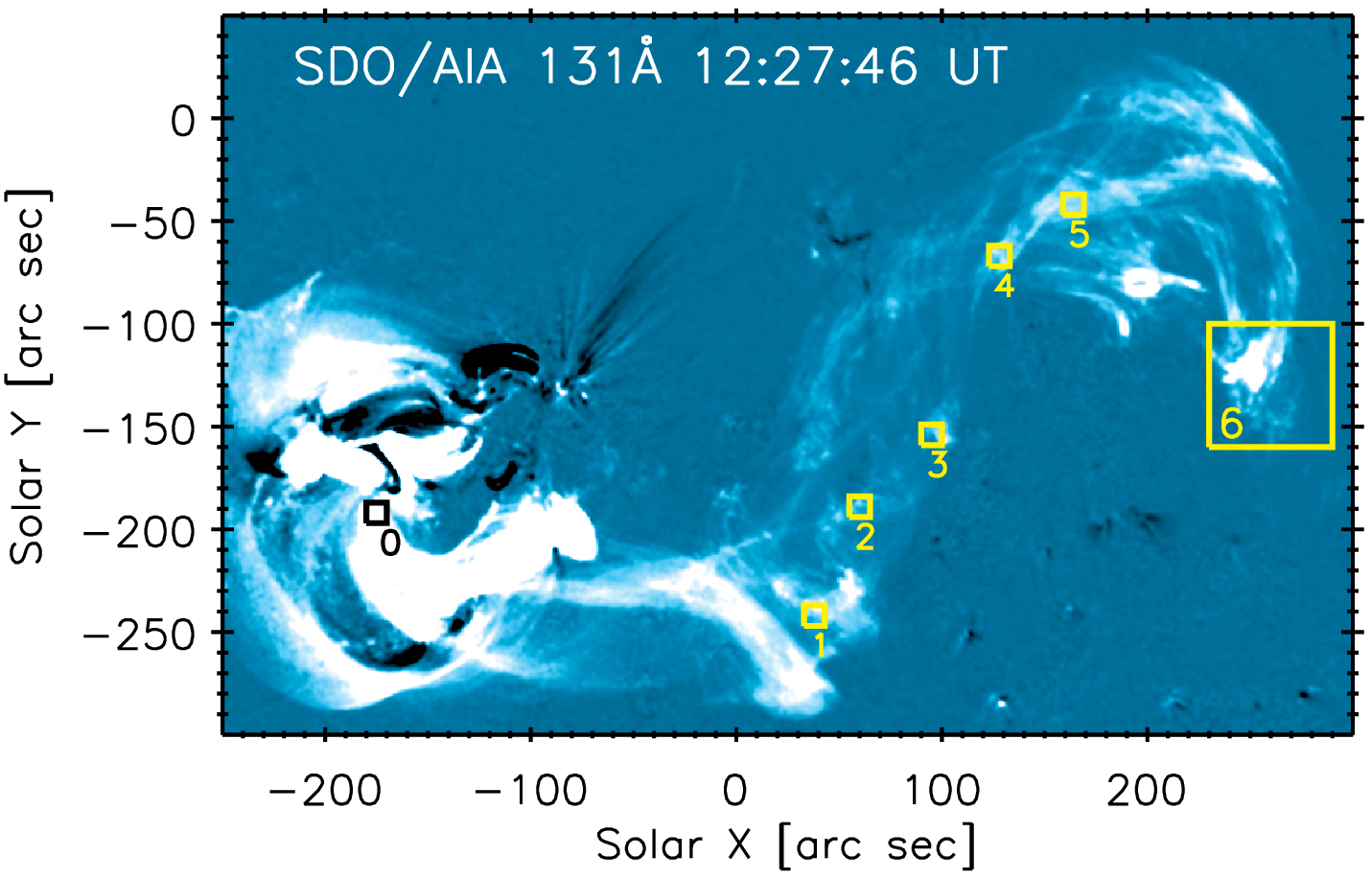}
        \put(-20,163){\textcolor{white}{\bf b)}}

\caption{Base difference of 131\,\AA~\textit{SDO}/AIA image at 12:27:46\,UT with
a base image taken at 11:30:10\,UT. The original image is in Figure~\ref{fig_evol1}(g).
Panel (a) shows newly created flare loops (arrows) and newly created twisted loops (red and yellow dots).
(b) The same image as (a) but showing positions of brightenings observed along the F (Regions 1--6) and
Region 0, which was located at the hook of R1.} \label{fig_fl}
\end{figure*}

Figure~\ref{fig_fl}(a) shows the base difference image of Figure~\ref{fig_evol1}(g). We depicted
there parts of the bright twisted threads of F by yellow and red dots. It seems that most likely
they connect Region 0 as far as to Region 6 (Figure~\ref{fig_fl}(b)). So, the observations suggest
that after magnetic reconnection occurred above the cusp, the twisted threads of the magnetic rope
F reached closer to the center of the AR  (Region 0). We also noticed further systems of flare loops
(Figure~\ref{fig_fl}(a)): within the AR (black arrow), below the cusp and also in the vicinity of Region
1 (yellow horizontal arrows). We used the observed activity in the preexisting
magnetic rope F to track the changes in connectivity. We do it by analysis of light curves (LCs) of Regions 1--6.

\begin{figure*}
       \centering
      \includegraphics[width=8.75cm,clip,viewport=0 0 485 328]{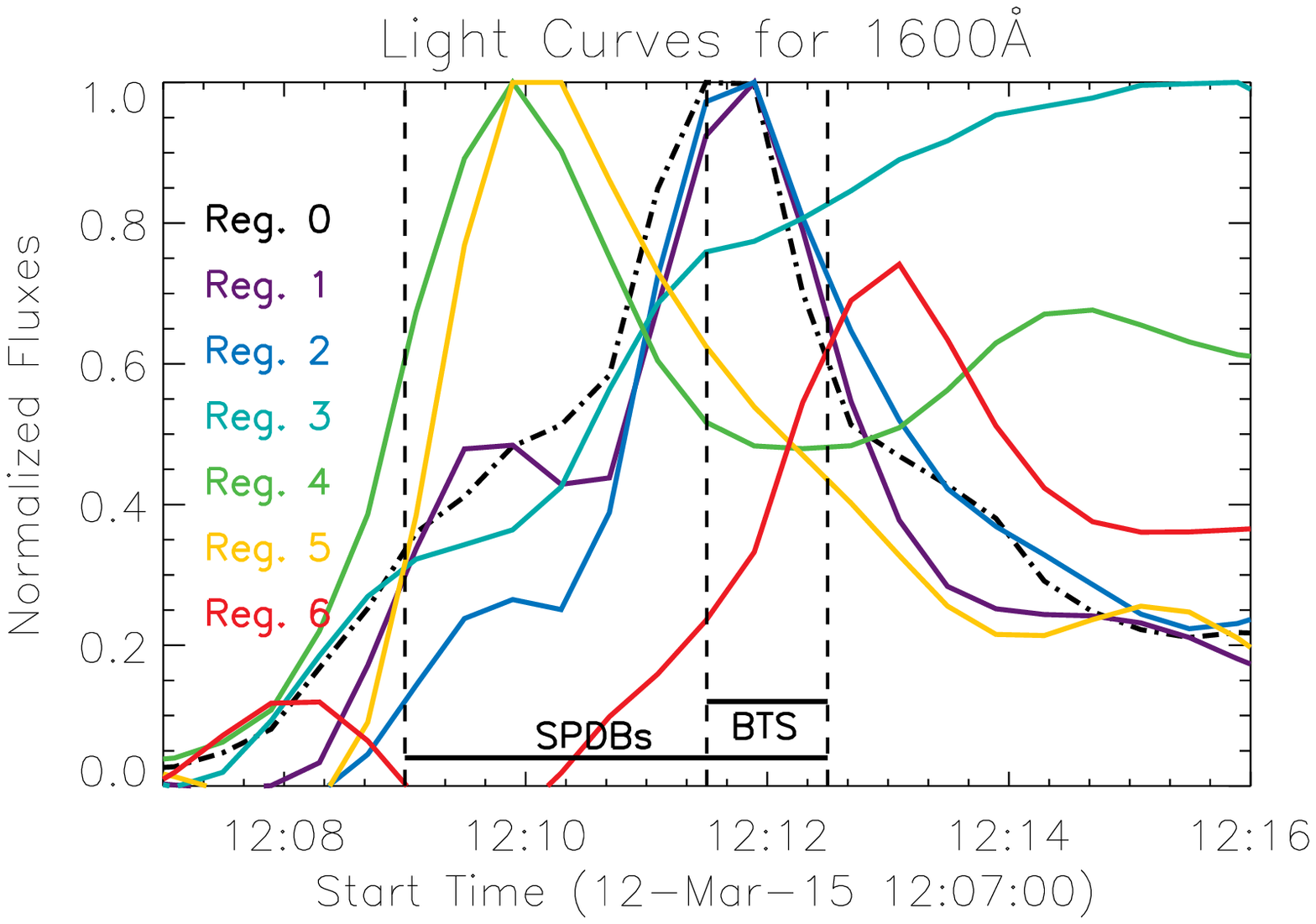}
      \put(-200,140){\textcolor{black}{\bf a)}}
      \includegraphics[width=8.75cm,clip,viewport=0 0 485 328]{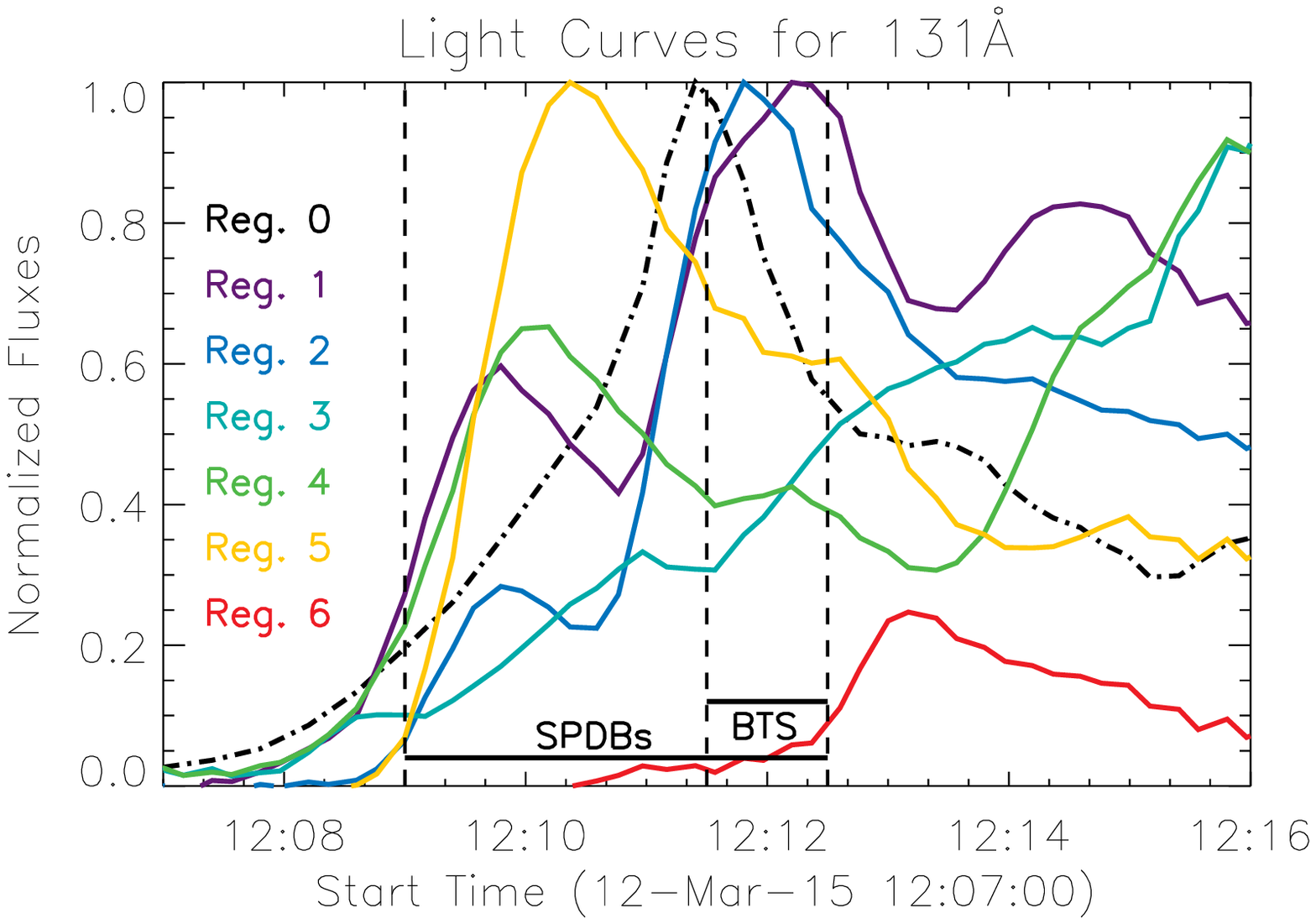}
       \put(-200,140){\textcolor{black}{\bf b)}}

      \includegraphics[width=8.75cm,clip,viewport=0 0 485 328]{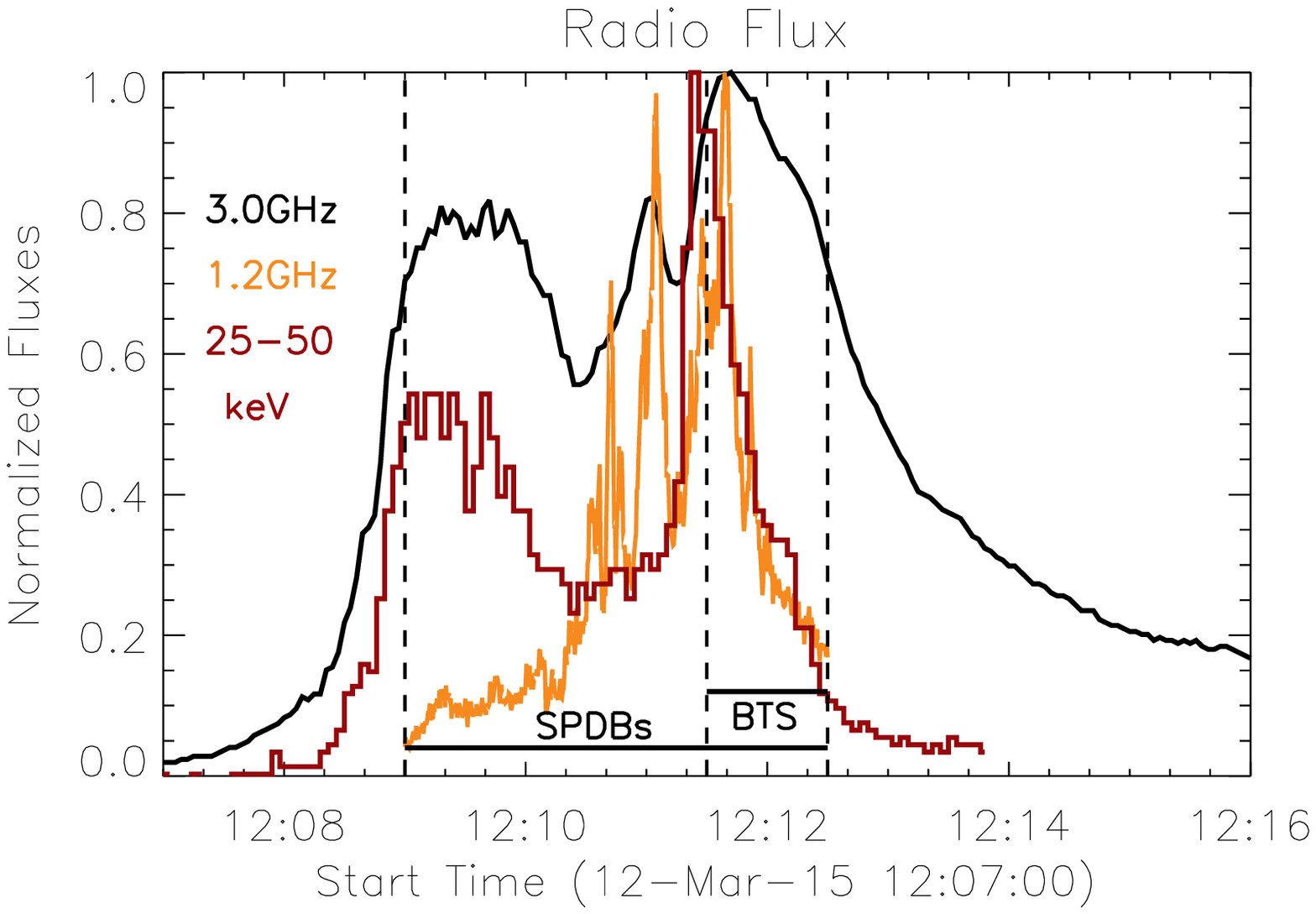}
     \put(-200,140){\textcolor{black}{\bf c)}}
     \includegraphics[width=8.75cm,clip,viewport=0 0 485 328]{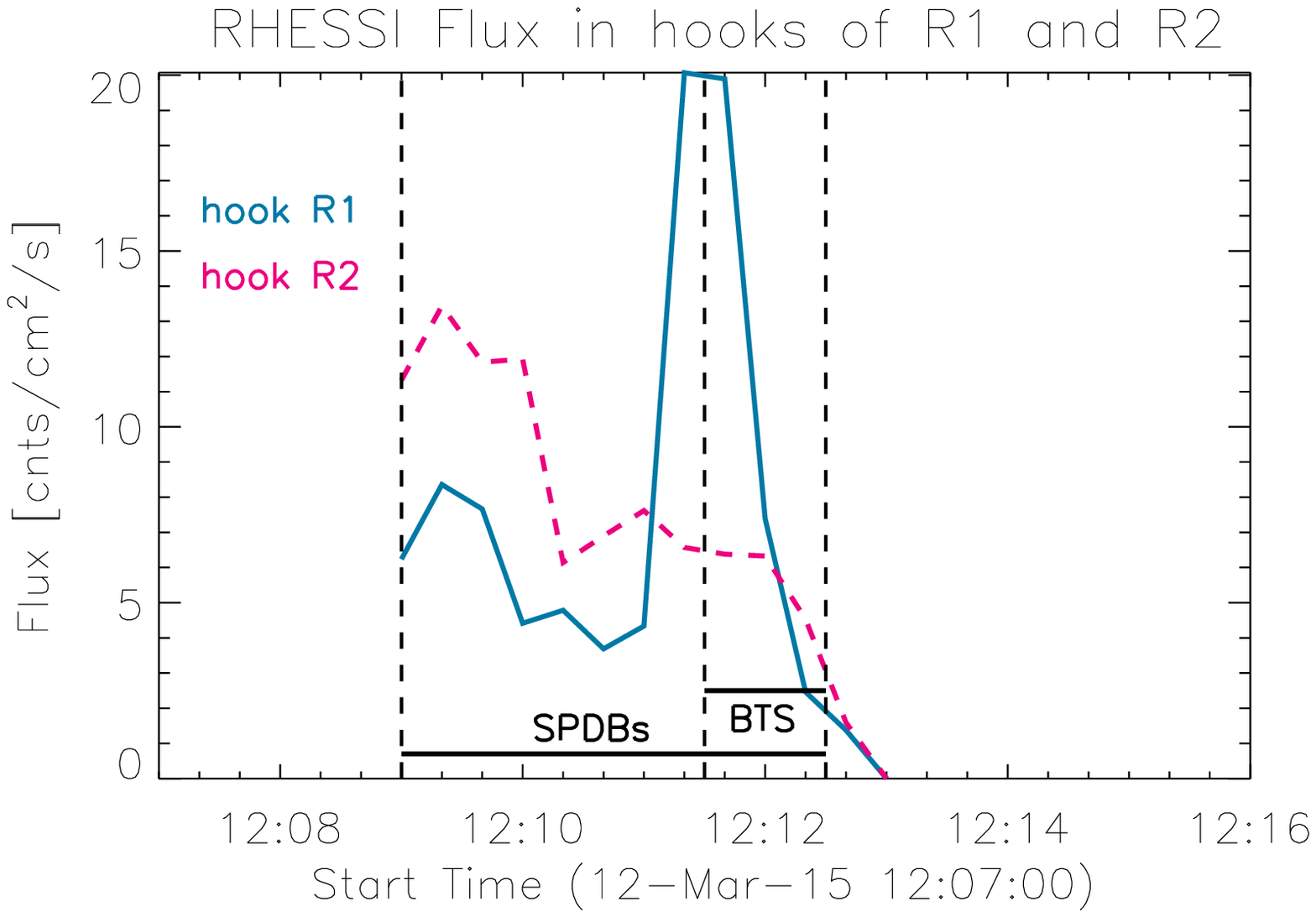}
     \put(-200,140){\textcolor{black}{\bf d)}}
        \caption{The light curves. (a) Normalized 1600\,\AA~light curves for brightenings located
         along the F and for the Region 0 located in the hook of R1.
         (b) The same as in (a) but for the 131\,\AA~filter. (c) Light curves in radio 3\,GHz (black), 1.2\,GHz
         (orange), and hard X-ray emission in 25--50\,keV (brown). (d) Light curves for hard X-ray
         sources observed in the hooks of R1 (blue) and  R2 (purple) in \textit{RHESSI} 25--50\,keV channel.
         Vertical dashed lines and their horizontal labels show time intervals when SPDBs (12:09:00--12:12:30\,UT)
         and BTS (12:11:30--12:12:30\,UT) were observed.}\label{fig_lc}
\end{figure*}

The UV/EUV LC for brightenings observed along the F are shown in Figures~\ref{fig_lc}(a) and (b). We added
there also an LC for Region 0 that was located approximately at the position of the hard X-ray source observed
inside the hook of R1 (Figure~\ref{fig_evol1}(b)). We did not constructed the UV/EUV LC for the hard X-ray
source observed in the hook of R2 because it moved a longer distance along the hook. Then in UV/EUV such
LC would not present a signal from a structure in a compact area as the other curves do. Each of the UV/EUV
light curves was constructed for the time interval 12:00--13:00\,UT and was normalized to its maximum within
this interval. Figure~\ref{fig_lc}(c) shows LCs for solar radio flux registered in 3\,GHz (gyrosynchrotron emission),
in 1.2\,GHz (plasma emission mechanism), and in hard X-rays at 25--50\,keV. Figure~\ref{fig_lc}(d) shows LCs
for individual hard X-ray sources observed in the hooks of ribbon R1 (blue) and R2 (red). At the bottom of each
panel in the Figure~\ref{fig_lc} we put labels to mark the times when SPDBs and narrow BTS were observed. It seems
that footpoints of the first flare loops, produced by early magnetic reconnection above the cusp, could be located
in Regions 1, 2, 4, and 5. Their UV (1600\AA) LCs (Figures~\ref{fig_lc}(a)) show maxima or local maxima about
12:10\,UT, which is about the first peak of 3\,GHz and 25--50\,keV emission (Figure~\ref{fig_lc}(c)). At this
time also weak SPDBs appear (Figure~\ref{fig_lc}(c), orange curve). This is valid also for EUV LCs
(Figures~\ref{fig_lc}(b)) but the aforementioned maxima or local maxima are slightly shifted in time due to
heating of plasma. About 12:12\,UT the UV/EUV LCs for Regions 1 and 2 (Figures~\ref{fig_lc}(a) and (b)) reached
their maxima. This is when the second peak in 3\,GHz occurs and when BTS forms and exists, and when the intensive
SPDBs (Figure~\ref{fig_lc}(c)) are observed. We have to note that at this time the Regions 1--5 were already
contaminated by emission of flare loops seen in EUV. Further, both hard X-ray sources located within the hooks
of R1 and R2 show small peaks before 12:10\,UT (Figure~\ref{fig_lc}(d)), during the first peak of 3\,GHz emission.
We can deduce that at this time of early reconnection they were still magnetically connected. But the hard
X-ray source located in the hook of R1 exhibits the maximum of the LC at 12:11:30\,UT and thus dominantly
contributes to the second peak of global hard X-ray emission in Figure~\ref{fig_lc}(c), appearing at the
time of the second 3\,GHz maximum. In UV/EUV the LC for the Region 0 corresponds to this source (Figures~\ref{fig_lc}(a)
and (b)). Finally, the UV/EUV LCs for Region 3 (Figures~\ref{fig_lc}(a) and (b)) have different character
compared to the other regions. They were continually rising and reached their maximum after 12:16\,UT when
plasma bulk passed through this region. It seems that this region likely represented a part of a loop.
Contrary to Region 3, both UV and EUV LCs for Region 6 show the local maximum about 12:13\,UT. This corresponds
to the appearance of the flare ribbon there (Figure~\ref{fig_evol1}(e) and Figure~\ref{fig_fl}(b)).

Summarizing the analysis of LCs, it comes out that at the beginning of the flare the hooks of R1 and R2
were connected by activated EUV loops (Figure~\ref{fig_evol1}(a)). Then, due to magnetic reconnection
within the region above the cusp new flare loops appeared in vicinity of Regions 1 and 2
(Figure~\ref{fig_evol1}(g)) and a new connectivity was also established between Regions 0 and 6
(Figure~\ref{fig_evol1}(g)). The latter two regions represent the hook of R1 and the far end of the magnetic
rope F (Figure~\ref{fig_fl}(b)). \\

\section{INTERPRETATIONS AND DISCUSSIONS}

\begin{figure*}
\begin{center}
\includegraphics[width=12cm]{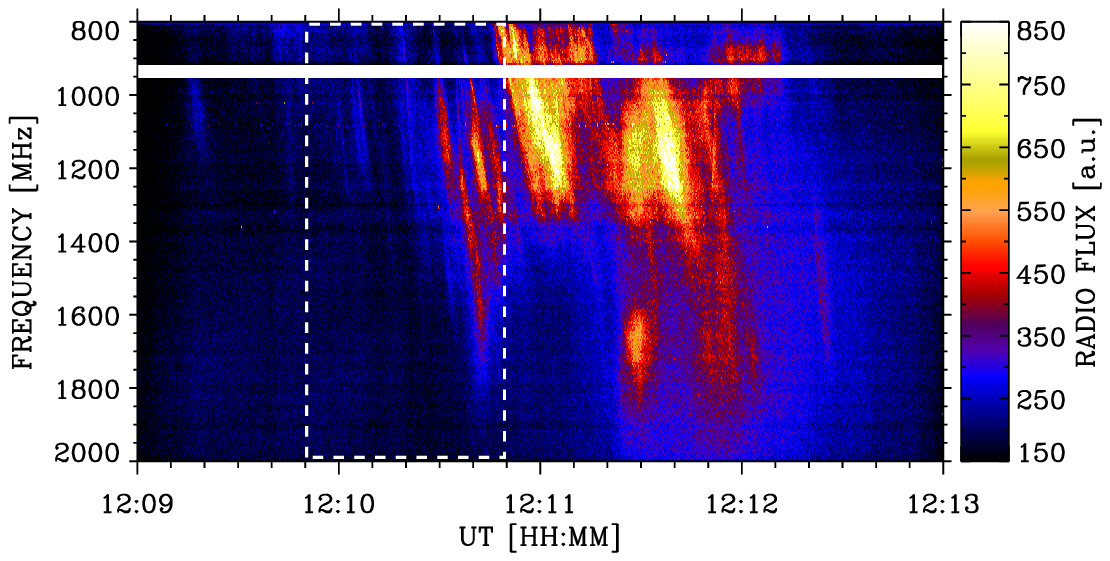}
\put(-295,35){\textcolor{white}{\bf a)}}

\includegraphics[width=12cm]{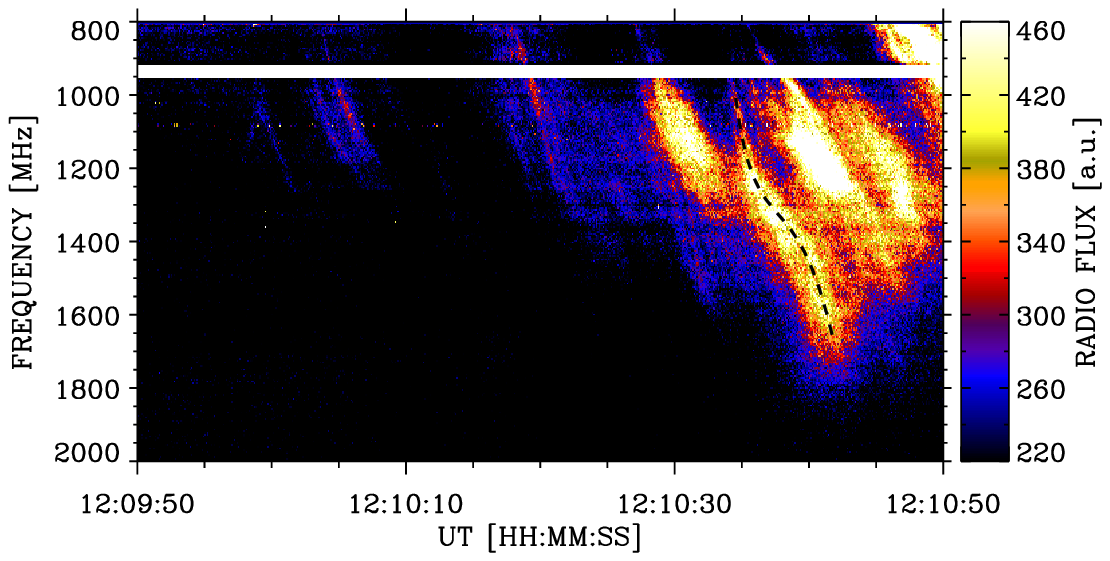}
\put(-295,35){\textcolor{white}{\bf b)}}
\put(-100,60){\textcolor{white}{\bf S}}
\includegraphics[width=3.3cm,clip,viewport=0 8 170 340]{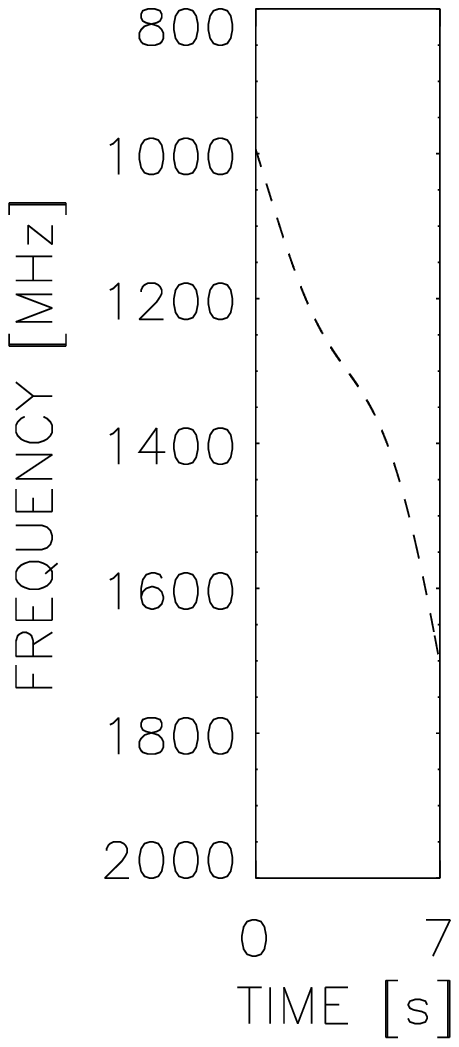}
\put(-24,155){\textcolor{black}{\bf c)}}
\end{center}
    \caption{Panel (a) shows the 800-2000 MHz radio spectrum observed at 12:09--12:13 UT
    during the 12 March 2015 flare by the Ond\v{r}ejov radiospectrograph.
    The white narrow horizontal band at about 950 MHz means no data. Panel (b) is the zoom
    of the spectrum during the 12:09:50 - 12:10:50 UT time interval (dashed-line box in Panel (a)).
    The dashed line and the letter S point out the S-like form of Burst S.
    Panel (c) shows the simulated S-like form of the Burst S.} \label{fig_radio}
\end{figure*}

In the previous section we analyzed the flare evolution with its expansion to the
preexisting magnetic rope (Figure~\ref{fig_evol1}) and phenomena accompanied
by unusual SPDBs (Figure~\ref{fig_radio}). This section is dedicated to
interpretation of our results and to their discussion.

Considering the existence of the hot cusp (Figure~\ref{fig_cusp}), ejection of
heated plasma and appearance of BTS (Figures~\ref{fig_evol2}), all accompanied
by SPDBs (Figure~\ref{fig_radio}), we argue that all those phenomena
are related to the process of magnetic reconnection. If the reconnection was
the case, then it had to occur in the vicinity of the cusp, most likely in the
region above it (Figure~\ref{fig_reco}). Then, accelerated particles
were ejected from the reconnection region and they followed the newly
created field lines. Hitting the lower atmosphere they produced new ribbons
(Figure~\ref{fig_evol1}(e) and (f)). In accordance with this proposed scenario, we
observed arcades of hot flare loops below the cusp (Figure~\ref{fig_evol1}(g))
and brightenings that appeared along the preexisting magnetic rope with the
embedded filament outside the AR (Figure~\ref{fig_evol1}). The bright twisted
threads (Figure~\ref{fig_fl}) document the involvement of the large filament F.

Now a question arises: what is the origin of SPDBs that appeared at the
beginning of the plasma expansion? According to their appearance in the radio
spectrum, they resembled the RS type III bursts. But their frequency drift is
much smaller. To learn more about these bursts, we analyzed the Burst S, the
one with the S-like form (Figure~\ref{fig_radio}(b)) in detail. Using the
Aschwanden's model of densities of the solar atmosphere \citep{Aschwanden2002}
and the assumption of the plasma emission on the fundamental frequency or on
the second harmonic, the velocity of the particle beam generating
Burst S is about 2040 and 3650 km s$^{-1}$, respectively. Note that this
velocity means the velocity in the vertical direction in the gravitationally
stratified solar atmosphere. The corresponding plasma densities are 1.23
$\times$ 10$^{10}$ -- 3.56 $\times$ 10$^{10}$ cm$^{-3}$ (for the emission on
the fundamental frequency) or 3.08 $\times$ 10$^{9}$ -- 8.92 $\times$ 10$^{9}$
cm$^{-3}$ (for the emission on the harmonic frequency). The estimated
velocities are too small for the electron beam considered in the (RS)
type III bursts. We note that in the present 800-2000 MHz frequency
range there is no systematic study of beam velocities for reverse type III
bursts due to their high frequency drift and limited time resolution of
radiospectrographs. Nevertheless, these velocities need to be greater than 0.15
c \citep{Aschwanden2002} and their typical velocity should be the same as for
type III bursts, i.e. 0.3 c \citep{Krueger1979}, where $c$ is the speed of
light, due to an expected symmetry in acceleration of beams in upward and
downward directions in the solar atmosphere.

However, based on the simultaneous start of SPDBs and first brightenings along the preexisting
magnetic rope, we assume that the beam propagates from the cusp region through the rope that
is nearly horizontally oriented. In such a case the frequency drift can correspond to much
higher beam velocities depending on the angle between the rope axis and the horizontal plane.
This interpretation is supported by an unusual S-like form of Burst S. Namely,
if the particle beam propagates along the helical structure of the nearly horizontal magnetic
rope then such an S-like form of the burst can be expected. It looks that the S-like burst
form corresponds to the beam trajectory with the 2$\pi$ twist angle.

We simulated this S-like form of the Burst S. We assumed the cylindrical magnetic rope with
the length 210000 km, which roughly corresponds to the distance between the cusp and the
detection point 5 in Figure~\ref{fig_evol1}(b).  The axis of the rope has some small angle
$\alpha$ to the photospheric plane. The rope is embedded in the gravitationally stratified
solar atmosphere, the density of which changes with the height scale corresponding to the
temperature T; H$_s$[m] = 50 T[K]~\citep{Priest1982}. We suppose that in this cylindrical
rope a particle beam moves downwards along the magnetic field line having the helical
form with the radius R around the rope axis. Along its trajectory the beam generates the
local electrostatic (Langmuir) waves by the beam-plasma instability. These waves are then
transformed by nonlinear processes to the electromagnetic (radio) waves observed as the
Burst S. Here, for simplicity we calculated beam trajectories only for the emission on the
fundamental frequency. Duration of Burst S is 7s which gives the beam velocity parallel
to the rope axis as $v_{\parallel}$=30000 km s$^{-1}$. The result of the beam induced
emission is shown in Figure~\ref{fig_radio}(c) by the dashed line. A qualitative agreement
with observations was found for the parameters shown in Table~\ref{tab1}. In all these cases
(A, B, and C) the same result (Figure~\ref{fig_radio}(c)) was obtained. This shows that
increasing the temperature $T$ in the solar atmosphere and thus the height scale requires an
increase of the radius R, the beam velocity $v_{\perp}$ perpendicular to the rope axis, and
increase of the angle $\alpha$ between the rope axis and horizontal plane. Due to no additional
information we cannot decide which case of A, B, and C is the most appropriate to observations.

\begin{table}
\centering
\begin{tabular}{c|c|c|c|c|c}
\hline\hline
Case & T (K) & Height scale H$_s$ (km) & $v_{\perp}/v_{\parallel}$ & R (km) & $\alpha$ (degree) \\
\hline
 A &1.5 $\times$ 10$^5$ &7500 &0.02 & 670  & 2.2   \\
 B &3.9 $\times$ 10$^5$ &19500 &0.05 & 1660 & 5.7   \\
 C &1 $\times$ 10$^6$ &50000 &0.1 & 3330 & 14.3   \\
\hline
\end{tabular}
\caption{Parameters of the magnetic rope corresponding to the S-like form of
Burst S. $T$ is the temperature in the solar atmosphere,
$v_{\parallel}$ = 30000 km s$^{-1}$ and $v_\perp$ is
the velocity parallel and perpendicular to the rope axis, R is the radius
around which the agent rotates in the helical magnetic rope, and $\alpha$
means the angle between the rope axis and horizontal plane. All these cases
give the same result (Figure~\ref{fig_radio}(c)).} \label{tab1}
\end{table}

Now, we first consider an electron beam as the beam generating Burst S. The
Burst S lasts 7 s and is produced in relatively high plasma densities, where
collisions are not negligible. Therefore, let us estimate the stop time for
electrons with the velocity 30000 km s$^{-1}$ in the plasma with the density of
2.4 $\times$ 10$^{10}$ cm$^{-3}$ (the mean density in the Burst S source for
the emission on the fundamental plasma frequency). (We note that the total
velocities in the cases A, B, and C (Table~\ref{tab1}) are slightly higher than
30000 km s$^{-1}$ due to the presence of the small component perpendicular to the
rope axis.) Using the relations \citep{Karlicky1990}
\begin{equation}
z_\mathrm{stop} = \frac{E_0^2}{3 K \Lambda n_e},
\label{eq1}
\end{equation}
\begin{equation}
t_\mathrm{stop} = \frac{2 z_\mathrm{stop}}{v_{0}},
\label{eq2}
\end{equation}
where $z_\mathrm{stop}$ and $t_\mathrm{stop}$ are the stop distance and stop
time for electrons with the initial energy $E_0 = 1/2 m_e v_0^2$, $m_e$ is the
electron mass, $v_0$ is the initial electron velocity, $K = 2 \pi e^4$, $e$ is
the electron charge, $\Lambda$ is the Coulomb logarithm, and $n_e$ is the
plasma density, we found that the stop time is 0.023 s. In the case, when the
emission of Burst S is on the harmonic frequency, i.e., the electrons propagate
in the plasma with the mean density of 6$\times$10$^9$ cm$^{-3}$, the stop time
is 0.093 s. In the both cases the duration of Burst S is much longer than the
beam stop time, that is why the electron beam would be destroyed in a
shorter time than the duration of the Burst S is, and thus the electron beam cannot
generate Burst S. Therefore, let us consider the proton or neutral (current
neutralized) beam. The neutral beam is the beam of protons and electrons having
the same velocity. It is formed in the beam-plasma system owing to the
inductive electromagnetic effects~\citep{Simnett1990}.

Because for the both proton and neutral beams the stop time is mainly given by
protons, we calculate the stop time only for the protons. Modifying the
relations~(\ref{eq1}) and~(\ref{eq2}) by the factor 1.5 $m_p/m_e$
\citep{Emslie1978}, where $m_p$ is the proton mass, the stop time for the
protons with the velocity 30000 km s$^{-1}$ in the plasma with the mean density
2.4 $\times$ 10$^{10}$ cm$^{-3}$ (the emission on the fundamental frequency) is
64.1 s and in the plasma with the mean density 6$\times$ 10$^9$ cm$^{-3}$ (the
emission on the harmonic frequency) is 256.3 s. Both the times are much longer
than the duration of Burst S (7s). This means that the proton or neutral
beam keeps its form during the time of Burst S (beams are not scattered during
this time). Thus, these beams can generate the Langmuir waves along their
trajectory downwards in the solar atmosphere in a similar way as the electron
beam \citep{Karlicky1998}. The Langmuir (electrostatic) waves are then
transformed by nonlinear processes into electromagnetic (radio) waves (Burst
S), similarly as in models of type III bursts. After the generation of Burst S, the
proton or neutral beam continues in its propagation downwards. However, in
lower atmospheric layers conditions for generation of the Langmuir waves and/or
their transformation to the radio waves and/or radio wave
absorption/propagation are changed and thus no further radio emission is
observed. Finally, when the proton or neutral beam enters to even deeper and
more dense atmospheric layers, these beams are collisionally stopped. In such a
way, we think that the proton or neutral beam can generate the Burst S. The
energy of the protons in this case is 4.69 MeV. This agrees with the results
presented by \cite{Benz1986}. For the generation of radio bursts, the beams
with the densities, which are several orders of magnitude smaller than the
density of the background plasma, are sufficient \citep{Melrose1980}. Therefore,
no additional observational effects of such weak beams can be expected.

Because we have no information about positions of the SPDB radio
sources and as during SPDBs several systems of flare loops were formed, a
question arises if SPDBs are not connected to any of them. We cannot rule out
this possibility. However, it is known that during an evolution of a flare, both 
shear \citep{Aulanier2012_I} and twist \citep{Sych2015} in the flare 
loops decrease and the magnetic field becomes more or less potential.
Thus, in the flare loops the helical structure of the magnetic
field, which is needed for the explanation of the S-like form of one SPDB, is
not very probable. Moreover, systems of flare loops are frequently observed
phenomenon but the SPDBs are rarely observed. Furthermore, first SPDBs appeared
simultaneously with the first brightenings in the magnetic rope. Therefore, we
think that the probability of locations of the SPDB sources in the flare loops
is low.

BTS (Figure~\ref{fig_evol2}(g), (h), and Figure~\ref{fig_vel}) is another
interesting aspect of this flare. It evolved while the plasma followed the
twisted loops of the magnetic flux rope (Figure~\ref{fig_evol1}(g)) and
it appeared during some SPDBs. When a hot plasma expands into the location of
preexisting magnetic rope, its expansion can be in principle described (a) as a
simple plasma expansion into the passive magnetic field structure of the
magnetic rope or (b) as the dissipation-spreading process \citep{Norman1978}. In
case (a) BTS could originate due to a projection effect along the line of sight.
For example, a narrow structure could appear when a bulk of heated plasma is
ejected along the highly twisted magnetic field lines of a flux rope, the axis
of which is oriented almost perpendicular to the line of sight. To verify this
possibility we estimated the angle of twisted loop compared to the rope axis
observed in Figure~\ref{fig_fl}(a) (yellow dots). We found the angle to be about 40
degrees and such an angle is not sufficient to produce a narrow structure in
projection. Moreover, a highly twisted magnetic flux rope would be unstable and
the observed one is stable. In case of (b), when considering the
dissipation-spreading process, physically it means that during the expansion of
hot flare plasma new energy-release processes appear at new locations within
the preexisting magnetic rope. The dissipation process can spread owing to
particle beams and/or magnetosonic waves \citep{Karlicky1989,Odstrcil1997}. In
this case BTS could be formed due to an additional magnetic reconnection
(triggered by a plasma expansion) between long loops of the magnetic rope and
shorter loops that are transverse to the rope axis and located below the rope
in the region close to BTS (Figure~\ref{fig_evol2}(b)--(c) and
(f)--(g)). The magnetic reconnection heats the shorter loops and forms BTS. The
kinetic pressure inside the loops of BTS increases and due to the ballooning
instability \citep{Shibasaki2001} BTS expands and disrupts. Then we see a motion of BTS
and plasma motion after BTS disruption. This interpretation can be supported by
the second 3 GHz maximum observed about 12:11:40\,UT, i.e., at the time of BTS,
which indicates some additional energy release as expected in the
dissipation-spreading process.

In \cite{Karlicky2015} an example of one SPDB was presented in association with
the falling bright EUV blob and H$\alpha$ chromospheric response. That SPDB was
interpreted as a radio signature of a thermal front. In the present case we find
some similarities with those observations. During the occurrence of several
strong SPDBs about 12:11--12:12\,UT we observed bright and multi-thermal BTS
(Figure~\ref{fig_multi_t}) moving along a cold filament. The BTS had a rather broad
EM(T) curve (Figure~\ref{fig_em}(b)) that peaked at $log\,T\,[K]=6.4$, i.e. at
T$\sim$2.5$\times$10$^6$ K. Furthermore, the average value of estimated projected
velocity of BTS, 240 $\pm$50 km\,s$^{-1}$, corresponded to the sound speed in plasma
with T$\sim$2.5$\times$10$^6$ K. However, in the present case, we did not
observe any particular cold preexisting loops into which a hot plasma was injected,
and few SPBDs, including Burst S, were  registered before the BTS was formed
(12:09:10--12:11:20\,UT, Figure~\ref{fig_radio}(a)). Moreover, it would be very difficult
to explain the SPDB with the S-like form (Burst S) by the thermal front. We
cannot exclude that the other SPDBs (except Burst S) were of different origin; however
we suggest that all SPDBs are produced by proton beams.\\

\section{CONCLUSIONS}

We report on a rare observation of slowly positively drifting radio bursts
(SPDBs) in the 0.8-1.2 GHz range. Since the radio spectrum
does not provide any information about the location of radio source,
we investigated the EUV and X-ray observations of the associated
flare in order to search for their origin.

The flare started close to the sunspot within the active region and then it
expanded into a huge preexisting magnetic rope with a cold filament. We found
that this expansion was accompanied by SPDBs and the bright transverse structure to the
rope axis (BTS).

We recognized the small cusp magnetic field structure during the impulsive
phase of the flare that looks to be a source of
the particle acceleration and plasma expansion into the rope.
The unique SPDB with S-like form that
occurred during this time was simulated by considering a propagation of the
particle beam in the helical magnetic field structure of the rope, which is
nearly horizontally oriented. This burst lasted 7 s in a relatively dense
plasma, where the particle collisions are not negligible. Considering these
collisions we interpret this burst as caused by the proton or neutral beam with
energies about 4.7\,MeV. Our interpretation is in a good agreement with the
works by \cite{Benz1986} and \cite{Simnett1995}, who explain the slowly
drifting type III bursts in a similar way.

Furthermore, during an expansion of the plasma into the preexisting magnetic
rope we found an unusual narrow EUV bright structure located transverse to
the rope axis BTS. This structure was observed in a broad range of temperatures
and moved along the filament with the velocity of about 240 km s$^{-1}$. The
formation and dissipation of this structure has been observed at times of some
SPDBs, but their mutual relation remains unclear. The bright EUV structure was
interpreted considering line-of-sight effects and the dissipation-spreading
process, which we found more likely to occur.

\acknowledgments
We acknowledge support from the project RVO:67985815 and Grants: 18-09072S, 19-09489S,
20-09922J and 20-07908S of the Grant
Agency of the Czech Republic. AIA and HMI data are provided courtesy of
NASA/SDO and the AIA and HMI science teams. We thank the RHESSI team for
providing the data and software support. Hinode is a Japanese mission
developed and launched by ISAS/JAXA, with NAOJ as domestic partner and NASA
and STFC (UK) as international partners. It is operated by these agencies
in co-operation with ESA and NSC (Norway). This work utilizes GONG data from
NSO, which is operated by AURA under a cooperative agreement with NSF and
with additional financial support from NOAA, NASA, and USAF.

\bibliographystyle{aasjournal}
\bibliography{mss_references}

\newpage

\end{document}